\documentclass{aa}  

\usepackage{graphicx}
\usepackage{txfonts}
\usepackage{hyperref}
\usepackage{subcaption}
\usepackage{xcolor}
\usepackage{gensymb}
\usepackage{array}
\usepackage[export]{adjustbox}
\usepackage{bm}

\newcounter{lenote}

\newcommand{\oiii}{[O\,\textsc{iii}]}
\newcommand{\nii}{[N\,\textsc{ii}]}
\newcommand{\sii}{[S\,\textsc{ii}]}
\newcommand{\oi}{[O\,\textsc{i}]}

\newcommand{\hb}{H$\beta$}
\newcommand{\ha}{H$\alpha$}
\newcommand{\feii}{[Fe\,\textsc{ii}]}

\begin{document}

   \title{MAGNUM survey: Compact jets causing large turmoil in galaxies}

   \subtitle{Enhanced line widths perpendicular to radio jets as tracers of jet-ISM interaction\thanks{Based on observations made with ESO Telescopes at the La Silla Paranal Observatory under program IDs 094.B-0321, 60.A-9339, 095.B-0532.}}

   \author{G. Venturi
          \inst{1}\fnmsep\inst{2},
          G. Cresci\inst{2},
          A. Marconi\inst{3}\fnmsep\inst{2},
          M. Mingozzi\inst{4},
          E. Nardini\inst{3}\fnmsep\inst{2},
          S. Carniani\inst{5}\fnmsep\inst{2},
          F. Mannucci\inst{2},
          A. Marasco\inst{2},
          R. Maiolino\inst{6}\fnmsep\inst{7}\fnmsep\inst{8},
          M. Perna\inst{9}\fnmsep\inst{2},
          E. Treister\inst{1},
          J. Bland-Hawthorn\inst{10}\fnmsep\inst{11}
          \and
          J. Gallimore\inst{12}
          }

   \institute{Instituto de Astrofísica, Facultad de Física, Pontificia Universidad Católica de Chile, Casilla 306, Santiago 22, Chile\\
              \email{gventuri@astro.puc.cl}
         \and
             INAF - Osservatorio Astrofisico di Arcetri, Largo E. Fermi 5, I-50125 Firenze, Italy\\
             \email{giacomo.venturi@inaf.it}
        \and
            Dipartimento di Fisica e Astronomia, Università degli Studi di Firenze, Via G. Sansone 1, 50019 Sesto Fiorentino, Firenze, Italy
        \and
            Space Telescope Science Institute, 3700 San Martin Drive, Baltimore, MD 21218, USA
        \and
            Scuola Normale Superiore, Piazza dei Cavalieri 7, I-56126 Pisa, Italy
        \and
            Cavendish Laboratory, University of Cambridge, 19 J. J. Thomson Ave., Cambridge CB3 0HE, UK
        \and
            Kavli Institute for Cosmology, University of Cambridge, Madingley Road, Cambridge CB3 0HA, UK
        \and
            Department of Physics and Astronomy, University College London, Gower Street, London WC1E 6BT, UK
        \and
            Centro de Astrobiología (CSIC-INTA), Departamento de Astrofísica, Cra. de Ajalvir Km. 4, 28850, Torrejón de Ardoz, Madrid, Spain
        \and
            Sydney Institute for Astronomy, School of Physics, The University of Sydney, Sydney, NSW 2006, Australia
        \and
            ARC Centre of Excellence for All Sky Astrophysics in Three Dimensions (ASTRO-3D), Canberra, ACT2611, Australia
        \and
            Department of Physics and Astronomy, Bucknell University, Lewisburg, PA 17837, USA
             }

   \date{Received September 15, 1996; accepted March 16, 1997}

 
  \abstract
   {Outflows accelerated by 
   active galactic nuclei (AGN) 
   are commonly observed in the form of coherent, mildly collimated high-velocity gas directed along the AGN ionisation cones and kinetically powerful ($\gtrsim\,$10$^{44-45}$~erg~s$^{-1}$) jets. 
   Recent works found that outflows can also be accelerated by low-power ($\lesssim\,$10$^{44}$~erg~s$^{-1}$) jets, and the most recent cosmological simulations indicate that these are the dominant source of feedback on sub-kiloparsec scales, but little is known about their effect on the galaxy host.
   }
   {We study the relation between radio jets and the distribution and kinematics of the ionised gas in \object{IC 5063}, \object{NGC 5643}, \object{NGC 1068,} and \object{NGC 1386} as part of our survey of nearby Seyfert galaxies called Measuring Active Galactic Nuclei Under MUSE Microscope (MAGNUM). All these objects host a small-scale ($\lesssim$\,1~kpc) low-power ($\lesssim\,$10$^{44}$~erg~s$^{-1}$) radio jet that has small inclinations ($\lesssim$\,45$\degree$) with respect to the galaxy disc.}
   {We employed seeing-limited optical integral field spectroscopic observations from the Multi Unit Spectroscopic Explorer (MUSE) at the Very Large Telescope (VLT) to obtain flux, kinematic, and excitation maps of the extended ionised gas. 
   We compared these maps with archival radio images and in one case, with {\it Chandra} X-ray observations.}
   {We detect a strong (up to $\gtrsim$\,800$-$1000~km/s) and extended ($\gtrsim$\,1~kpc) emission-line velocity spread perpendicular to the direction of the AGN ionisation cones and jets in all four targets. The gas excitation in this region of line-width enhancement is entirely compatible with shock ionisation. These broad and symmetric line profiles are not associated with a single coherent velocity of the gas.  A `classical' outflow component with net blueshifted and redshifted motions is also present, but is directed along the ionisation cones and jets.
   %
   %
   }
   {We interpret the observed phenomenon as due to the action of the jets perturbing the gas in the galaxy disc. These intense and extended velocity spreads perpendicular to AGN jets and cones are indeed currently only observed in galaxies hosting a low-power jet whose inclination is sufficiently low with respect to the galaxy disc to impact on and strongly affect its material. 
   In line with cosmological simulations, our results demonstrate that low-power jets are indeed capable of affecting the host galaxy.
   }

    
    \titlerunning{MAGNUM survey: Compact jets causing \textit{\textup{large}} turmoil in galaxies}           
    \authorrunning{G. Venturi et al.}

   \maketitle
%

\begin{table*}[t]
        \caption{VLT/MUSE data employed.}
        \setlength{\extrarowheight}{0.5pt}
        \begin{tabular*}{\textwidth}{l @{\extracolsep{\fill}} c c c c}
                \hline\hline \\[-1em]
                Name                            &  Program ID / P.I.   &  $t_{\textrm{exp}}$\,$\times$\,n$_{\textrm{exp}}$\,=\,{\bf $\bm{T_{\textrm{exp}}}$ [s]}\tablefootmark{a}  & $t_{\textrm{exp}}^\textrm{sky}$\,$\times$\,n$_{\textrm{exp}}^\textrm{sky}$ [s]\tablefootmark{b} \\ \\[-1.1em]
                \hline \\[-0.9em]
                {\bf IC 5063}              &    60.A-9339 / Marconi/Hawthorn/Salvato            &               600\,$\times$\,4\,=\,{\bf 2400}           &        60\,$\times$\,2         \\
                {\bf NGC 5643}     &    095.B-0532 / Carollo    &               900\,$\times$\,4\,=\,{\bf 3600}           &        120\,$\times$\,2  \\
                    & 60.A-9339 / Marconi/Hawthorn/Salvato & 500\,$\times$\,4\,=\,2000\tablefootmark{c} & 100\,$\times$\,4 \\
                {\bf NGC 1068}     &    094.B-0321 / Marconi    &               500\,$\times$\,4\,+\,100\,$\times$\,8\,=\,{\bf 2800}\tablefootmark{d}     & \ \ 100\,$\times$\,4\,+\,100\,$\times$\,8  \\
                {\bf NGC 1386}     &    094.B-0321 / Marconi    &               500\,$\times$\,8\,=\,{\bf 4000}           &        100\,$\times$\,8  \\
                \hline
        \end{tabular*}
    \tablefoot{
    \tablefoottext{a}{Total exposure time on object $T_{\textrm{exp}}$, given by the combination of the single n$_{\textrm{exp}}$-times repeated exposures having duration of $t_{\textrm{exp}}$ each.}
        \tablefoottext{b}{Exposure time of each dedicated sky exposure times the number of exposures.}
        \tablefoottext{c}{After comparing the two datasets from program 60.A-9339 and 095.B-0532 for NGC 5643, we employed the observations from program 095.B-0532 here because their seeing is superior to that from program 60.A-9339.}
        \tablefoottext{d}{The first OB was in the central spaxels affected by emission lines that were saturated or in the non-linear response regime of the instrument. The second observing block (OB) was then acquired with shorter exposure times (100s instead of 500s), and only this second OB was adopted in the central spaxels.}
        }
\label{table:tb}
\end{table*}

\begin{table*}[t]
        \caption{Basic information about the four galaxies presented.}
        \setlength{\extrarowheight}{0.5pt}
        \begin{tabular*}{\textwidth}{l @{\extracolsep{\fill}} c c c}
                \hline\hline \\[-1em]
                Name                            &  Distance [Mpc]\tablefootmark{a} &  FOV extent [kpc$^2$]\tablefootmark{b}  & Spatial scale [pc/arcsec]\tablefootmark{c} \\ \\[-1.1em]
                \hline \\[-0.9em]
                {\bf IC 5063} & 46\,$\pm$\,7 & 14$\,\times\,$14 & 220 \\ 
                {\bf NGC 5643} & 16\,$\pm$\,7 & 5$\,\times\,$5 & 78 \\ 
                {\bf NGC 1068} & 10.5\,$\pm$\,1.7 & 3.3$\,\times\,$3.3 & 51 \\ 
                {\bf NGC 1386} & 16.4\,$\pm$\,0.8 & 5.1$\,\times\,$5.1 & 80 \\ 
                \hline
        \end{tabular*}
        \tablefoot{
        \tablefoottext{a}{Distance of the galaxy from Earth, obtained from \href{HyperLeda}{http://leda.univ-lyon1.fr/} best distance modulus, i.e. the weighted average between the redshift distance modulus corrected for infall of the Local Group towards Virgo and the weighted average of the published redshift-independent distance measurements.}
    \tablefoottext{b}{Central portion of the galaxy covered by the MUSE $\sim$1$'\times$1$'$ FOV.}
    \tablefoottext{c}{Spatial scale at the distance of the galaxy.}
        }
\label{table:fov}
\end{table*}

\section{Introduction}
Outflows and jets accelerated by active galactic nuclei (AGN) are considered to have an important role in galaxy evolution (feedback effect; e.g. \citealt{Fabian:2012aa} for a review). In the standard picture of AGN-driven winds, the AGN radiation pressure and/or magnetic fields can accelerate outflows that are able to expel large quantities of gas from galaxies and consequently deplete the gas reservoir needed to form stars (so-called radiative or quasar-mode feedback), leaving a red and dead galaxy, while jets in a following and longer phase keep the gas in the galaxy halo hot, preventing re-accretion and re-ignition of star formation (so-called kinetic or radio-mode feedback). However, strong jets can also cause the acceleration of powerful outflows by pushing the gas in their direction of propagation (e.g. \citealt{Nesvadba:2008aa}, \citealt{Vayner:2017aa}). Some studies (e.g. \citealt{Combes:2013aa}, \citealt{Garcia-Burillo:2014aa}, \citealt{Cresci:2015aa}, \citealt{Harrison:2015aa}, \citealt{Morganti:2015aa}, \citealt{Jarvis:2019aa}, \citealt{Molyneux:2019aa}) revealed that this phenomenon can occur not only in the traditional radio-loud objects\footnote{The radio loudness is commonly defined based on $R$~=~$L_{\nu_\mathrm{5\,GHz}}/ L_{\nu_{\mathrm{4400\,\AA}}}$, the ratio of the monochromatic radio luminosity at 5 GHz and the optical luminosity at 4400 \AA. An object is defined as radio loud when $R$\,$>$\,10 and as radio quiet otherwise (e.g. \citealt{Kellermann:1989aa}). Alternatively, objects are classified as radio quiet when their radio luminosity (either at 5 GHz, e.g. \citealt{Kellermann:1994aa}, \citealt{Xu:1999aa}, or at 1.4 GHz, e.g. \citealt{Padovani:2017aa}, \citealt{Wylezalek:2018aa}) is $\lesssim\,$10$^{24}$~W/Hz.}, which have powerful jets, but also in galaxies hosting low-power\footnote{Throughout the paper, we mean the kinetic power of the jet when we write 'jet power' if not stated otherwise, not the power (luminosity) of its radio emission.}, compact radio jets, which are typically classified as radio-quiet. 
Some of these works showed an interaction between the low-power jet and the gas in the galaxy disc (e.g. \citealt{Garcia-Burillo:2014aa}, \citealt{Morganti:2015aa}, \citealt{Cresci:2015aa}).
The most recent cosmological simulations indeed suggest that the more common low-power jets may be the dominant source of feedback on sub-kiloparsec scales (e.g. \citealt{Weinberger:2017aa}, \citealt{Pillepich:2018aa}).

In this paper we present data of four nearby ($\lesssim$\,50 Mpc) Seyfert galaxies hosting low-luminosity radio jets obtained with the optical and near-IR spectrograph Multi Unit Spectroscopic Explorer (MUSE; \citealt{Bacon:2010aa}) at the Very Large Telescope (VLT), which is part of European Southern Observatory's (ESO) Paranal Observatory. In these four objects we find that the low-power jets appear to strongly interact with the gas in the disc, driving a peculiar turbulent phenomenon perpendicularly to their direction of propagation.

The four sources presented in this work belong to our survey called Measuring Active Galactic Nuclei Under MUSE Microscope (MAGNUM), which targets  nearby Seyfert galaxies with MUSE at VLT. An overview of the survey, which already produced a number of publications (\citealt{Cresci:2015aa}, \citealt{Venturi:2018aa}, \citealt{Mingozzi:2019aa}), can be found in \cite{Venturi:2017aa}. A detailed presentation of the survey will be the subject of a forthcoming paper (Venturi et al., in prep.).

\section{MUSE data description and analysis}
The MUSE data of the four galaxies belong to different programs, as detailed in Table \ref{table:tb}. Each exposure was dithered by 1$''$ and/or rotated by 90$\degree$ relative to one another. All the observations were acquired in seeing-limited Wide Field Mode (WFM), covering 1$'$\,$\times$\,1$'$ with a sampling of 0.2$''$/spaxel, and nominal mode, spanning the spectral range 4750$-$9350~\AA. The side of the MUSE field of view (FOV hereafter) spans a range of $\sim$3.3 to 14 kpc in these four galaxies.

The data reduction and exposure combination were carried out using the ESO MUSE pipeline (\citealt{Weilbacher:2020aa}) version v1.6. Depending on the galaxy, this was done either making use of self-made scripts executing the Common Pipeline Library (CPL; \citealt{Banse:2004aa}, \citealt{ESOCPL2014}) reduction recipes with EsoRex (ESO Recipe Execution Tool; \citealt{ESOCPL2015}) or employing ESO Reflex (Recipe flexible execution workbench, \citealt{Freudling:2013aa}), which gives a graphical and automated way to perform the reduction (still operated by EsoRex using the CPL recipes), within the Kepler workflow engine (\citealt{Altintas:2004aa}). For NGC 5643 we adopted the reduced cube provided by the ESO Quality Control Group. The different reduction softwares used do not give any measurable difference on the final reduced cubes because given the default reduction strategy (i.e. correction for instrumental and atmospheric effects and sky subtraction) they all rely on the same recipes run with EsoRex. Because the source emission is extended and fills the entire MUSE FOV, dedicated offset sky observations pointed towards regions free of galaxy emission were employed (see Table \ref{table:tb}).

The data analysis was executed using custom Python scripts, following an approach similar to that described in \cite{Venturi:2018aa}, \cite{Mingozzi:2019aa}, and \cite{Marasco:2020aa}. We briefly summarise it in the following. A more detailed explanation of the analysis procedure employed for all the sources in the MAGNUM survey will be given in a forthcoming paper (Venturi et al., in prep.).

We first fitted and subtracted the stellar continuum from each spaxel. To do so, we initially performed a Voronoi adaptive binning (\citealt{Cappellari:2003aa}) on the cube to achieve a minimum signal-to-noise ratio (S/N) per bin on the continuum. 
We set the minimum S/N per bin to an average value of 50 per wavelength channel, considering the signal and associated noise below 5530 \AA, where stellar absorption features are more prominent, and excluded gas emission lines. We then fitted the stellar continuum in each bin through the \textsc{ppxf} code (penalized pixel-fitting; \citealt{Cappellari:2004aa}) in the range 4770$-$6800~\AA\ using a linear combination of \cite{Vazdekis:2010aa} synthetic spectral energy distributions (SEDs) for single-age, single-metallicity stellar populations (SSPs). In this step, the stellar templates are shifted and broadened to accommodate a varying velocity and velocity dispersion in each bin.
We fitted the continuum together with the main gas emission lines, that is, \hb, \oiii$\lambda\lambda$4959,5007, \oi$\lambda\lambda$6300,6364, \ha, \nii$\lambda\lambda$6548,6584, and \sii$\lambda\lambda$6716,6731\footnote{Unless specified otherwise, hereafter we denote with \oiii\ and \nii\  only the \oiii$\lambda$5007 and \nii$\lambda$6584 strongest lines of each respective doublet, and with \sii\ the entire \sii$\lambda\lambda$6716,6731 line doublet.}, with the only goal of better constraining the underlying stellar continuum.
We then subtracted the fitted stellar continuum spaxel by spaxel after having rescaled the modelled continuum to the median of the observed continuum in each spaxel. 

At this point we fitted the gas emission lines described above from the continuum-subtracted cube using the \textsc{mpfit} routine (\citealt{Markwardt:2009aa}). 
We adopted one, two, or three Gaussians to reproduce the emission line profiles, 
as determined by a reduced-$\chi^2$ criterion, so as to use multiple components only in case of complex profile shapes (a more detailed description will be given in Venturi et al., in prep.).
For each separate Gaussian component, we required all the emission lines to have the same velocity and velocity dispersion, and we fixed the flux ratios between the strongest and the faintest lines of the \oiii, \oi,\ and \nii\ doublets to the theoretical value of 3.

We finally produced emission line maps for the emission lines from the total modelled profile made up of the sum of the fitted Gaussians. The maps are presented in Sect. \ref{sec:muse_maps}.



\section{MUSE maps\label{sec:muse_maps}}
In this section we show the maps obtained from the analysis of the MUSE data of IC 5063, NGC 5643, NGC 1068, and NGC 1386, which, as introduced before, reveal a peculiar phenomenon that we present and discuss in the following. A S/N cut of 3 per spaxel on the emission lines involved in the maps was applied\footnote{The S/N is defined here as the ratio of the peak flux of the modelled line profile and the standard deviation of the data minus model residuals around that line.}. 
We also excluded deviant spaxels in which an incorrect fit resulted in extremely broad wings fitting the noise by setting a maximum value on the fitted velocity dispersion (varying from 350$-$450 km/s depending on the galaxy) in spaxels with low S/N in the wings.
This was done so as to carefully exclude only spurious high velocity dispersions and not real ones. 


\subsection{IC 5063}\label{ssec:ic5063}

\begin{figure*}
\centering
        \hfill
        \begin{subfigure}[t]{0.01\textwidth}
        \textbf{a}
        \end{subfigure}
        \begin{subfigure}[t]{0.29\textwidth}
        \centering \includegraphics[width=0.9\textwidth,valign=b,trim={-1cm -6cm -1.1cm 0},clip]{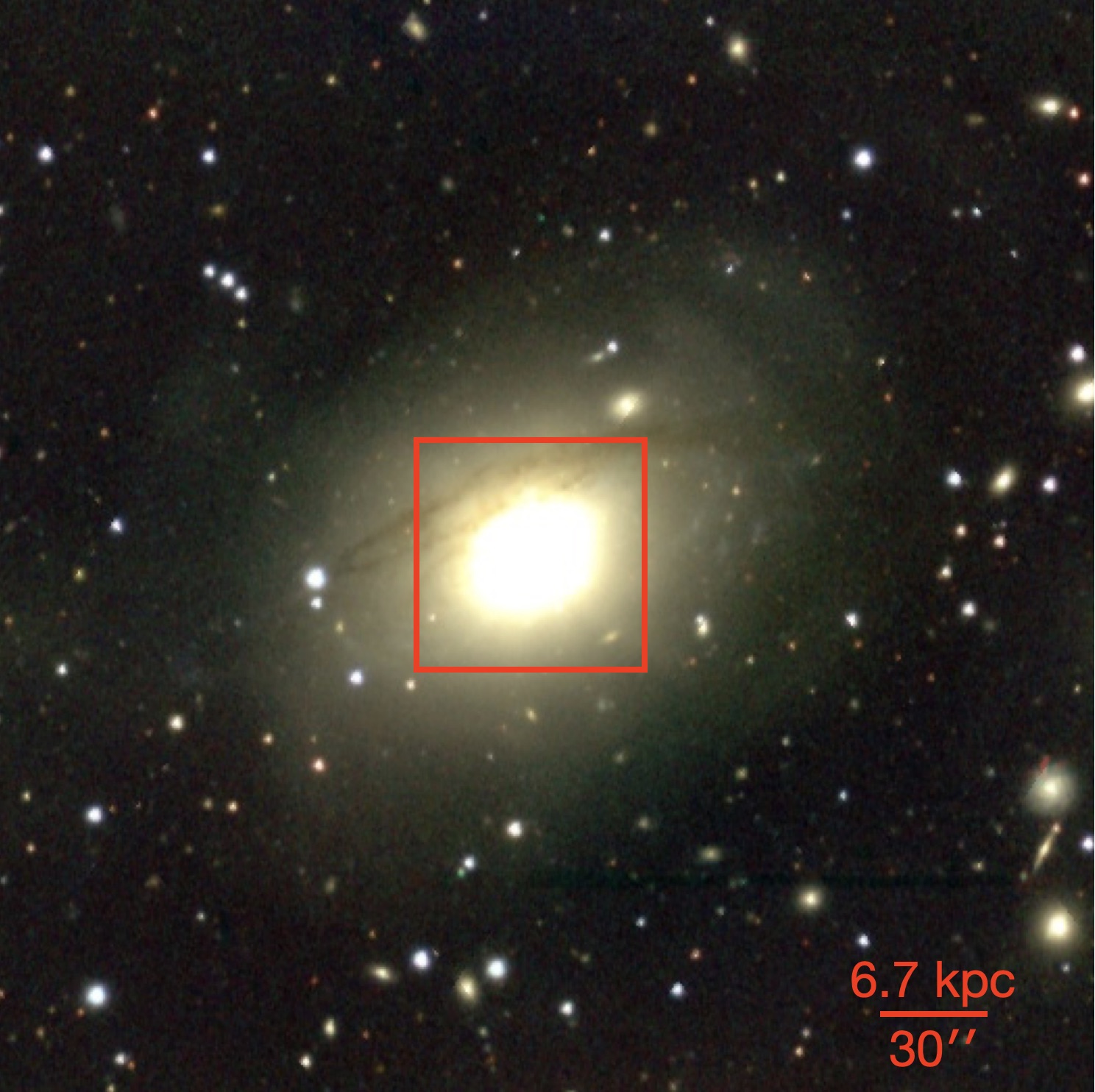}
        \end{subfigure}
        \hfill
        \begin{subfigure}[t]{0.01\textwidth}
        \textbf{b}
        \end{subfigure}
        \begin{subfigure}[t]{0.29\textwidth}
        \centering \includegraphics[width=\textwidth,valign=b,trim={2.3cm 0.2cm 4cm 3.8cm},clip]{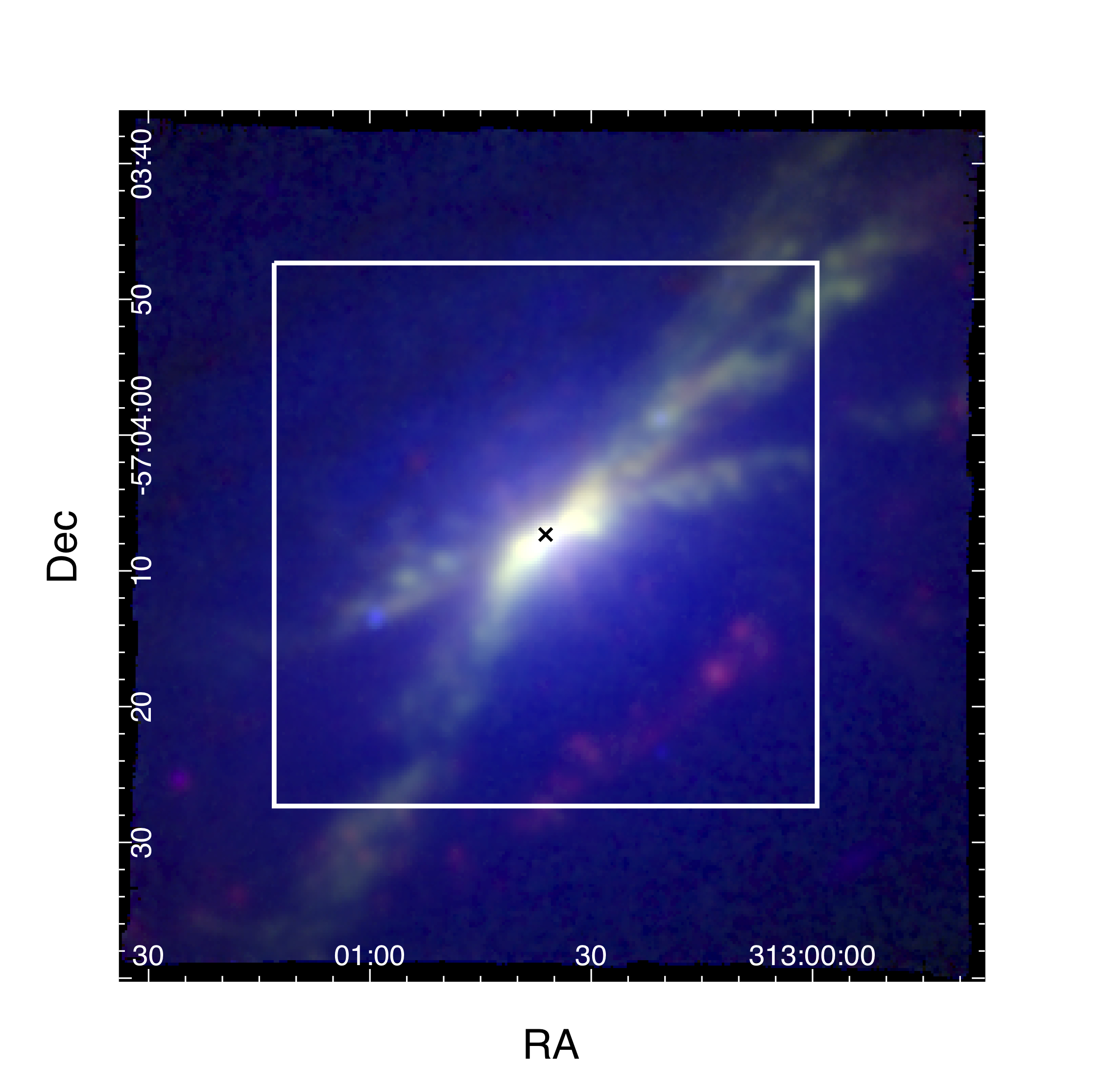}
        \end{subfigure}
        \hfill
        \begin{subfigure}[t]{0.01\textwidth}
        \textbf{c}
        \end{subfigure}
        \begin{subfigure}[t]{0.36\textwidth}\includegraphics[width=\textwidth,trim={2.6cm 0.5cm 2.3cm 0.5cm},clip]{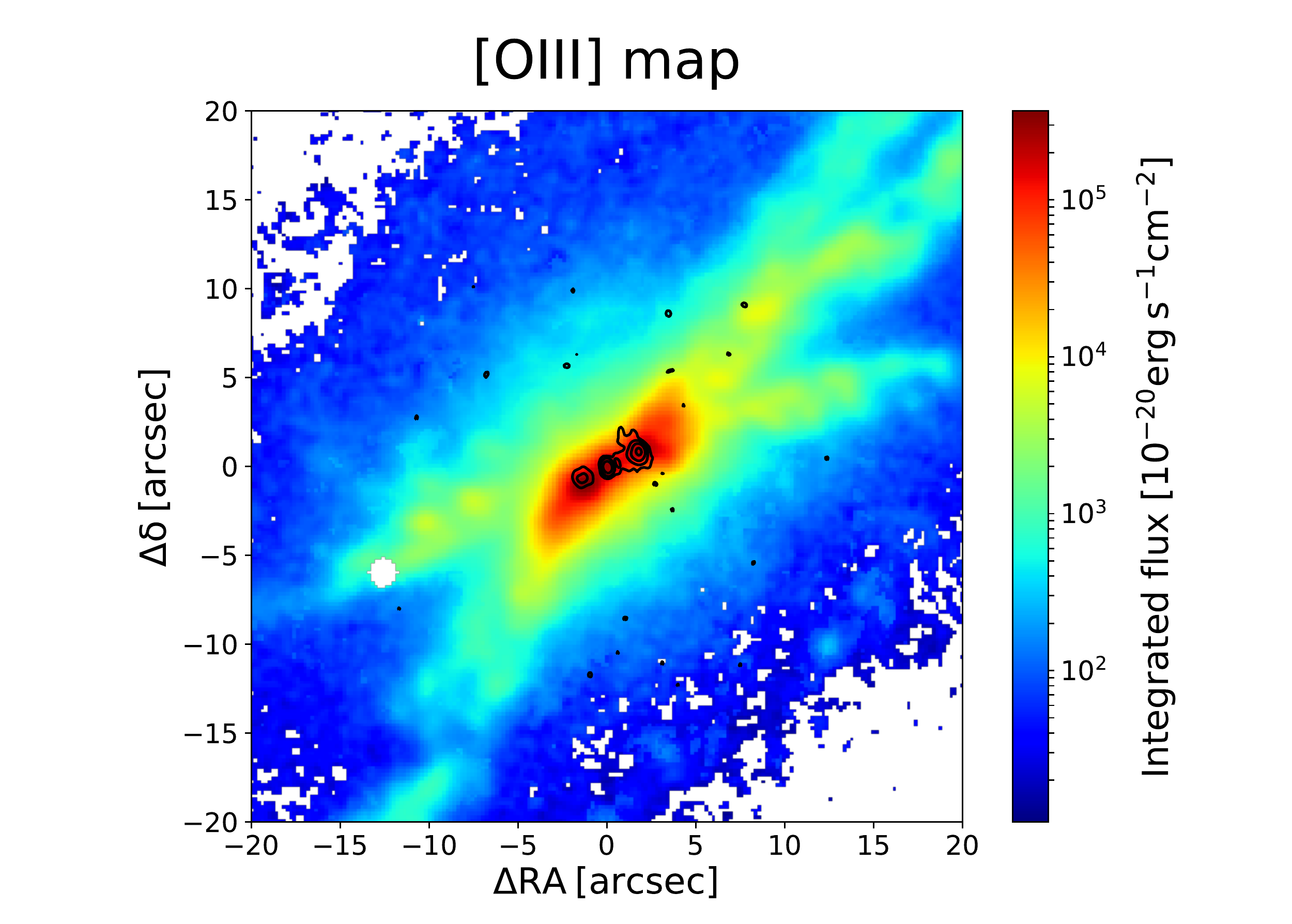}
        \end{subfigure}
        \hfill\null
        \hfill\null
        \caption{IC 5063. {\bf (a)} Coloured image of IC 5063 in $g$, $r$, $i$ filters from Dark Energy Survey (DES) DR1 LIneA (\citealt{Abbott:2018aa}).
        The red box shows the FOV of our MUSE map in panel b, whose side spans $\sim$14 kpc. {\bf (b)} Three-colour image from MUSE, showing [O\,\textsc{iii}] (green), H$\alpha$ (red), and stellar continuum (blue) collapsed in the spectral range $\sim$5100--5800 \AA. [O\,\textsc{iii}] and H$\alpha$ fluxes are obtained from modelling the stellar continuum-subtracted, 1\,spaxel-$\sigma$ smoothed data cube. [O\,\textsc{iii}] is also separately reported in {\bf (c)} in a 40$''$\,$\times$\,40$''$ zoomed region (white box). The reported flux is per pixel. A S/N cut of 3 has been applied. Circular masked regions mark spaxels that were excluded because Galactic foreground stars disturbed the data analysis. The ATCA 17.8 GHz radio contours from \cite{Morganti:2007aa} are superimposed. North is up and west is right.}
        \label{fig:maps_ic5063_1}
        
\vspace*{\floatsep}

        \centering
        \hfill
        \begin{subfigure}[t]{0.01\textwidth}
        \textbf{a}
        \end{subfigure}
        \begin{subfigure}[t]{0.36\textwidth}
        \includegraphics[width=\textwidth,trim={2.6cm 0.5cm 2.3cm 0.5},clip]{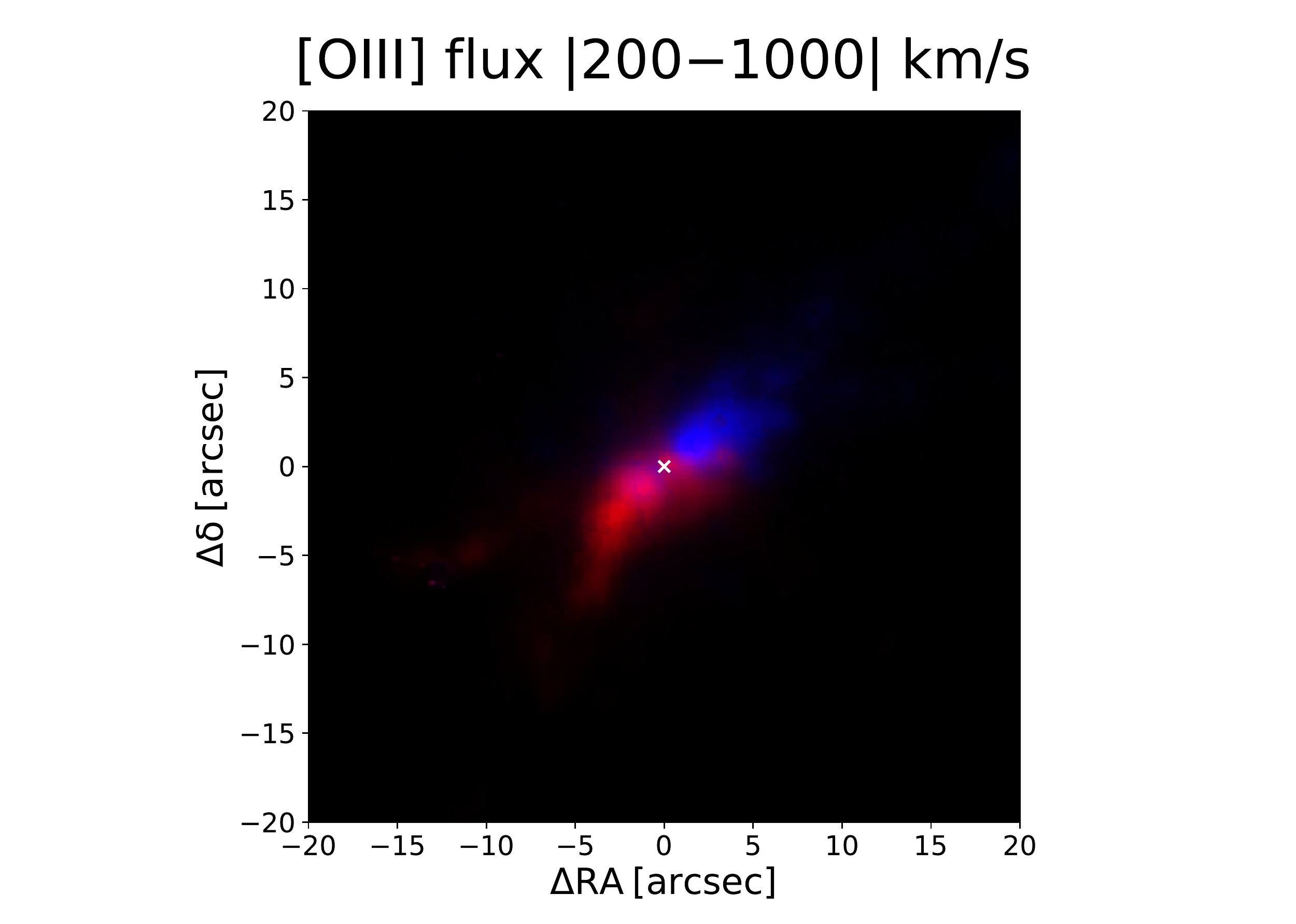}
        \end{subfigure}
        \hfill
        \begin{subfigure}[t]{0.01\textwidth}
        \textbf{b}
        \end{subfigure}
        \begin{subfigure}[t]{0.36\textwidth}
        \includegraphics[width=\textwidth,trim={2.6cm 0.5cm 2.3cm 0.5},clip]{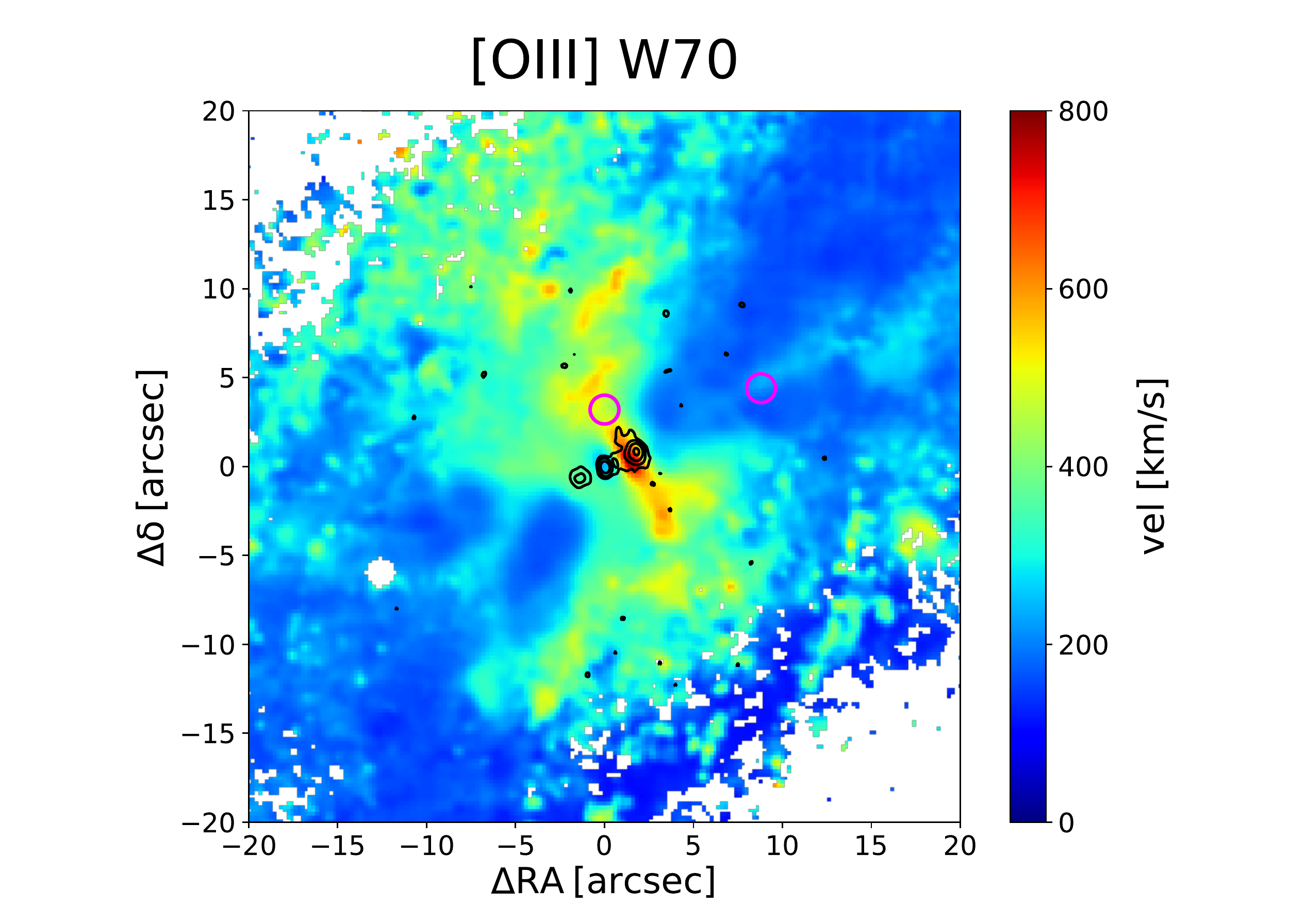}
        \end{subfigure}
        \hfill\null\\
        \hfill
        \begin{subfigure}[t]{0.01\textwidth}
        \textbf{c}
        \end{subfigure}
        \begin{subfigure}[t]{0.36\textwidth}
        \includegraphics[width=\textwidth,trim={11cm 1.5cm 11cm -1cm},clip]{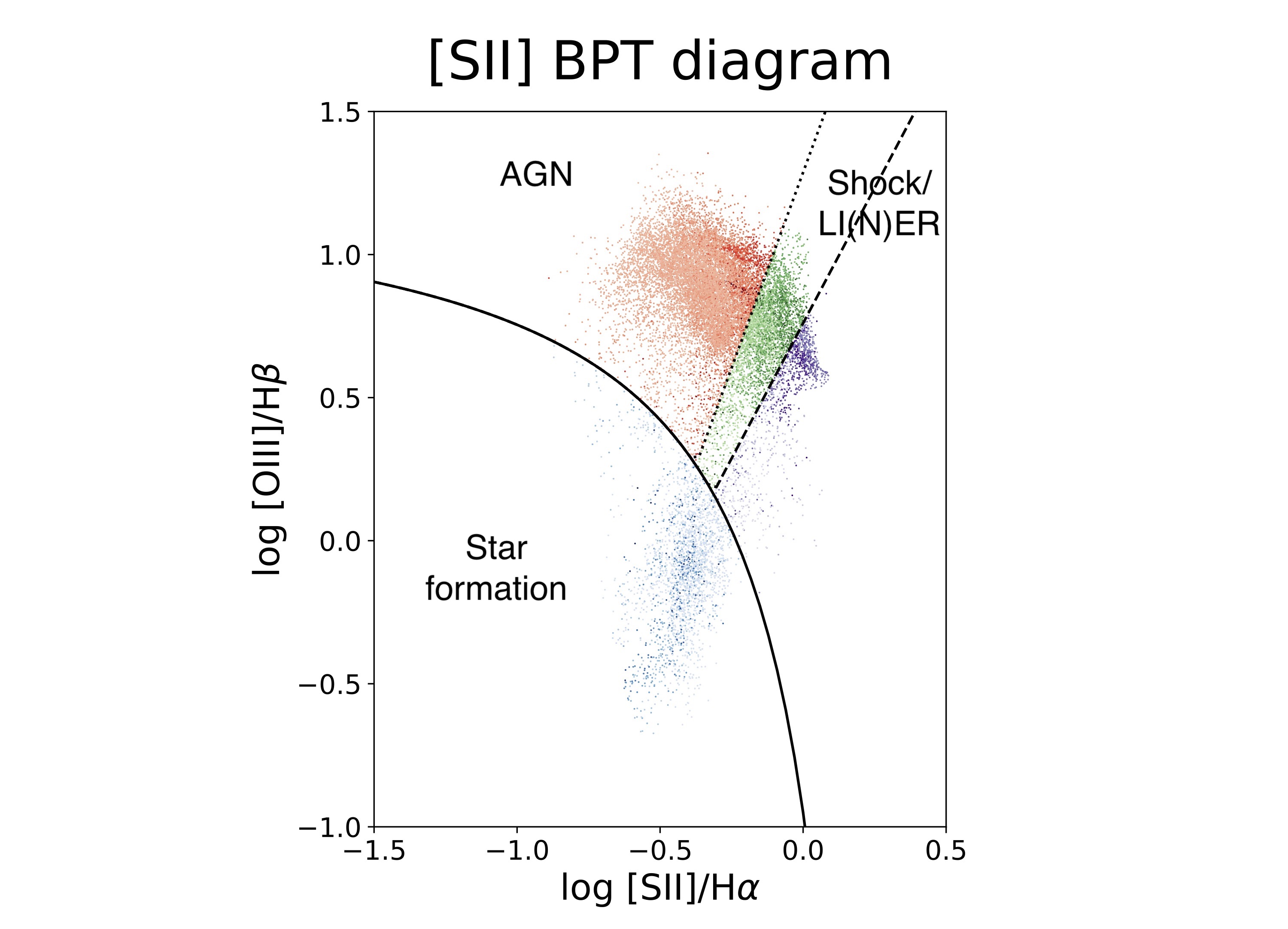}
        \end{subfigure}
        \hfill  
        \begin{subfigure}[t]{0.01\textwidth}
        \textbf{d}
        \end{subfigure}
        \begin{subfigure}[t]{0.36\textwidth}
        \includegraphics[width=\textwidth,trim={3.7cm 0.5cm 1.3cm 0.5},clip]{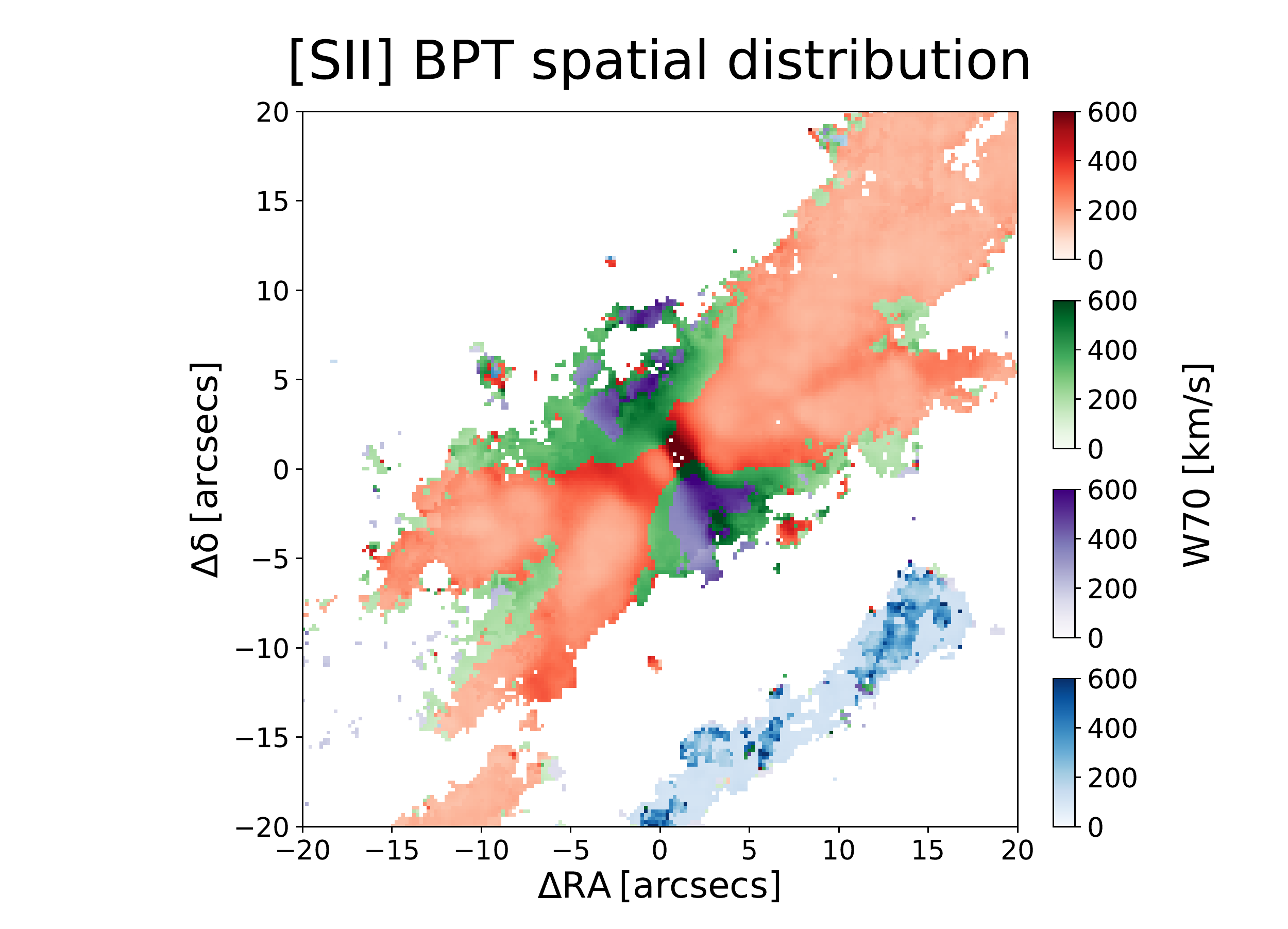}
        \end{subfigure}
        \hfill\null
        \caption{\small IC 5063. {\bf (a)} 
        [O\,\textsc{iii}] flux integrated from the fitted profile in the velocity range $\pm$|200--1000| km/s (blue when negative, i.e. approaching, red when positive, i.e. receding) with respect to the stellar velocity in each pixel (to avoid including the contribution to the flux from the gas that rotates in the disc). It shows the bulk of the high-velocity outflow.
        {\bf (b)} [O\,\textsc{iii}] W70 map, i.e. the difference between the 85th and 15th percentile velocities of the modelled line profile. The map has been re-smoothed with a Gaussian kernel having $\sigma$ = 1 pixel for better visual clarity. The radio contours are the same as in Fig. \ref{fig:maps_ic5063_1}c. Magenta circles indicate 4-spaxel radius extraction apertures of the spectra displayed in Fig. \ref{fig:spectra_ic5063}.
        {\bf (c)} [O\,\textsc{iii}]$\lambda$5007/H$\beta$ vs. \sii$\lambda\lambda$6716,6731/H$\alpha$ BPT diagram and {\bf (d)} associated spatial distribution. The colour intensity in the diagram and the map is coded according to the [O\,\textsc{iii}] W70. The solid curve defines the theoretical upper bound for pure star formation (\citealt{Kewley:2001ab}). The dashed line represents the \cite{Kewley:2006aa} demarcation between Seyfert galaxies and shocks or LI(N)ERs, while the dotted line is the \cite{Sharp:2010aa} bisector line between AGN- and shock-ionisation. SF-dominated regions are then marked in blue, AGN-dominated ionisation in red, and purple and green mark shock-ionised or LI(N)ER regions. As in Fig. \ref{fig:maps_ic5063_1}c, the maps are zoomed in the central 40$''$\,$\times$\,40$''$ compared to the full-FOV image in Fig. \ref{fig:maps_ic5063_1}b. A S/N cut of 3 has been applied to the maps, together with a cut on high velocity dispersions resulting from poor fits (see text for details).
        }
        \label{fig:maps_ic5063_2}
\end{figure*}

\begin{figure*}
    \centering
    \includegraphics[width=0.56\textwidth,trim={0 0 0 1.3cm},clip]{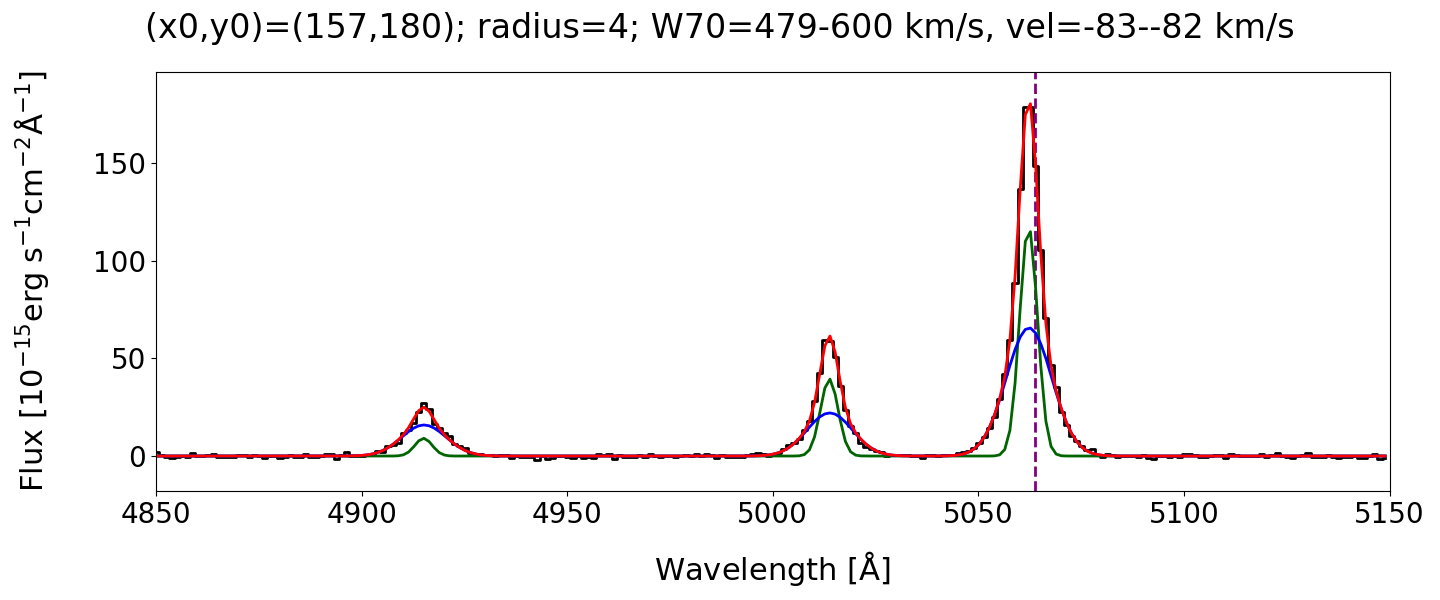}
    \includegraphics[width=0.56\textwidth,trim={0 0 0 1.3cm},clip]{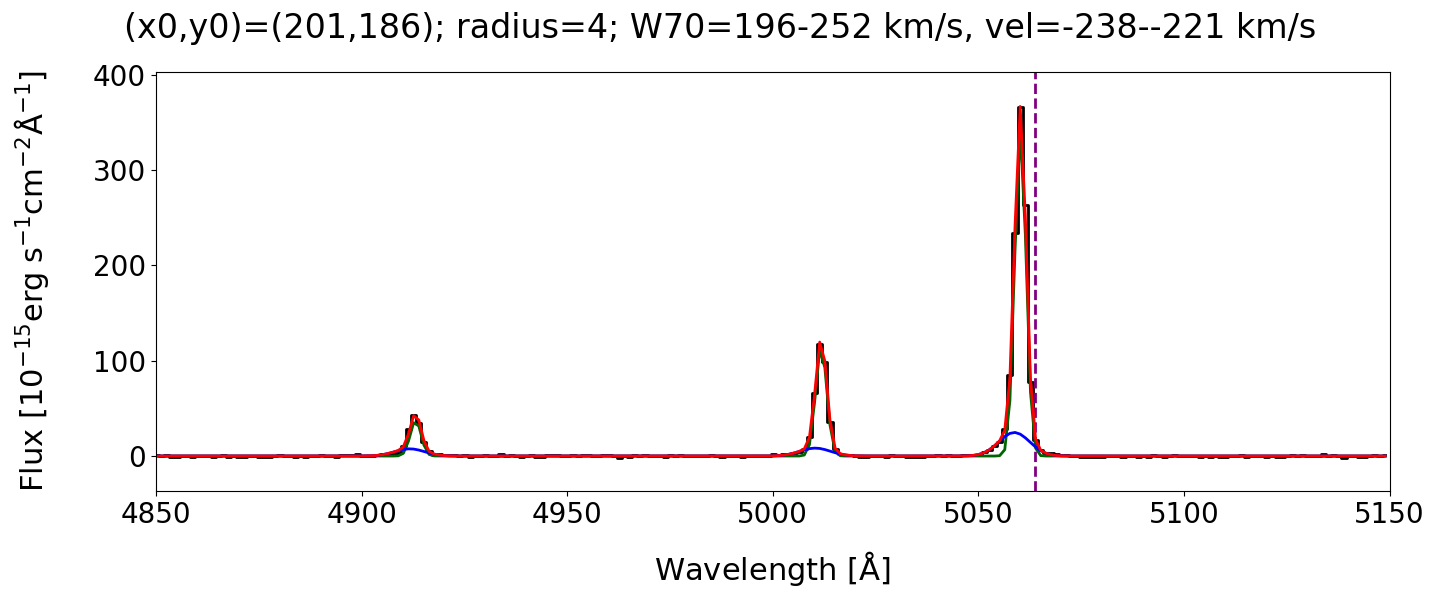}
    \caption{Spectra of IC 5063 in the \oiii\ doublet (right) and \hb\ (left) spectral range extracted from the apertures shown in Fig. \ref{fig:maps_ic5063_2}b. The spectrum in the top panel is representative of the region of line velocity width enhancement, exhibiting wide, fairly symmetric line profiles with no clear net velocity, while the spectrum in the bottom panel comes from the direction of the radio jet and ionisation cones and presents narrower profiles with asymmetric wings and a higher overall net velocity of the gas.
    The dashed purple lines indicate the galaxy systemic velocity.
    The wavelengths are reported in the observed frame.
    }
    \label{fig:spectra_ic5063}
\end{figure*}

IC 5063 is an S0 early-type galaxy residing at a distance of $\sim$46 Mpc from Earth (1$''$ $\sim$ 220 pc).
With a radio power of $P_\mathrm{1.4~GHz}$ = $3 \times 10^{23}$~W~Hz$^{-1}$ (\citealt{Tadhunter:2014aa}), it is one of the brightest Seyfert galaxies in the radio, although it is still radio quiet. 
IC 5063 hosts a radio jet limited to the inner $\sim$1 kpc of the galaxy along its major axis ($\sim$0.5 kpc per side), which drives an outflow of neutral atomic (\citealt{Morganti:1998aa}, \citealt{Oosterloo:2000aa}), molecular (\citealt{Morganti:2013aa}, \citealt{Tadhunter:2014aa}, \citealt{Morganti:2015aa}, \citealt{Dasyra:2016aa}, \citealt{Oosterloo:2017aa}), and ionised gas (\citealt{Morganti:2007aa}, \citealt{Dasyra:2015aa}, \citealt{Congiu:2017aa}) on the same scales along the jet direction by creating a cocoon of swept off gas.
The jet is known to lie close to the plane of the disc, as clear signs of jet-interstellar medium (ISM) interaction are observed along the entire jet in HI (\citealt{Oosterloo:2000aa}), CO (\citealt{Morganti:2015aa}) and H$_2$ (\citealt{Tadhunter:2014aa}).

In Figs. \ref{fig:maps_ic5063_1} and \ref{fig:maps_ic5063_2} we report our MUSE flux and kinematic maps for IC 5063, respectively.
The FOV of the MUSE observations (red box in Fig. \ref{fig:maps_ic5063_1}a) covers a significant fraction of the galaxy, $\sim$14$\times$14 kpc$^2$.
In the three-colour image in Fig. \ref{fig:maps_ic5063_1}b we report the [O\,\textsc{iii}] (green), H$\alpha$ (red) and stellar continuum emission (blue).
The [O\,\textsc{iii}] (also Fig. \ref{fig:maps_ic5063_1}c) traces the large-scale X-shaped AGN ionisation cone extending up to $\sim$10 kpc per side along the galaxy major axis (see also \citealt{Colina:1991aa}, \citealt{Morganti:1998aa}, \citealt{Sharp:2010aa}).
The brightest line emission stems from the active nucleus and from two lobes a few arcsec to the E and W of the nucleus ($\sim$500 pc away per side), spatially consistent with the radio (e.g. \citealt{Morganti:1998aa} and Australia Telescope Compact Array (ATCA) 17.8 GHz contours in Fig. \ref{fig:maps_ic5063_1}c, from \citealt{Morganti:2007aa}) and X-ray lobes (\citealt{Gomez-Guijarro:2017aa}). 

To isolate the high-velocity outflowing gas, we report in Fig. \ref{fig:maps_ic5063_2}a the [O\,\textsc{iii}] flux integrated in the velocity ranges [200,1000] km/s (receding) and [$-$1000,$-$200] km/s (approaching) of the fitted profile. 
We adopted the fitted stellar velocity in each pixel as zero velocity, so as to minimise the contribution to the flux from gas regularly rotating in the disc and maximise that of the high-velocity outflowing gas (following the same approach as \citealt{Mingozzi:2019aa}).
We point out that the stellar velocity dispersion varies from $\sim$100~km/s to a maximum of $\sim$170~km/s across the FOV, therefore $\sim$200~km/s represents a fairly appropriate threshold to separate rotating from outflowing material.
The map shows that the ionised material is flowing out with a high bulk velocity in the same direction as the AGN ionisation cones and the radio jet.


However, by inspecting the map of the [O\,\textsc{iii}] line velocity width (measured through W70\footnote{W70 = $v_{85} - v_{15}$, difference between the 85th percentile and 15th percentile velocities of the fitted line profile.}) in Fig. \ref{fig:maps_ic5063_2}b, we note an intriguing feature. The map shows an elongated region of enhanced line width (up to $\sim$800 km/s at its centre; see also \citealt{Lopez-Coba:2020aa}) perpendicular to the radio jet, AGN ionisation cones, X-ray emission, and ionised and molecular outflow. This perpendicular enhanced line width is also very extended, spanning about 7 kpc ($\sim$3.5~kpc per side), while the radio jet only extends over the central $\sim$1 kpc.
On very small scales ($\sim$1--2$''$), \cite{Dasyra:2015aa} also reported in a few near-IR emission lines an indication of velocity dispersion enhancement perpendicular to the radio jet and to the line-emission major axis. Our maps reveal that this small-scale enhancement, limited by S/N to the inner $\sim$1--2$''$ in \cite{Dasyra:2015aa}, traces the base of a much larger (up to $\sim$35$''$) and unambiguous velocity dispersion enhancement.

In Fig. \ref{fig:spectra_ic5063} we report two representative spectra extracted from the region of line-width enhancement and from the direction of radio jet and ionisation cones. The first shows a broad, symmetric emission line profile, with W70 of $\sim$500 km/s (\oiii) to $\sim$600 km/s (\hb), close to the galaxy systemic velocity. The second instead presents a narrower ($\sim$200 km/s) profile with higher net velocity shift and asymmetric wings.
We note that the observed W70 enhancement is indeed generally not associated with a coherent differential gas motion in the receding or approaching direction, 
which is instead observed in the direction of jet and ionisation cones, where the bulk of ionised gas with significant net velocity is located (see Fig. \ref{fig:maps_ic5063_2}a).
This does not mean that the line profiles in the W70-enhanced region always have a centroid velocity comparable with the galaxy systemic velocity.

Moreover, in Figs. \ref{fig:maps_ic5063_2}c and \ref{fig:maps_ic5063_2}d we show the [O\,\textsc{iii}]/H$\beta$ versus [S\,\textsc{ii}]/H$\alpha$ spatially resolved BPT diagnostic diagram (\citealt{Veilleux:1987aa}; hereafter [S\,\textsc{ii}]-BPT), which is commonly employed to identify the dominant gas ionisation mechanism between AGN, star formation processes, and shocks or low ionisation (nuclear) emission-line regions (LI(N)ERs). Each point in the diagram (panel c) corresponds to a pixel in the associated map (panel d). In addition to colour-coding the BPT diagram and the map according to the dominant ionisation mechanism, we also set the intensity of the colour proportional to the [O\,\textsc{iii}] W70 (i.e. the line velocity width). The BPT map reveals that while the AGN ionisation dominates in the bi-cones along the galaxy major axis, shock or LI(N)ER ionisation is present in the direction perpendicular to the jet and ionisation cones, where we observe the line-width enhancement. 
This is consistent with \cite{Mingozzi:2019aa}, who found that the regions of the galaxy with the highest \sii/\ha\ and \nii/\ha, perpendicular to the AGN cones, have the highest line velocity widths. 
This clearly also stands out in the \sii-BPT maps produced from the same MUSE data separately for the low- and high-velocity gas in Fig. D.3 of \cite{Mingozzi:2019aa}, where the spaxels along the galaxy minor axis (perpendicular to the AGN cones) populate the area of the BPT diagram at high \sii/\ha\ ratios (as also shown in \citealt{Lopez-Coba:2020aa}), which is indicative of shock or LI(N)ER ionisation.
The observed line ratios were compared with shock models from \cite{Allen:2008aa} comprising shock velocities in the range 100$-$1000 km/s  (details in \citealt{Mingozzi:2019aa}), which reproduced the high ratios (up to values $\sim$0.3 in log).
These shock velocities are compatible with the W70s that we observe (up to $\sim$800 km/s) and with the velocity dispersions resulting from the jet-ISM interaction simulations by \cite{Mukherjee:2018ab} tailored to IC 5063 (a few 100s km/s).
Consistent with the excess of \ha\ (compared to \oiii) that we observe perpendicularly to the \oiii\ cones in Fig. \ref{fig:maps_ic5063_1}b and with the low \oiii-to-Balmer and high \sii/\ha\ ratios arising there in our work and in \cite{Mingozzi:2019aa} and \citealt{Lopez-Coba:2020aa}, a recent work by \cite{Maksym:2020ab} resolved a filamentary loop in \sii\, and possibly \ha\ (but not \oiii) in Hubble Space Telescope (HST) data to the south of the nucleus in the inner $\sim$2$''$. This filament corresponds to the lower resolution plume-like feature that can be seen in \ha\ in our MUSE map in Fig. \ref{fig:maps_ic5063_1}b, immediately south of the nucleus.

In the spatially resolved BPT diagram in Fig. \ref{fig:maps_ic5063_2}d, we also note an extended (by $\sim$\,15--20$''$) star-forming stripe to the south-west of the nucleus, oriented perpendicular to the direction of the galaxy minor axis. This is also visible in \ha\  in Fig. \ref{fig:maps_ic5063_1}b, showing a clumpy morphology. Its velocities are consistent with those of the rotating disc (see the disc-outflow separation in \citealt{Mingozzi:2019aa}).  Interestingly, this star-forming stripe lies at the south-western edge of the line-width enhancement (Fig. \ref{fig:maps_ic5063_2}b), perpendicular to its direction and broadly centred on its axis.
Their relative spatial location is more clearly visible in Fig. \ref{fig:ic5063_posfeedb} in the appendix, where we report the contours of \oiii\ W70 superimposed on the map of \ha\ emission for visual clarity.
In principle, the turbulent gas exhibiting the large emission line widths might be responsible for favouring the observed star formation through compression and fragmentation of the impacted gas clouds, as in the positive-feedback mechanism from outflows and jets (e.g. \citealt{Silk:2013aa}, \citealt{Zubovas:2013aa}, \citealt{Cresci:2015aa}, \citealt{Cresci:2015ab}, \citealt{Santoro:2016aa}).
However, the study of this possible case of positive feedback goes beyond the scope of this paper and would require a focused deeper investigation.

Interestingly, we note that low spatial resolution 1.4 GHz radio data from \cite{Morganti:1998aa} show extended emission on scales of $\sim\,$40$''$ in the same direction as the W70 enhancement that we observe, perpendicular to the small-scale high-resolution radio jet. The nature of this extended perpendicular radio emission is unclear. Given its direction and scale, it might be related to the line-width enhancement that we observe.

A recent work by \cite{Maksym:2020aa} also discovered fan-shaped dark radial rays (similar to the crepuscular rays observed at sunset on Earth) in the direction perpendicular to the galaxy major axis and AGN ionisation cones. These extend on scales of $\sim$\,1 arcmin, comparable with those of the MUSE observations presented in our work. Among the different possible interpretations for the dark rays, the authors propose dusty reflection of AGN emission escaping in the direction perpendicular to the ionisation cones, with the dark rays either produced by the lack of reflecting dust or by the excess of absorbing dust.
In this case, the phenomenon observed in \cite{Maksym:2020aa} and the enhanced emission line widths that we observe in the same direction and scales might be part of the same phenomenon (possibly also together with the large-scale radio emission from \citealt{Morganti:1998aa} mentioned above), with the dust entrained in the turbulent material,  traced by the enhanced line widths, giving rise to the dark rays.


\subsection{NGC 5643}

\begin{figure*}
\centering
        \hfill
        \begin{subfigure}[t]{0.01\textwidth}
        \textbf{a}
        \end{subfigure}
        \begin{subfigure}[t]{0.29\textwidth}
        \centering \includegraphics[width=0.9\textwidth,valign=b,trim={-1cm -6cm -1cm 0},clip]{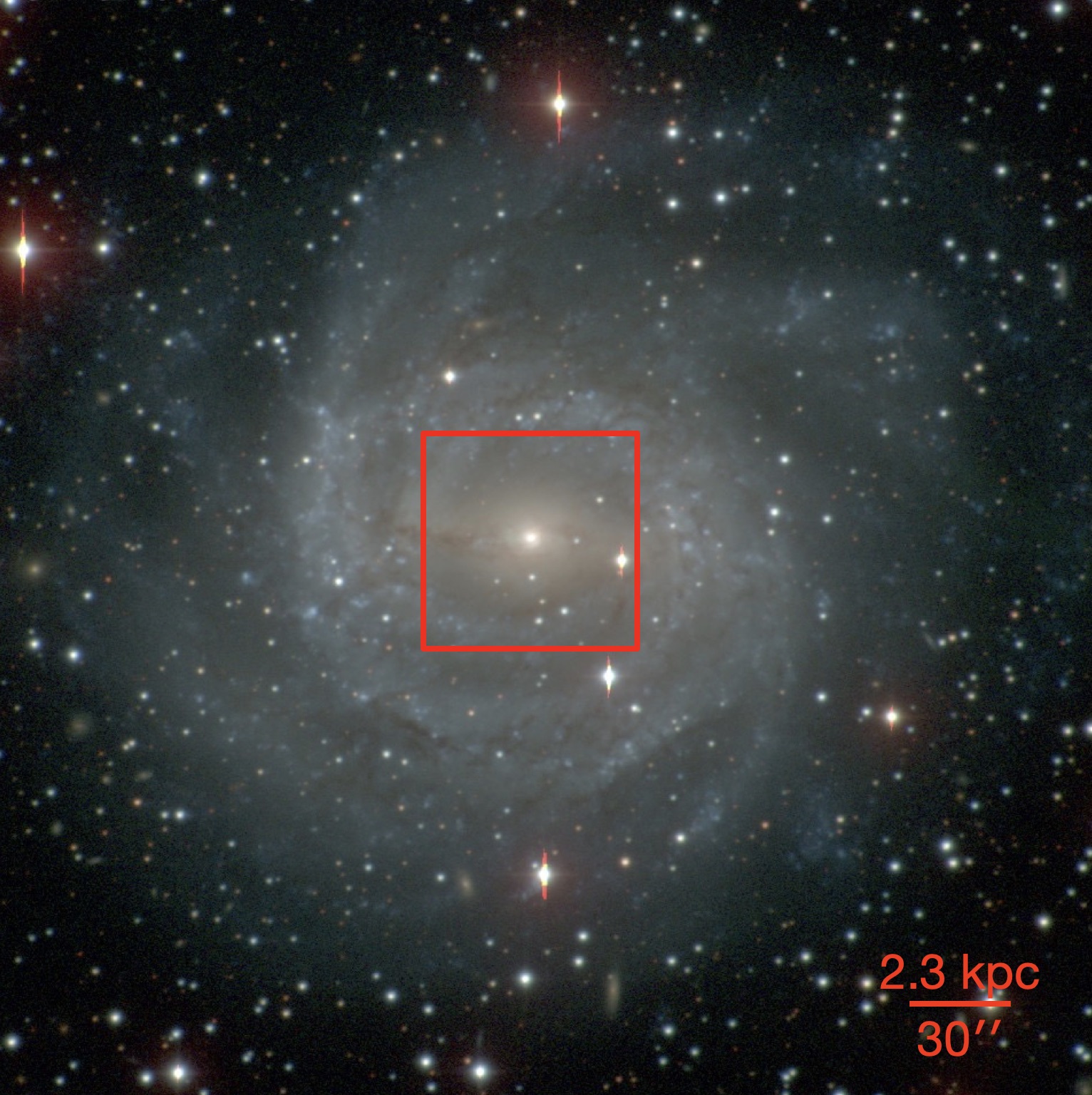}
        \end{subfigure}
        \hfill
        \begin{subfigure}[t]{0.01\textwidth}
        \textbf{b}
        \end{subfigure}
        \begin{subfigure}[t]{0.29\textwidth}
        \centering \includegraphics[width=\textwidth,valign=b,trim={2.3cm 0.2cm 4cm 3.8cm},clip]{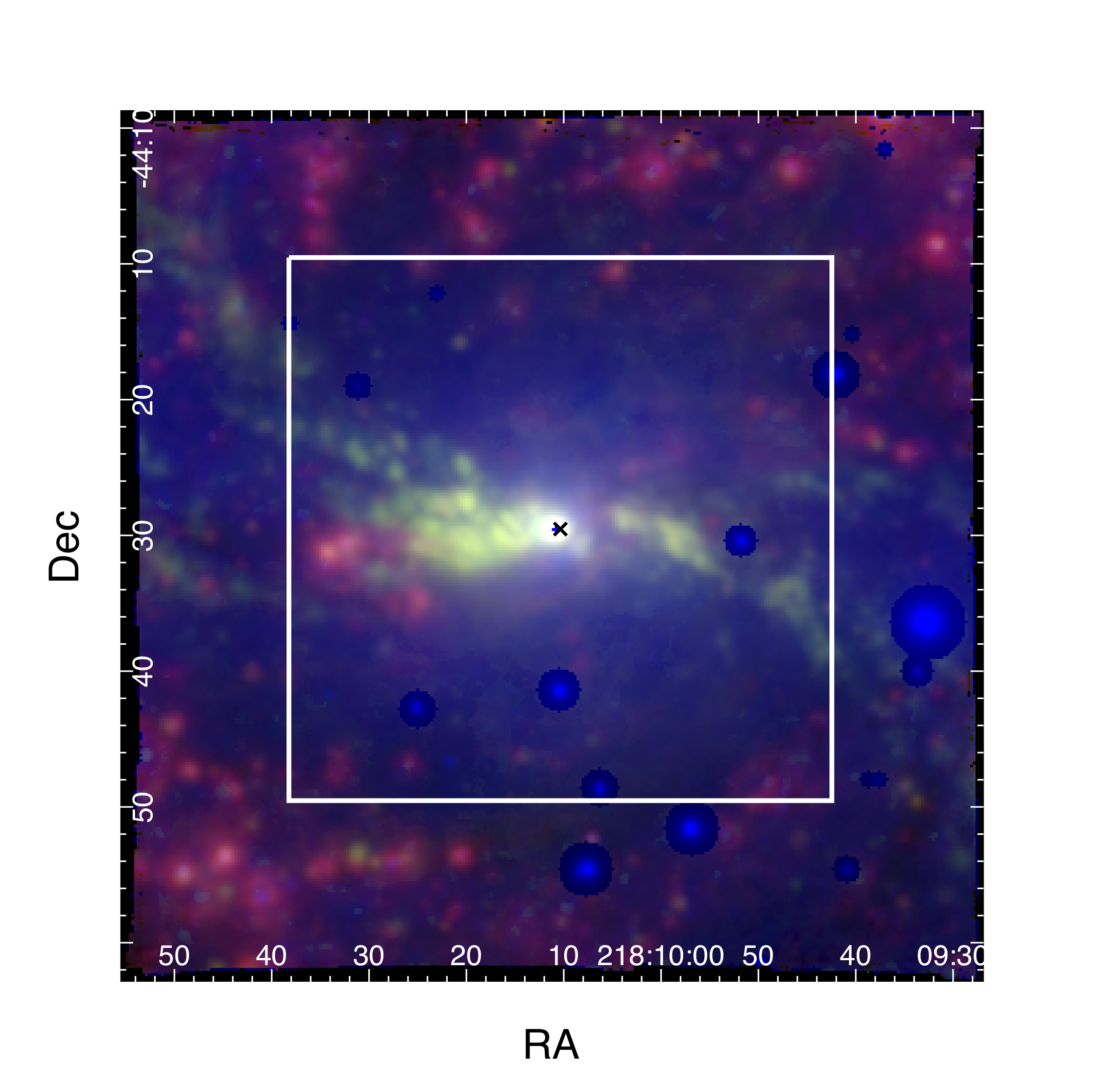}
        \end{subfigure}
        \hfill
        \begin{subfigure}[t]{0.01\textwidth}
        \textbf{c}
        \end{subfigure}
        \begin{subfigure}[t]{0.36\textwidth}\includegraphics[width=\textwidth,trim={2.6cm 0.5cm 2.3cm 0.5cm},clip]{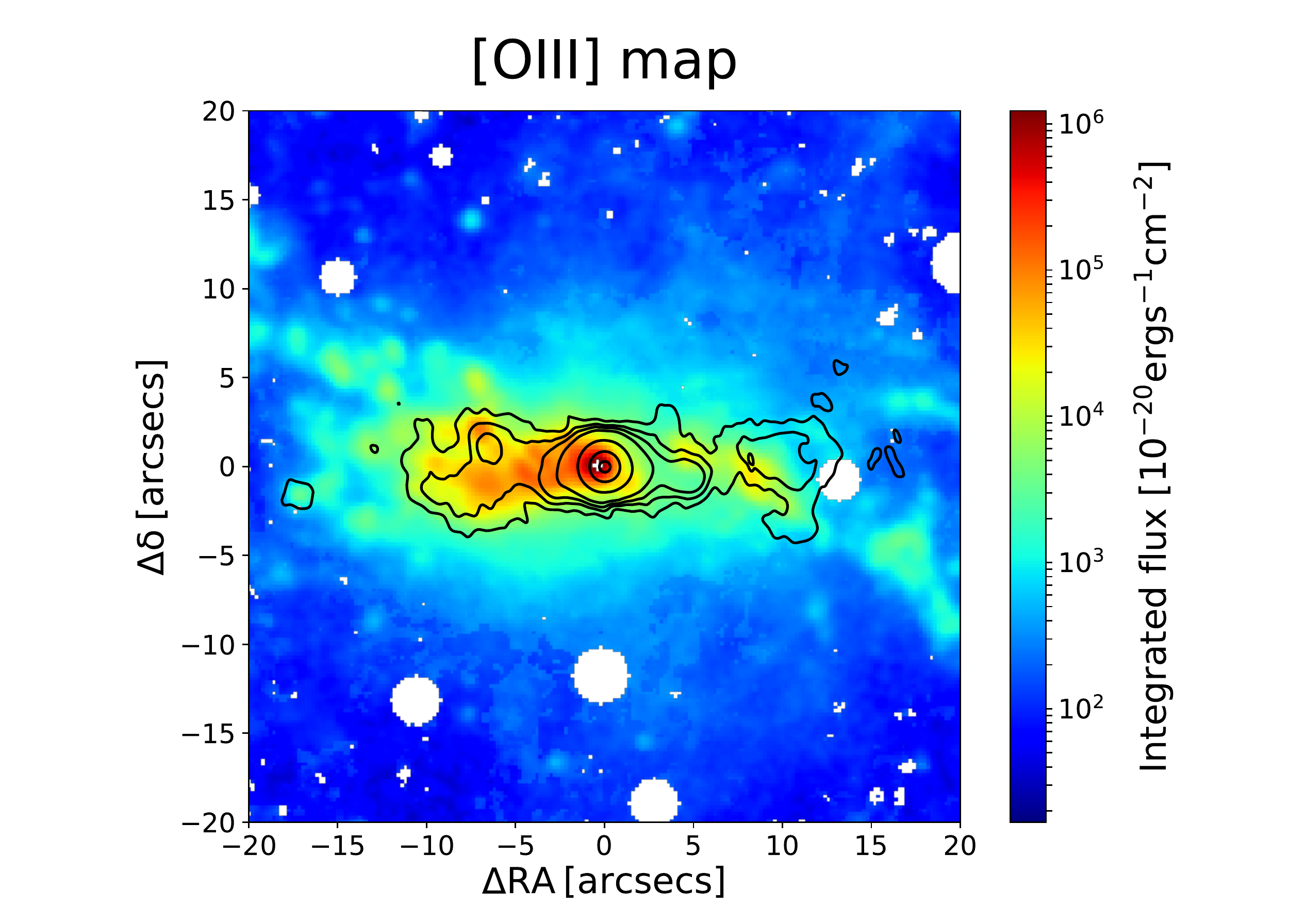}
        \end{subfigure}
        \hfill\null
        \hfill\null\\
        \caption{NGC 5643. {\bf (a)} Three-colour image of NGC 5643 ($B$ band in blue, $V$ band in green, and $I$ band in red) obtained with the 2.5 m du Pont Telescope at Las Campanas Observatory for the Carnegie-Irvine Galaxy Survey (CGS, \citealt{Ho:2011aa}). The red box shows the FOV of our MUSE map in panel b, whose side spans $\sim$5 kpc. Same as in Fig. \ref{fig:maps_ic5063_1} for {\bf (b)} and {\bf (c)}. The contours display the 8.4 GHz VLA radio observations from \cite{leipski2006}.}
        \label{fig:maps_n5643_1}
        
        \vspace*{\floatsep}

        \centering
        \hfill
        \begin{subfigure}[t]{0.01\textwidth}
        \textbf{a}
        \end{subfigure}
        \begin{subfigure}[t]{0.36\textwidth}
        \includegraphics[width=\textwidth,trim={2.6cm 0.5cm 2.3cm 0.5},clip]{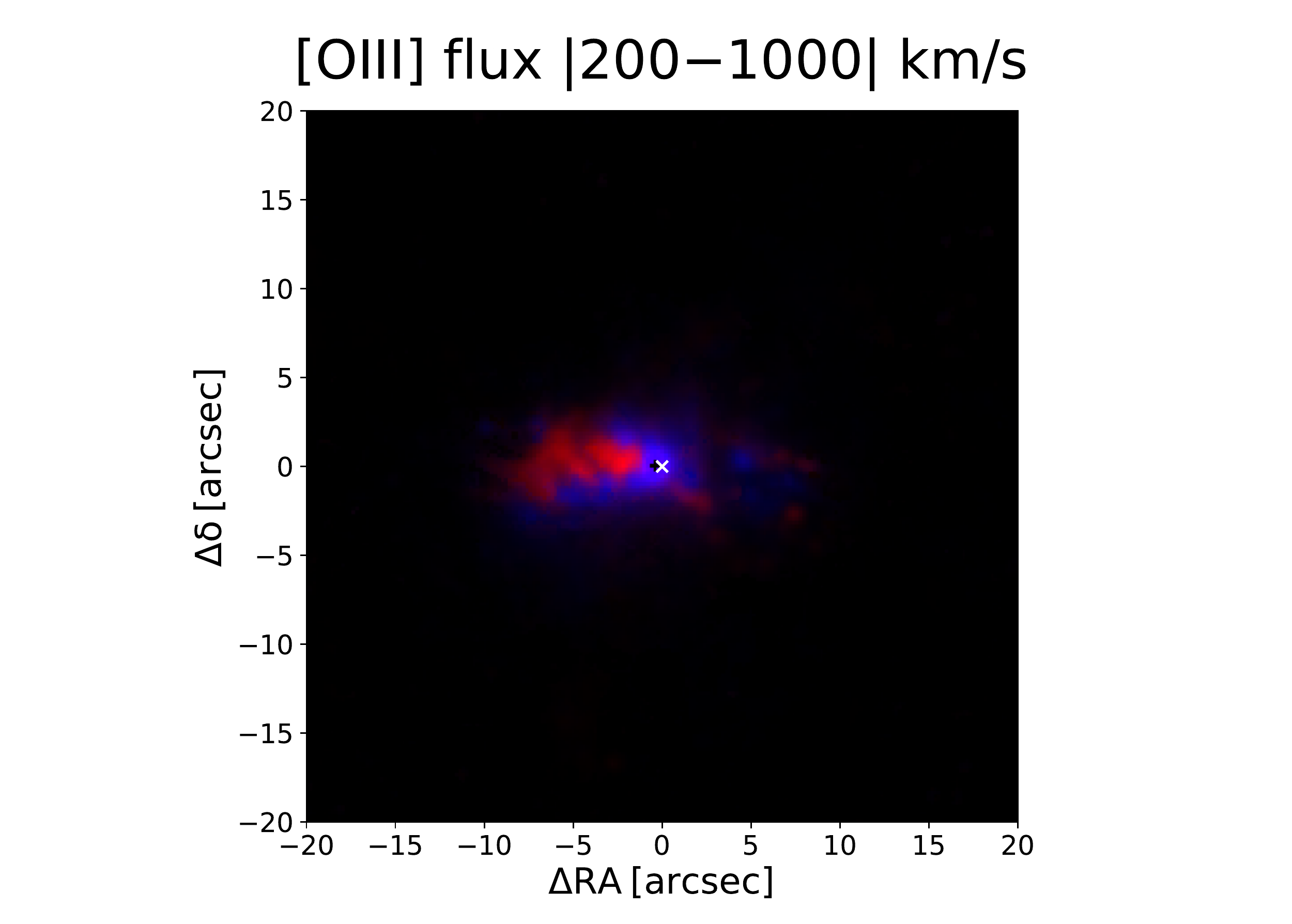}
        \end{subfigure}
        \hfill
        \begin{subfigure}[t]{0.01\textwidth}
        \hfill
        \textbf{b}
        \end{subfigure}
        \begin{subfigure}[t]{0.36\textwidth}
        \includegraphics[width=\textwidth,trim={2.6cm 0.5cm 2.3cm 0.5},clip]{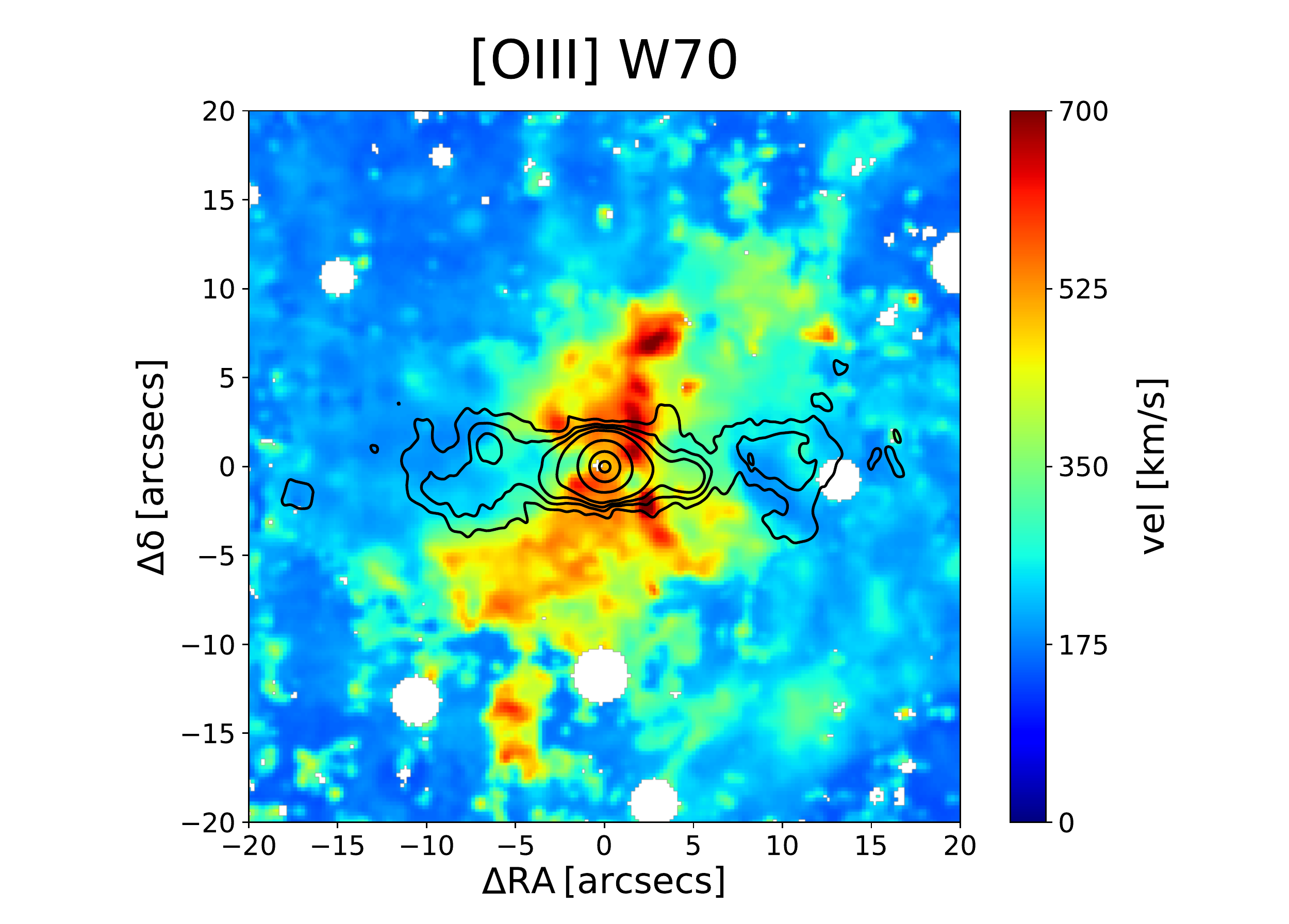}
        \end{subfigure}
        \hfill\null\\
        \hfill
        \begin{subfigure}[t]{0.01\textwidth}
        \textbf{c}
        \end{subfigure}
        \begin{subfigure}[t]{0.36\textwidth}
        \includegraphics[width=\textwidth,trim={11cm 1.5cm 11cm -1cm},clip]{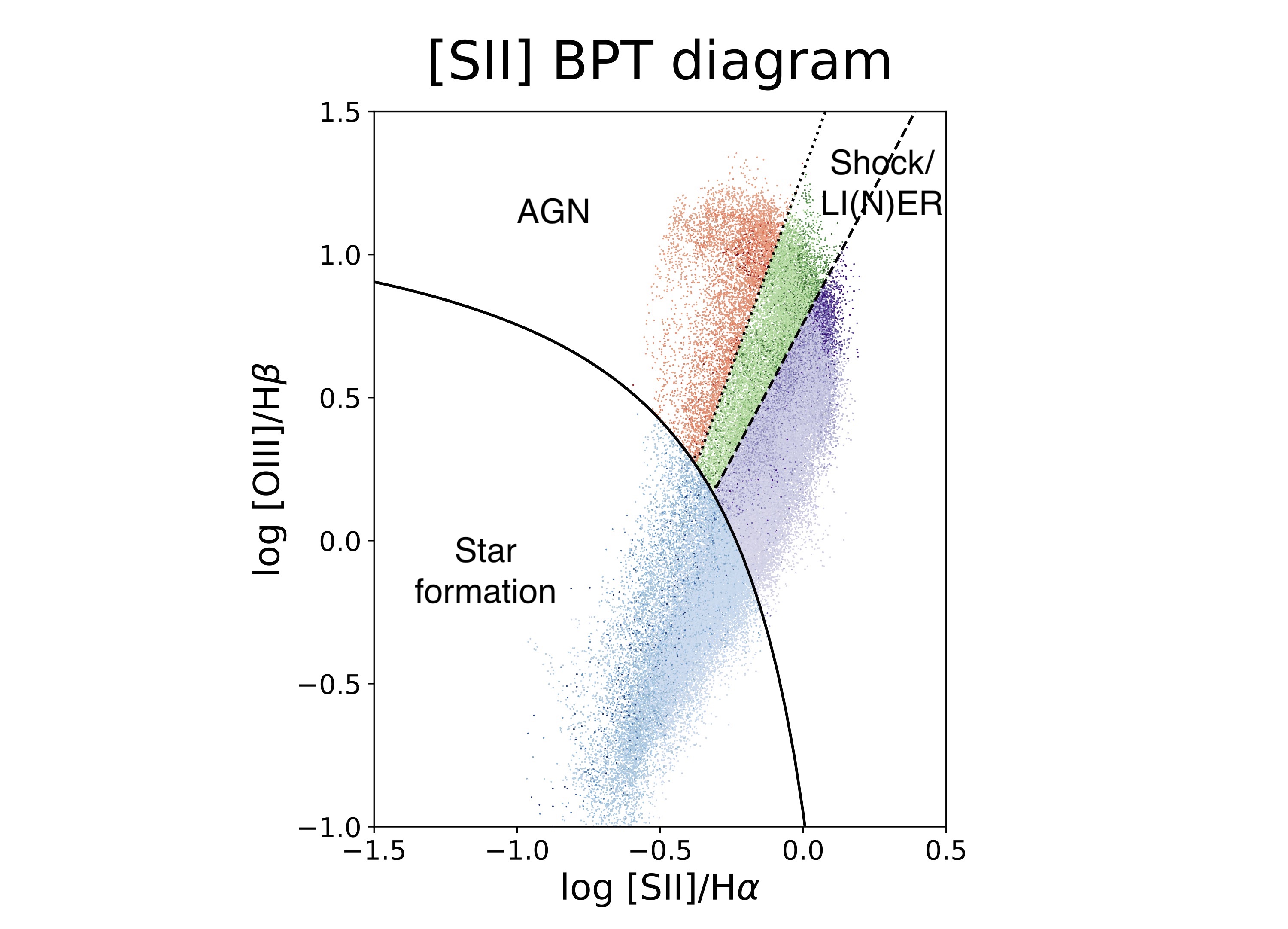}
        \end{subfigure}
        \hfill  
        \begin{subfigure}[t]{0.01\textwidth}
        \textbf{d}
        \end{subfigure}
        \begin{subfigure}[t]{0.36\textwidth}
        \includegraphics[width=\textwidth,trim={3.7cm 0.5cm 1.3cm 0.5},clip]{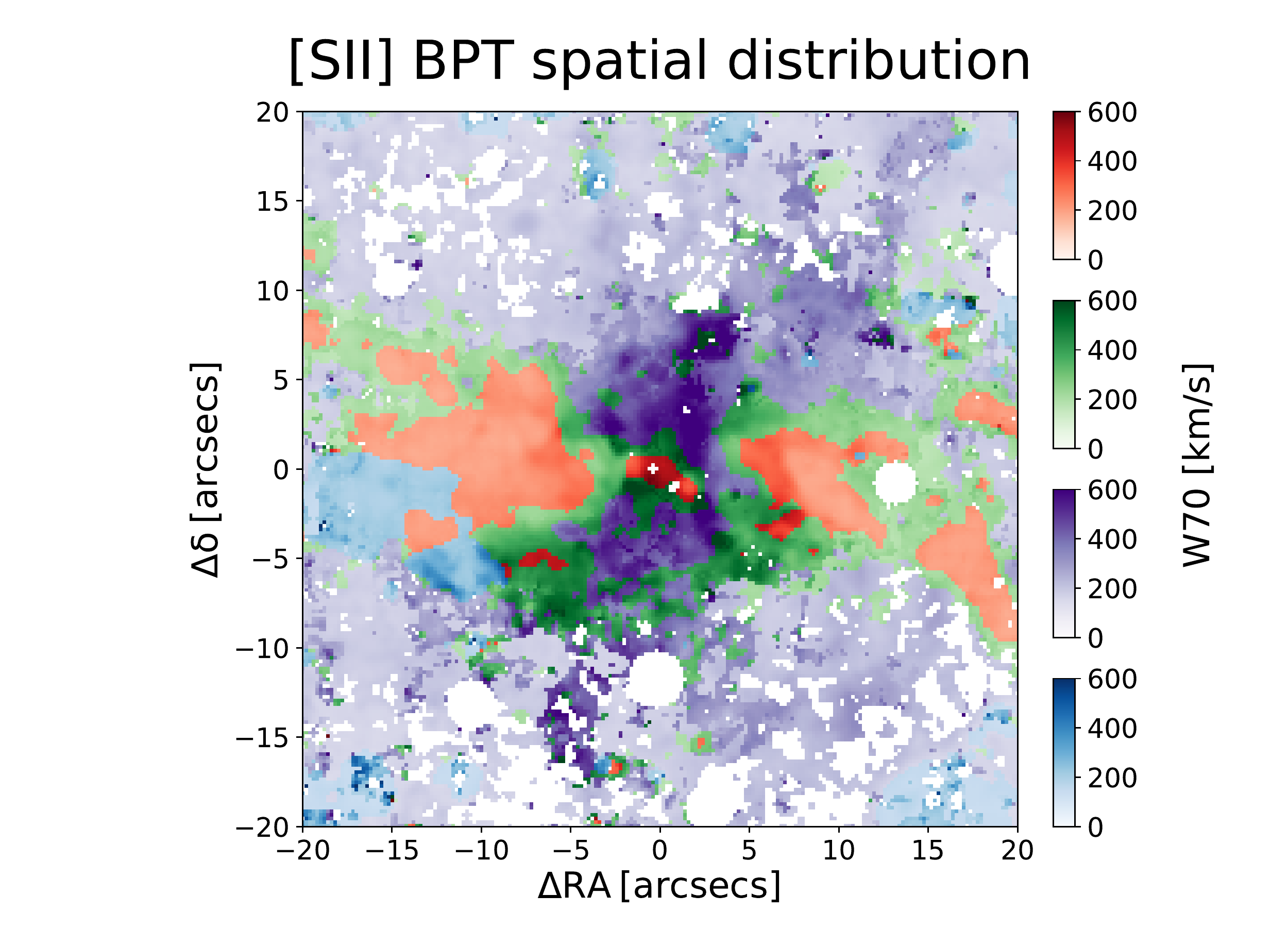}
        \end{subfigure}
        \hfill\null
        \caption{NGC 5643. Same as in Fig. \ref{fig:maps_ic5063_2}.}
        \label{fig:maps_n5643_2}
\end{figure*}

NGC 5643 ($D$\,$\sim$\,16 Mpc; 1$''$\,$\sim$\,78 pc) is a barred radio-quiet (\citealt{leipski2006}) Seyfert 2 galaxy seen almost face-on. It hosts an ionisation cone extending over a few kiloparsec eastward of the nucleus (e.g. \citealt{schmitt1994}, \citealt{simpson1997}) and inclined by $\sim$40$\degree$ to the galaxy disc, according to the model by \cite{fischer2013}. An outflow along the cone was found in [O\,\textsc{iii}] with MUSE by \cite{Cresci:2015aa} (as part of our MAGNUM survey), who also detected the western side of the ionisation bi-cone.
The galaxy also hosts a low-luminosity bipolar radio jet aligned with the ionisation cones and on similar scales, spanning $\sim$25--30$''$ ($\sim$2 kpc; \citealt{morris1985}, \citealt{leipski2006}), which is interacting with the galaxy disc (\citealt{Cresci:2015aa}, \citealt{Alonso-Herrero:2018aa}, \citealt{Garcia-Bernete:2020aa}).

Figs. \ref{fig:maps_n5643_1} and \ref{fig:maps_n5643_2} display our MUSE flux and kinematic maps for NGC 5643, respectively. Fig. \ref{fig:maps_n5643_1}a contains an image of the entire galaxy where the FOV of our MUSE maps, covering the central $\sim$5$\times$5 kpc$^2$, is reported in red. In the three-colour MUSE image in Fig. \ref{fig:maps_n5643_1}b we show the bright [O\,\textsc{iii}] ionisation bi-cone (green), stronger to the E of the nucleus than to the W, and H$\alpha$ emission (red), tracing star formation.
In Fig. \ref{fig:maps_n5643_1}c we report the [O\,\textsc{iii}] emission together with the 8.4 GHz radio contours from the Karl G. Jansky Very Large Array (VLA) presented in \cite{leipski2006}. The [O\,\textsc{iii}] cones and the radio jet are co-spatial and aligned in the E-W direction.


Fig. \ref{fig:maps_n5643_2}a reports the $\pm$|200--1000| km/s high-velocity ionised gas, showing that the bulk of the fast outflowing material is aligned co-spatially with the AGN bi-cone and the radio jet as discussed in \cite{Cresci:2015aa}.
The stellar velocity dispersion ranges from $\sim$60~km/s to $\sim$100~km/s across the FOV, making 200~km/s a safe threshold to exclude the contribution from rotating material.

In contrast, high values of the [O\,\textsc{iii}] line width (Fig. \ref{fig:maps_n5643_2}b), up to W70 $\sim$ 700 km/s, are observed in the N-S direction, perpendicular to the ionisation cones and radio jet, over a distance of $\sim$1.5 kpc per side ($\sim$3 kpc in total).
In Fig. \ref{fig:maps_n5643_2}c,d we also display the spatially resolved [S\,\textsc{ii}]-BPT diagram, which shows that shock or LI(N)ER ionisation dominates in the direction of enhanced line velocity width perpendicular to radio jet and AGN ionisation cones (as found by \citealt{Mingozzi:2019aa}), as in IC 5063.

\subsection{NGC 1068}\label{sec:n1068}

\begin{figure*}
\centering
        \hfill
        \begin{subfigure}[t]{0.01\textwidth}
        \textbf{a}
        \end{subfigure}
        \begin{subfigure}[t]{0.29\textwidth}
        \centering \includegraphics[width=0.9\textwidth,valign=b,trim={-1cm -6.5cm -1.5cm 0},clip]{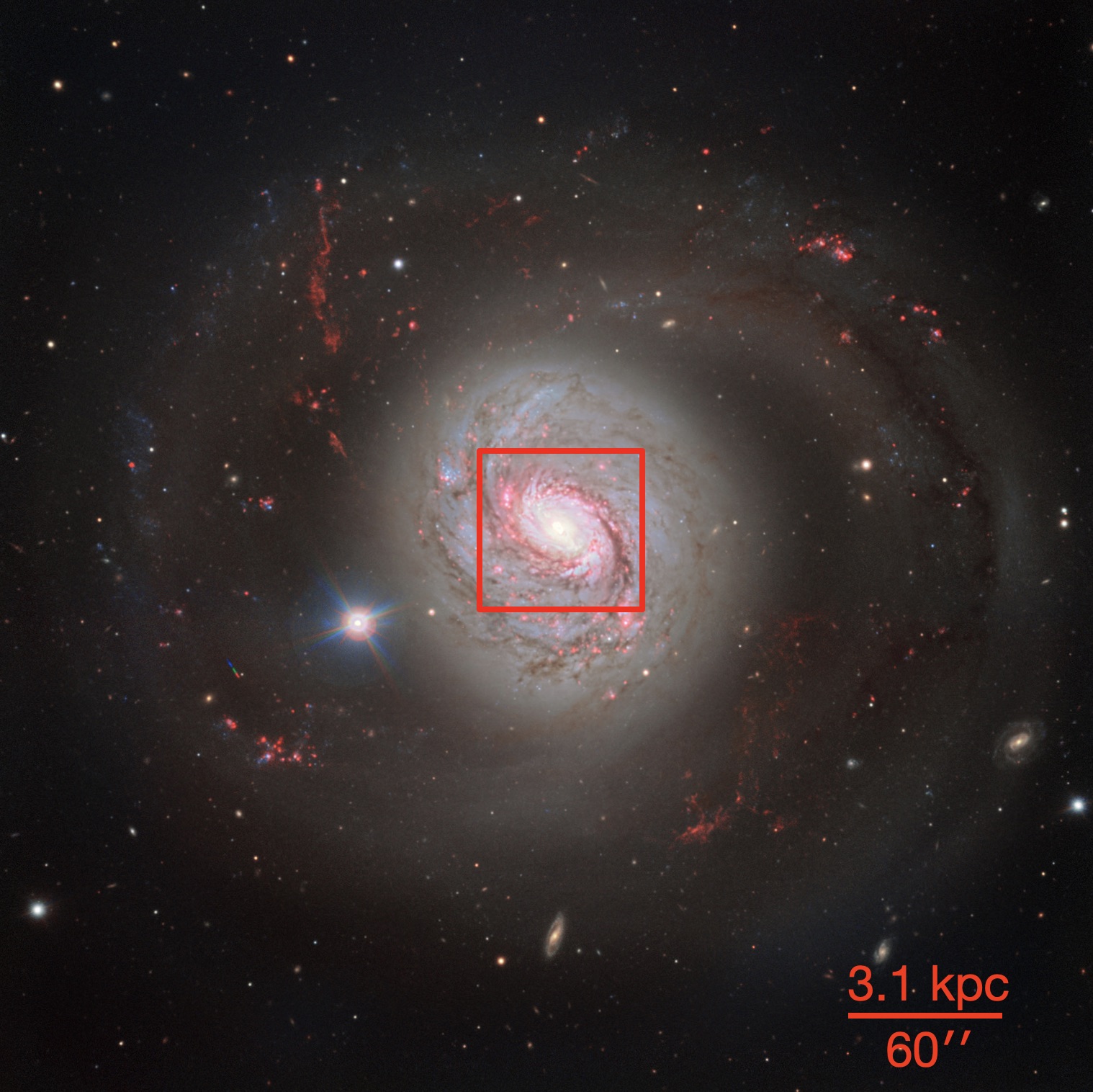}
        \end{subfigure}
        \hfill
        \begin{subfigure}[t]{0.01\textwidth}
        \textbf{b}
        \end{subfigure}
        \begin{subfigure}[t]{0.29\textwidth}
        \centering \includegraphics[width=\textwidth,valign=b,trim={3.5cm 1.4cm 5.2cm 8cm},clip]{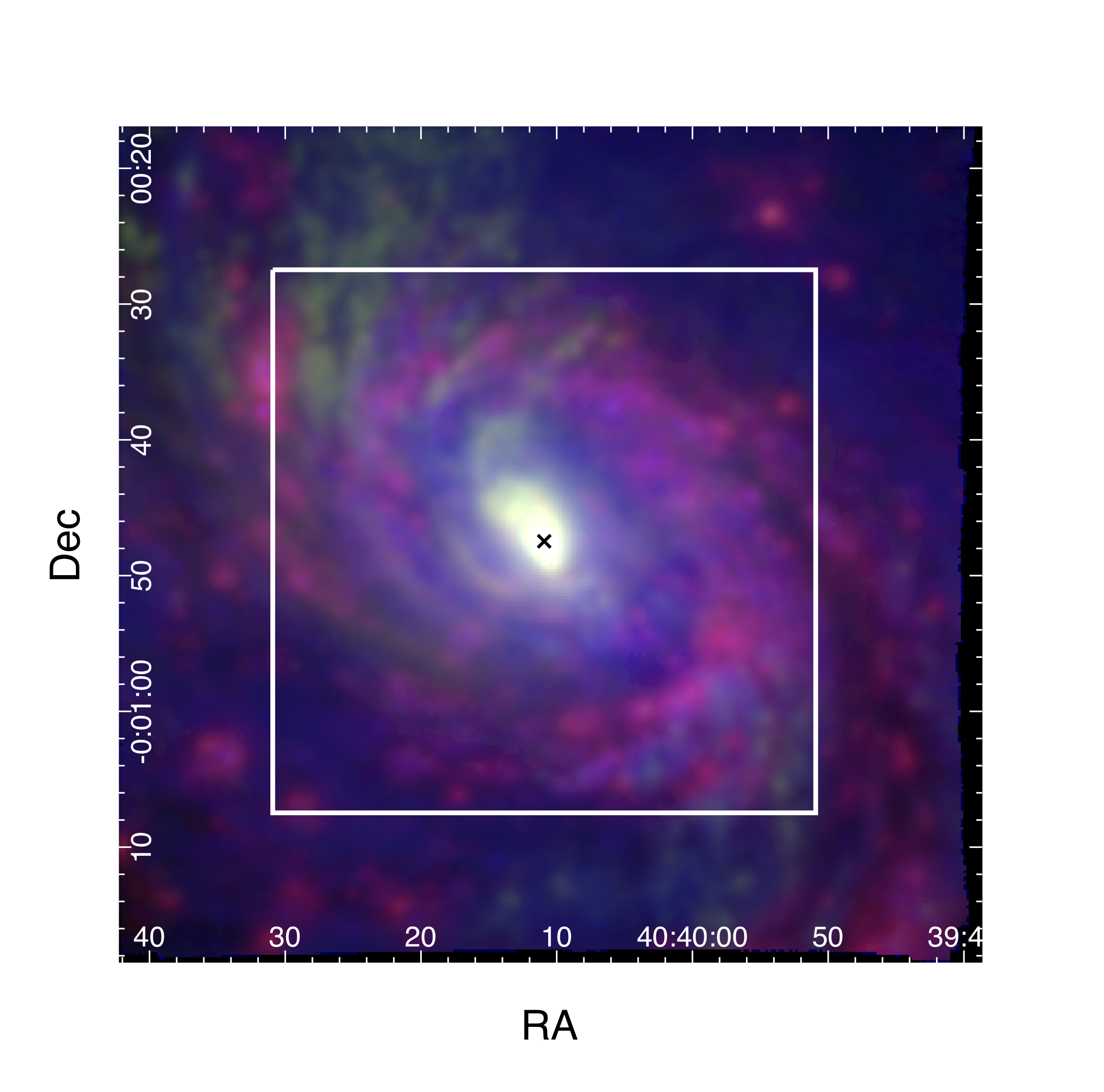}
        \end{subfigure}
        \hfill\null
        \begin{subfigure}[t]{0.01\textwidth}
        \textbf{c}
        \end{subfigure}
        \begin{subfigure}[t]{0.36\textwidth}\includegraphics[width=\textwidth,trim={2.5cm 0.5cm 2.25cm 0.5cm},clip]{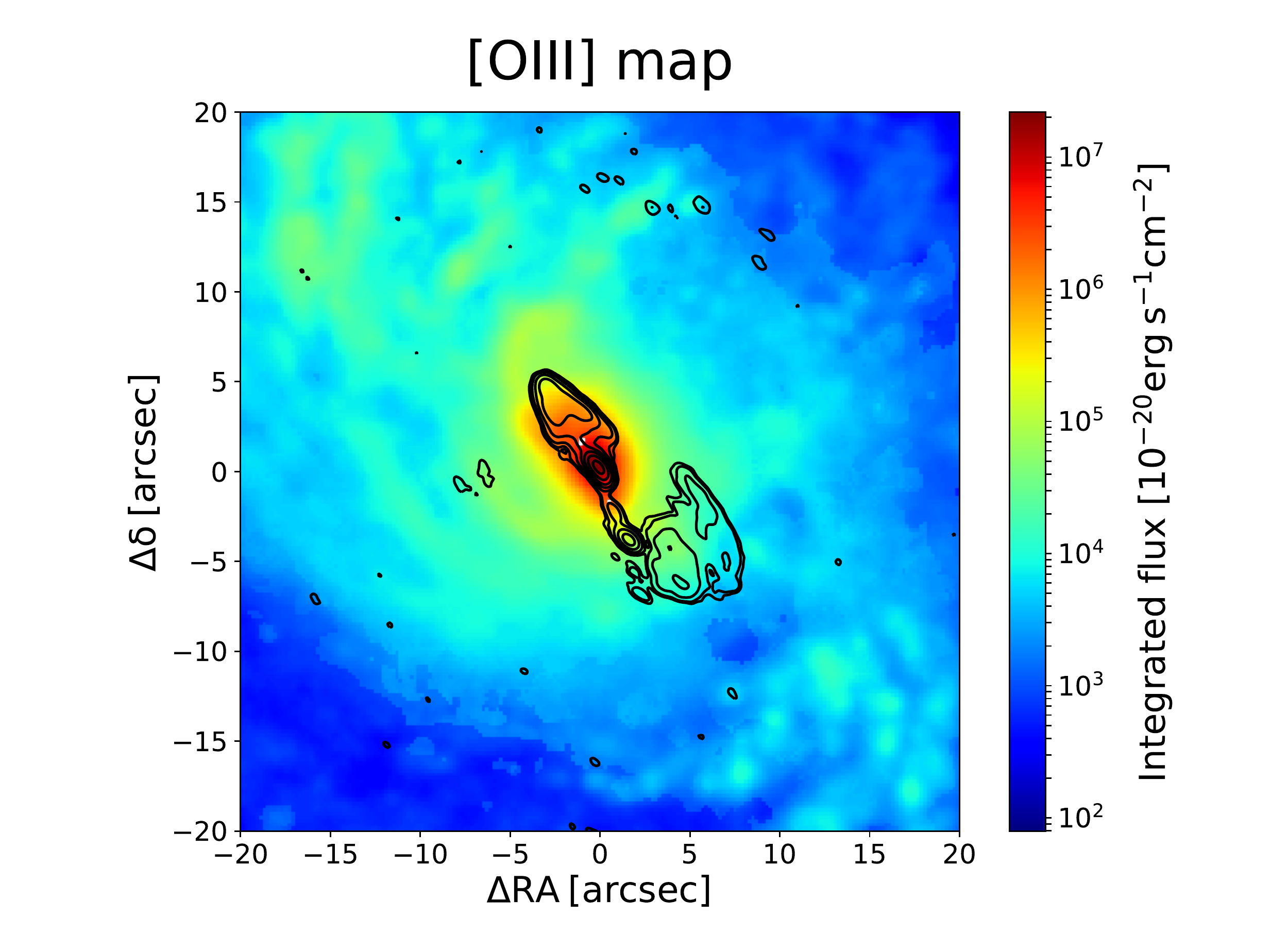}
        \end{subfigure}
        \hfill\null
        \caption{NGC 1068. {\bf (a)} Four-colour image of NGC 1068 from FORS2 at the VLT, where emission in {\it b}, {\it v,} and {\it R} filters is reported in blue, green, and orange, respectively, and H$\alpha$ in red; credit: ESO. The red box shows the FOV of our MUSE map in panel b, whose side spans $\sim$3.3 kpc. Same as in Fig. \ref{fig:maps_ic5063_1} for {\bf (b)} and {\bf (c)}. The contours in {\bf (c)} are the VLA 5 GHz (C band) A-array radio data from \cite{Gallimore:1996aa}.}
        \label{fig:maps_n1068_1}

\vspace*{\floatsep}

        \centering
        \hfill
        \begin{subfigure}[t]{0.01\textwidth}
        \textbf{a}
        \end{subfigure}
        \begin{subfigure}[t]{0.365\textwidth}\includegraphics[width=\textwidth,trim={2.5cm 0.5cm 2cm 0.5cm},clip]{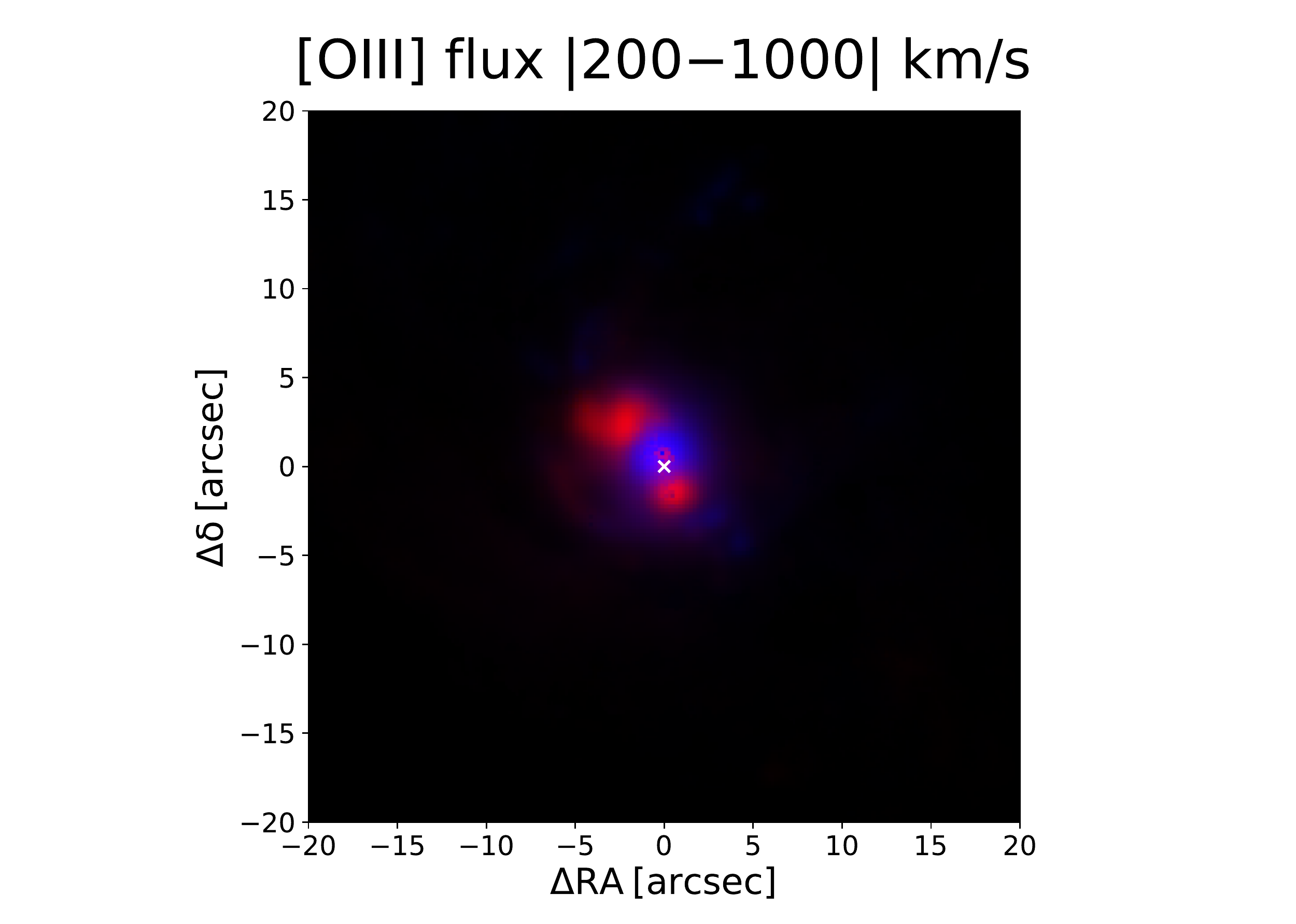}
        \end{subfigure}
        \hfill
        \begin{subfigure}[t]{0.01\textwidth}
        \textbf{b}
        \end{subfigure}
        \begin{subfigure}[t]{0.365\textwidth}\includegraphics[width=\textwidth,trim={2.5cm 0.5cm 2cm 0.5cm},clip]{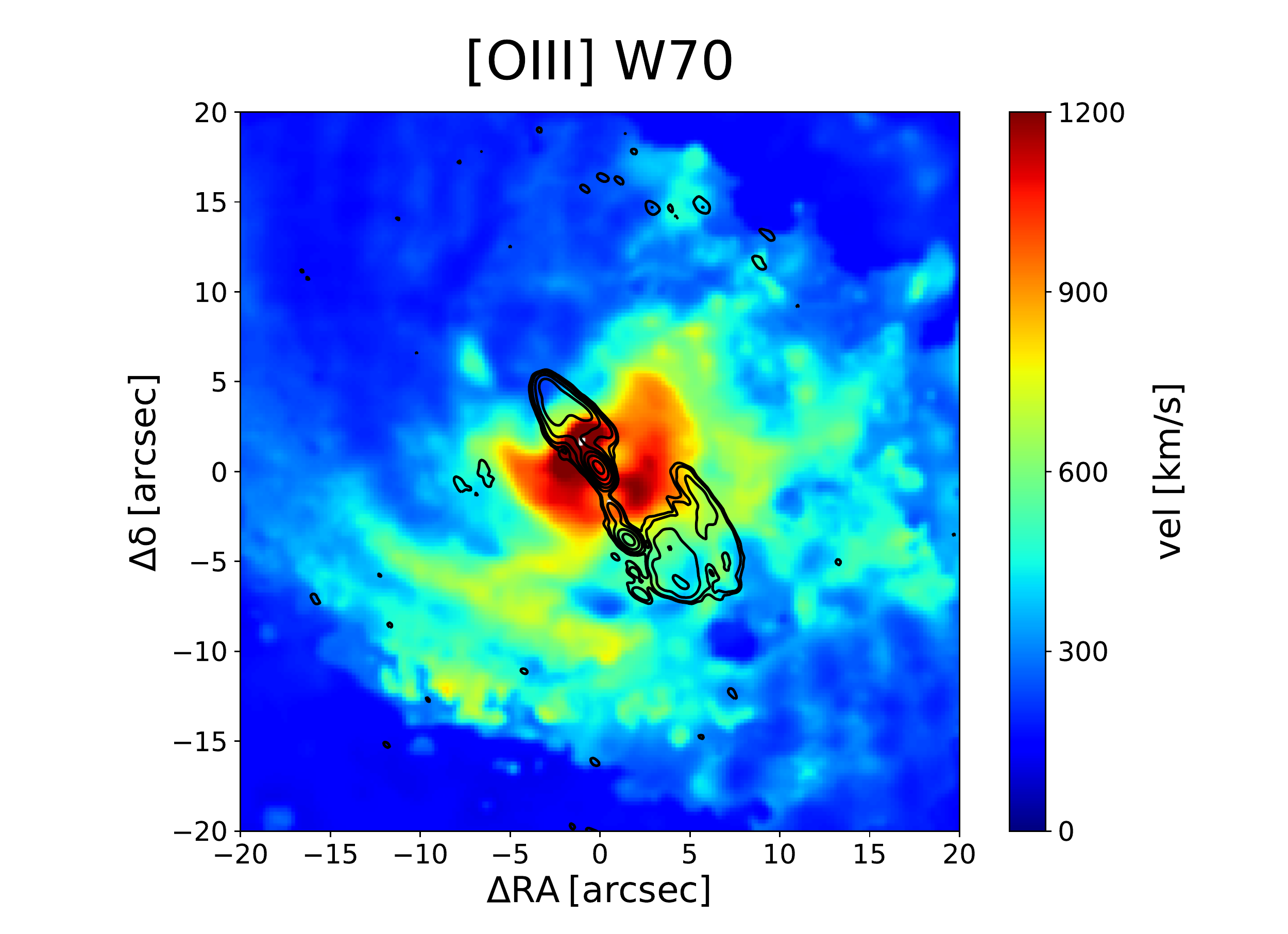}
        \end{subfigure}
        \hfill\null\\
        \hfill
        \begin{subfigure}[t]{0.01\textwidth}
        \textbf{c}
        \end{subfigure}
        \begin{subfigure}[t]{0.365\textwidth}\includegraphics[width=\textwidth,trim={11cm 1.5cm 11cm -1cm},clip]{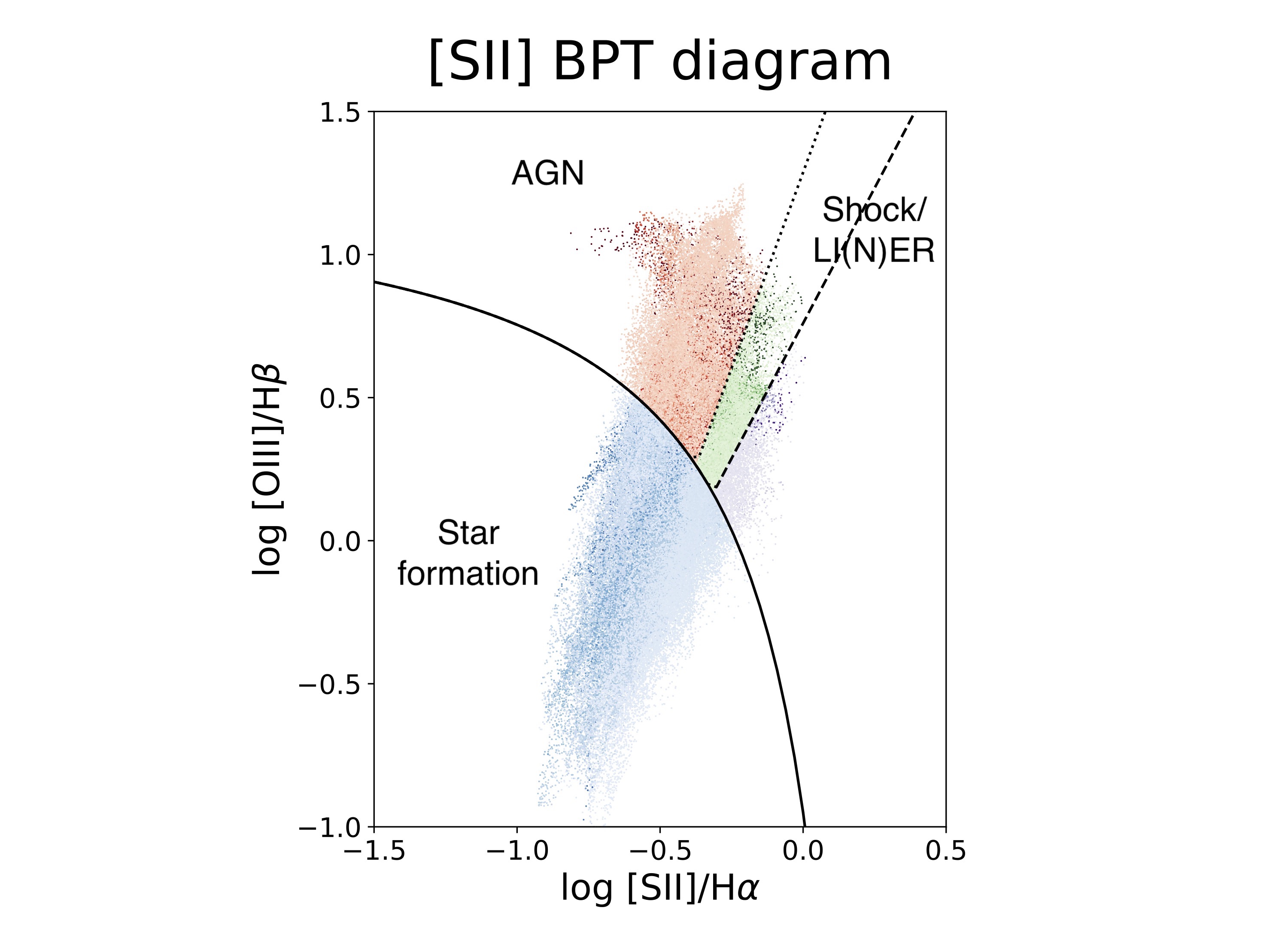}
        \end{subfigure}
        \hfill  
        \begin{subfigure}[t]{0.01\textwidth}
        \textbf{d}
        \end{subfigure}
        \begin{subfigure}[t]{0.365\textwidth}\includegraphics[width=\textwidth,trim={3.8cm 0.5cm 1cm 0.5},clip]{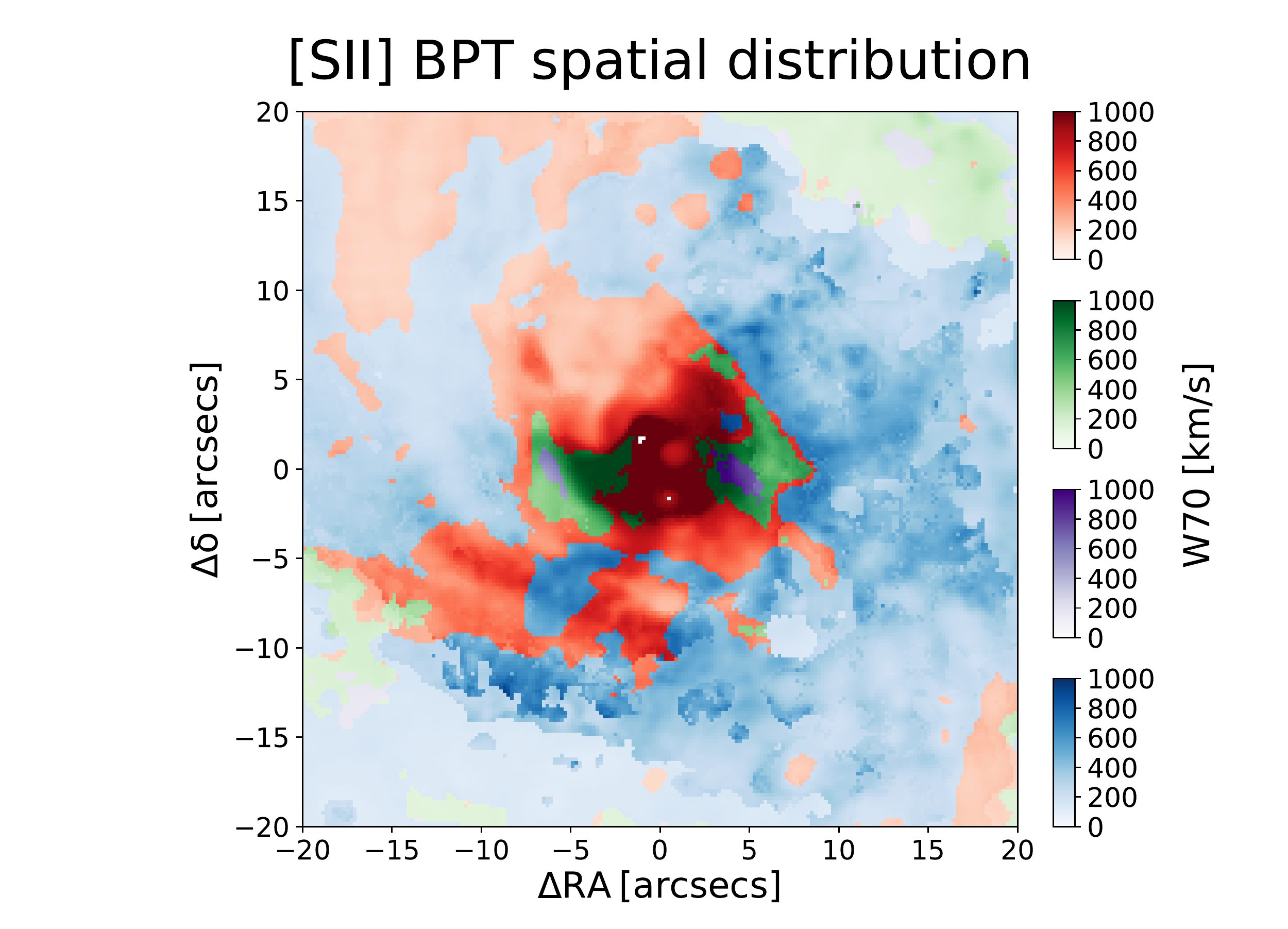}
        \end{subfigure}
        \hfill\null
        \caption{NGC 1068. Same as in Fig. \ref{fig:maps_ic5063_2}.}
        \label{fig:maps_n1068_2}
\end{figure*}

\begin{figure*}
    \centering
    \begin{subfigure}[t]{0.01\textwidth}
        \textbf{a}
        \end{subfigure}
        \begin{subfigure}[t]{0.47\textwidth}
        \centering
    \includegraphics[width=0.65\textwidth]{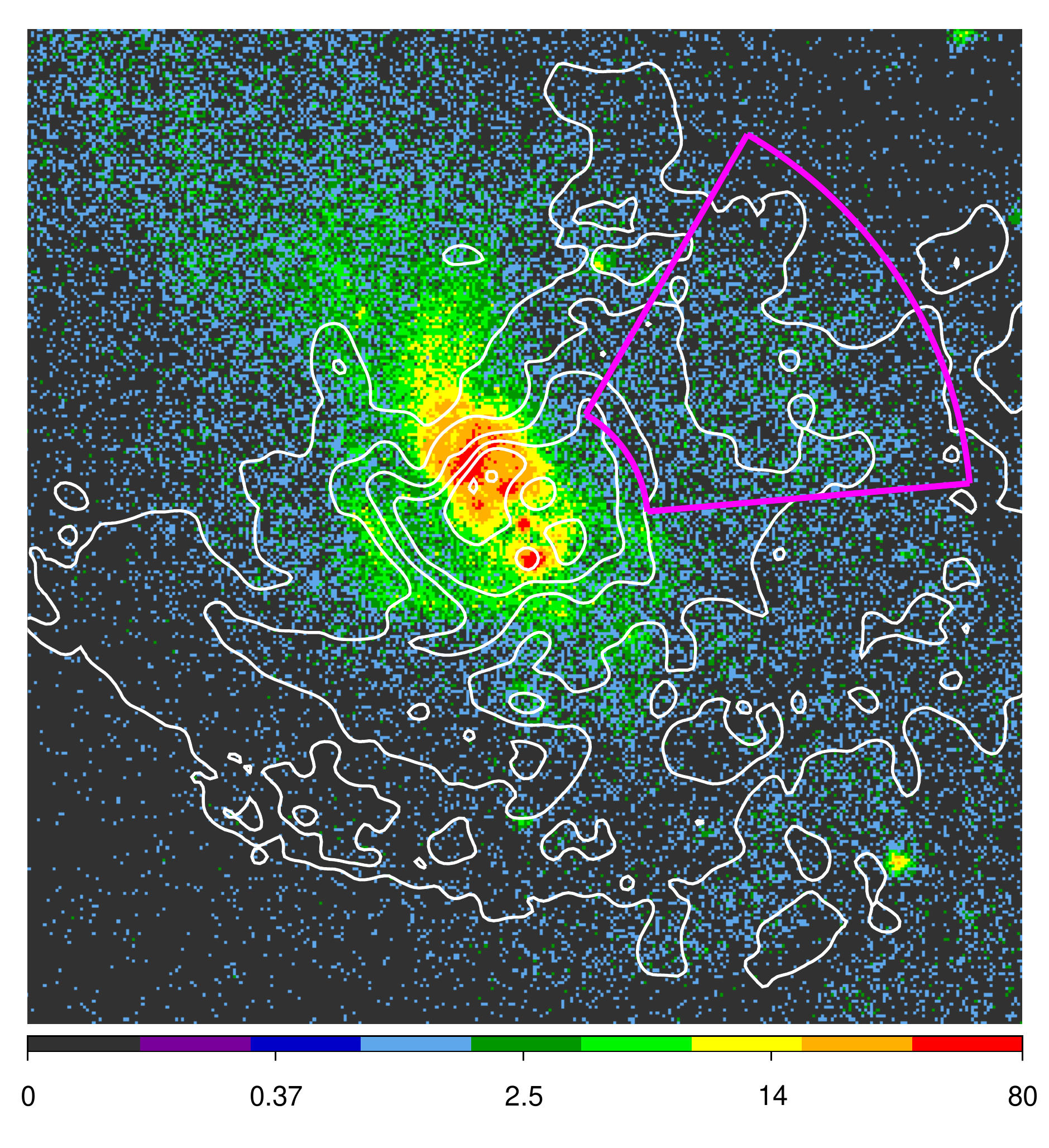}
    \end{subfigure}
    \hfill
    \begin{subfigure}[t]{0.01\textwidth}
        \textbf{b}
        \end{subfigure}
        \begin{subfigure}[t]{0.47\textwidth}
        \centering
    \includegraphics[width=0.9\textwidth,trim=0 -0.2cm 0 0 ]{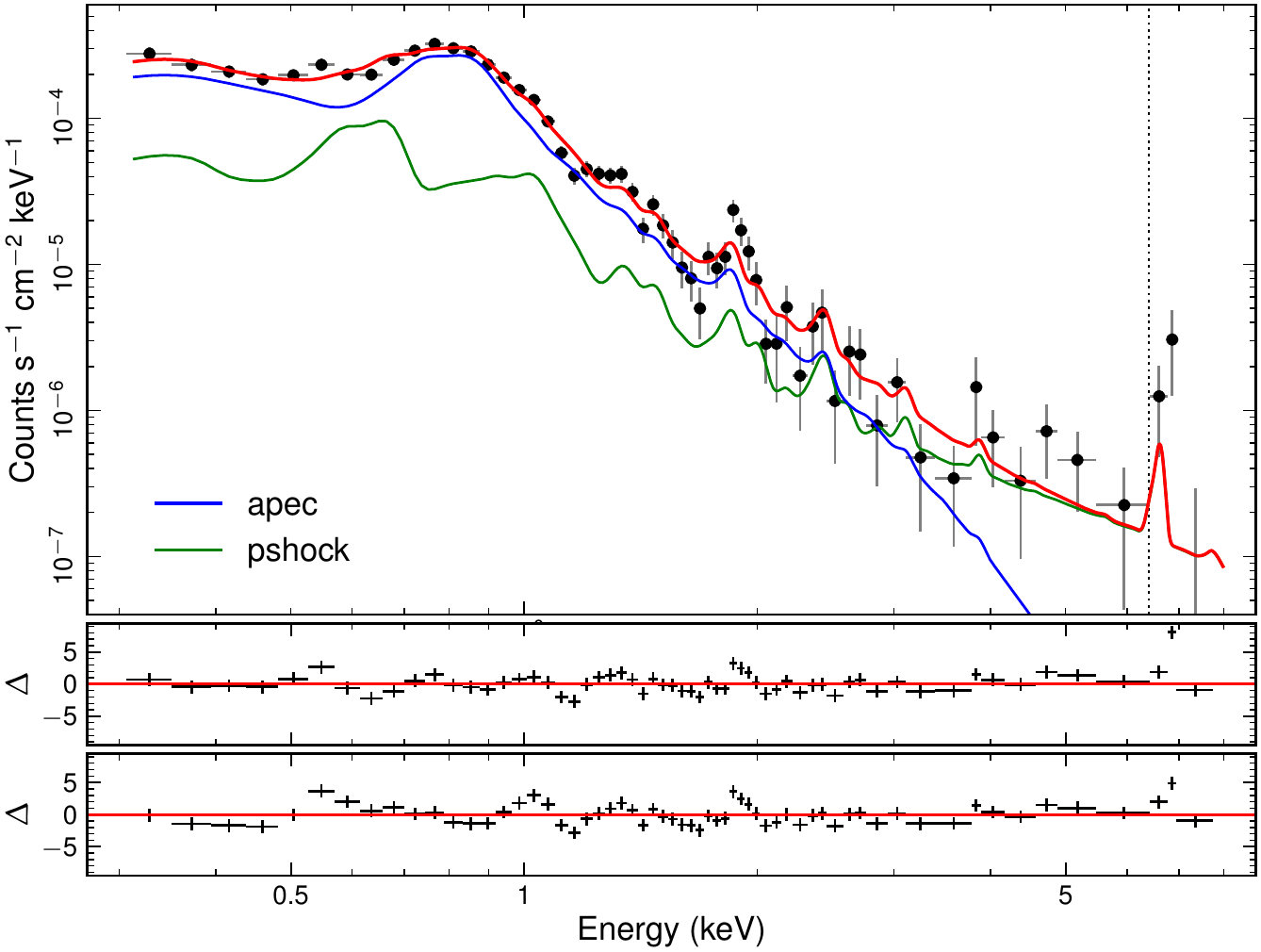}
    \end{subfigure}
    \caption{{\bf (a)} {\it Chandra} ACIS-S 0.3--8 keV X-ray image of NGC 1068 (in the same 40$''$\,$\times$\,40$''$ central region of the MUSE maps from Figs. \ref{fig:maps_n1068_1} and \ref{fig:maps_n1068_2}; reported values are photon counts). Contours of \oiii\ W70 (from Fig. \ref{fig:maps_n1068_2}b) are reported in white. {\bf (b)} Top panel: Spectrum (black points) extracted from the magenta region reported in panel {\bf (a)}. The  \texttt{apec} + \texttt{pshock} best-fit model is shown in red, and its separate components in other colours. The vertical dotted line marks 6.4 keV. The residuals of the \texttt{apec} + \texttt{pshock} and \texttt{apec} + \texttt{apec} fitting procedures (see text for further details), with $\Delta$ = (data -- model)/error, are reported for comparison in the middle and bottom panels, respectively.}
    \label{fig:n1068_xray}
\end{figure*}

NGC 1068 is a nearby prototypical Seyfert 2 galaxy located at a distance of $\sim$\,10.5 Mpc from Earth (1$''$\,$\sim$\,51 pc). 
The galaxy also shows powerful starburst activity, mainly concentrated in a prominent starburst ring of $\sim$1--1.5 kpc radius (e.g. \citealt{schinnerer2000}, \citealt{Garcia-Burillo:2014aa}) elongated in the NE-SW direction.
Like IC 5063, it is one of the brightest Seyfert galaxies at radio wavelengths ($P_\mathrm{1.4~GHz}$ = $2 \times 10^{23}$~W~Hz$^{-1}$, \citealt{Ulvestad:1984aa}), although it is still radio quiet (e.g. \citealt{Prieto:2010aa}, \citealt{Teng:2011aa}).
NGC 1068 hosts a radio jet spanning up to $\sim$\,800 pc in the NE-SW direction (e.g. \citealt{Gallimore:1996aa}, \citealt{Krips:2006aa}). The jet and the AGN ionisation cones, oriented in the same direction, are inclined at $\sim$45$^\circ$ with respect to the plane of the disc, such that in this case as well, the gas in the disc is illuminated by the AGN radiation and interacts with the jet (e.g. \citealt{cecil1990}, \citealt{bland-hawthorn1997}, \citealt{Garcia-Burillo:2014aa}). An outflow indeed propagates in the same direction, detected in the ionised (e.g. \citealt{Axon:1997aa}, \citealt{Crenshaw:2000aa}, \citealt{Cecil:2002aa}, \citealt{barbosa2014}) and molecular gas (\citealt{Garcia-Burillo:2014aa}, \citealt{Gallimore:2016aa}), driven by the AGN (\citealt{Garcia-Burillo:2014aa}). Extended X-ray emission is observed in the same direction on either side of the nucleus, following the shape of the AGN cones (\citealt{Bauer:2015aa}).

In Figs. \ref{fig:maps_n1068_1} and \ref{fig:maps_n1068_2} we show our MUSE flux and kinematic maps for NGC 1068, respectively.
Fig. \ref{fig:maps_n1068_1}a shows the FOV of our MUSE maps (red box) on a large-scale image of the galaxy, covering the central $\sim$3.3$\times$3.3 kpc$^2$.
The MUSE three-colour image in Fig. \ref{fig:maps_n1068_1}b displays the H$\alpha$ emission (red) that follows the spiral arms and the large circumnuclear star-forming ring.
[O\,\textsc{iii}] emission (green) is instead prominent in the known bright inner ionisation cone (e.g. \citealt{Macchetto:1994aa}) and in fainter outer lobes extending spirally from the inner one up to scales of $\sim$4.5 kpc to the NE (brighter) and to the SW (fainter; see also \citealt{bland-hawthorn1997}, \citealt{Lopez-Coba:2020aa}). This is consistent with the ionisation cones illuminating the spiralling gas in the disc, above it in case of the NE brighter cone, below it in case of the SW fainter cone. The spiral-like \oiii\ emission roughly traces the extended X-ray emission (\citealt{Bauer:2015aa}). Moreover, the [O\,\textsc{iii}] emission extends in the same direction as the radio jet (spanning about 5 arcsec per side), whose VLA 5 GHz (C band) A-array contours (from \citealt{Gallimore:1996aa}) are reported in Fig. \ref{fig:maps_n1068_1}c.

Similarly to IC 5063 and NGC 5643, Fig. \ref{fig:maps_n1068_2}a shows that the bulk of the high-velocity gas is elongated in the same NE-SW direction as the [O\,\textsc{iii}] inner cone and the radio jet, tracing an outflow that is spatially consistent with the jet.
The stellar velocity dispersion ranges from $\sim$60~km/s to $\sim$170~km/s across the FOV, therefore 200~km/s is an appropriate demarcation between rotating and outflowing material.

The [O\,\textsc{iii}] line velocity width map (Fig. \ref{fig:maps_n1068_2}b) instead shows a prominent and extended enhancement (over $\sim$30$'' \!\sim\,$1.5 kpc) perpendicular to the jet and ionisation cones, reaching values above W70 $\sim$ 1000~km/s around the centre. This enhancement is also tentatively observed in \ha\ by \cite{Lopez-Coba:2020aa}, although on a smaller scale and magnitude than \oiii\ reported here.

In this case,  the resolved \sii-BPT diagram (Fig. \ref{fig:maps_n1068_2}c,d) also exhibits shock or LI(N)ER-like ionisation to the east and west of the nucleus, approximately in the direction of the enhanced line velocity widths. This also clearly stands out in the \sii-BPT maps produced from the same MUSE data separately for the low- and high-velocity gas in Fig. D.3 of \cite{Mingozzi:2019aa}, where the spaxels in an elongated region to the east and west of the nucleus populate the area of the BPT diagram at high \sii/\ha\ ratios, which is indicative of shock or LI(N)ER ionisation.

\subsubsection*{Extended X-ray emission along the line-width enhancement}
To further investigate the presence of shocks perpendicularly to the AGN ionisation cones and jets, as inferred from spatially resolved BPT diagrams, we examined the extended X-ray emission in this galaxy through {\it Chandra} ACIS-S data. NGC 1068 is the only target of the four we analysed in this work whose archival {\it Chandra} observations have significant statistics along the minor axis of the AGN cones, in the region of line velocity width enhancement.

We retrieved the highest quality {\it Chandra} observation of NGC 1068 (ObsID: 344) from the archive, performed in February 2000. The data were reprocessed following the standard procedure with \textsc{ciao} v4.12 (\citealt{Fruscione:2006aa}).
We extracted the spectrum from an annular sector lying to the NW of the nucleus (see Fig. \ref{fig:n1068_xray}), centred on the hard X-ray nuclear source and defined from 5$\arcsec$ to 18$\arcsec$ and position angles (from W to N) from 5$\degr$ to 60$\degr$. This is spatially coincident with a portion of the W70 enhancement region (white contours). The circumnuclear X-ray emission of NGC 1068 is rather complex, and it was analysed in detail by \cite{Young:2001aa}. Similarly to the wider `west' region considered by these authors (their Fig. 1), the spectrum from our annular sector shows a hard (2--8 keV) continuum component, and tentative evidence of an iron emission line. We modelled the spectrum within \textsc{xspec} v12.11 (\citealt{Arnaud:1996aa}) with two thermal components from collisionally ionised gas (\texttt{apec}; \citealt{Smith:2001aa}) and obtained a barely acceptable fit with a $C$-statistics (\citealt{Cash:1979aa}) of 251/181. The soft, colder component has $kT$ = 0.72$\,\pm\,$0.02 keV and prefers a low metallicity (0.12$\,\pm\,$0.02 solar); the hard, hotter one ($kT$ $>$ 5 keV, poorly constrained) instead requires solar abundances in order to account for the prominent iron line. Clear residuals are seen, especially in the O\,\textsc{vii} and Si\,\textsc{xiii} He$\alpha$ bands. We then replaced the hot thermal component with a shock component (\texttt{pshock}; \citealt{Borkowski:2001aa}). The physical parameters of both components remain the same, but the fit significantly improves down to $C$-stat $=$ 218/181. While this is not conclusive proof of the presence of shocks in the W70 enhancement region, we note that the iron emission line falls in the Fe\,\textsc{xxv}--\textsc{xxvi} K$\alpha$ band, which is not compatible with fluorescence from cold gas. 

In conclusion, both MUSE BPT diagrams and {\it Chandra} X-ray data suggest the presence of shocks in the region of enhanced W70 perpendicular to the AGN ionisation cones and jet.

We also estimated the density of the X-ray-emitting gas, $n$, from its emission measure (EM\,$\sim$\,$n^2 V \eta$, where $V$ is the geometrical volume, assumed to be conical, of the extraction region and $\eta$ is a filling factor). For the two adopted models, we obtained EM $\sim$ 3.5--$5 \times 10^{62}$ cm$^{-3}$, which translates into a density $n$\,$\sim$\,0.2$-$0.3 $\eta^{-1/2}$ cm$^{-3}$. We employ this density estimate later in the discussion of the origin of the W70 enhancement (Sect. \ref{ssec:discussion}).

\subsection{NGC 1386}

\begin{figure*}
\centering
        \hfill
        \begin{subfigure}[t]{0.01\textwidth}
        \textbf{a}
        \end{subfigure}
        \begin{subfigure}[t]{0.29\textwidth}
        \centering \includegraphics[width=0.9\textwidth,valign=b,trim={-1cm -6cm -1cm 0},clip]{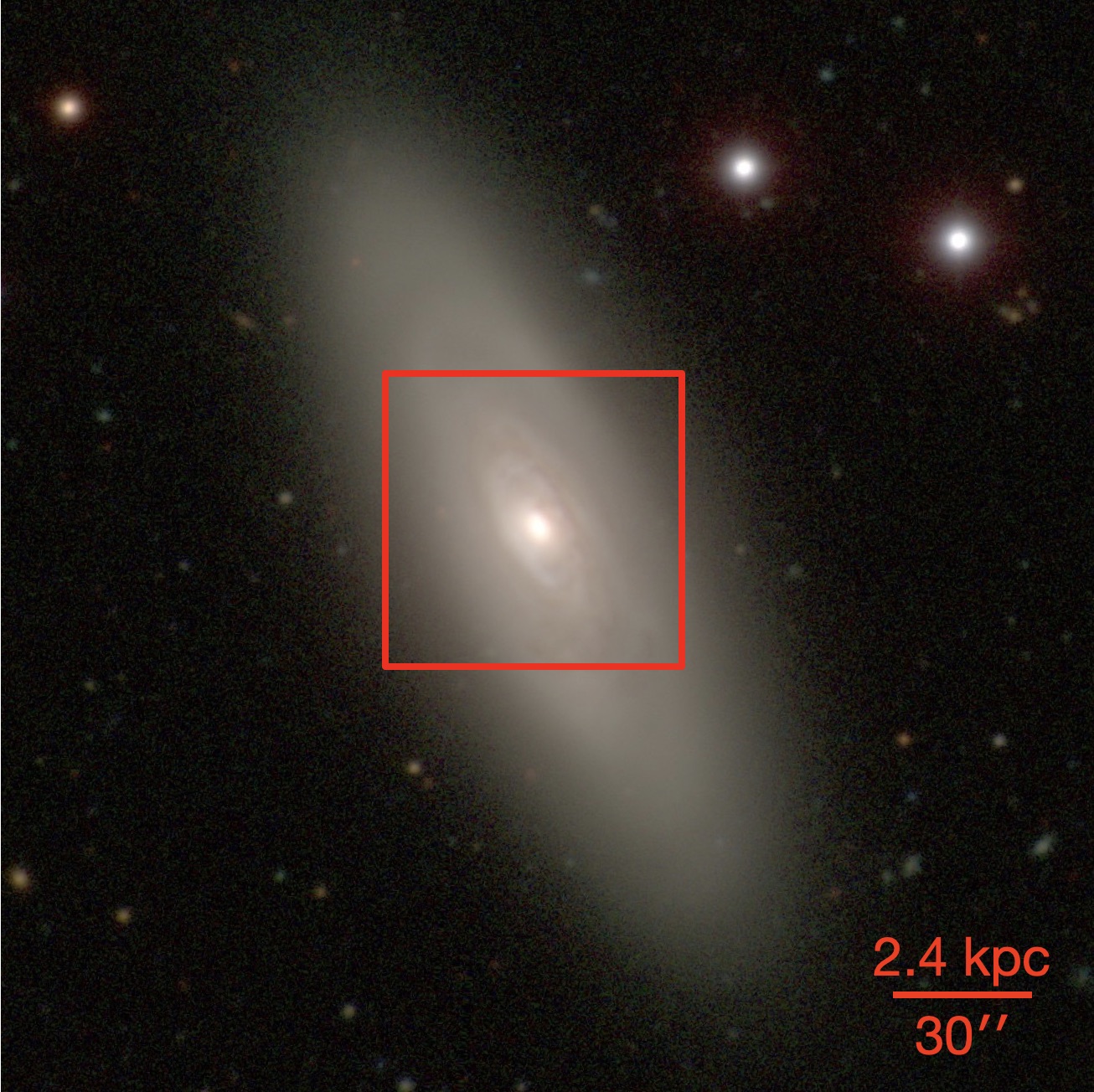}
        \end{subfigure}
        \hfill
        \begin{subfigure}[t]{0.01\textwidth}
        \textbf{b}
        \end{subfigure}
        \begin{subfigure}[t]{0.29\textwidth}
        \centering \includegraphics[width=\textwidth,valign=b,trim={3cm 0cm 4cm 3cm},clip]{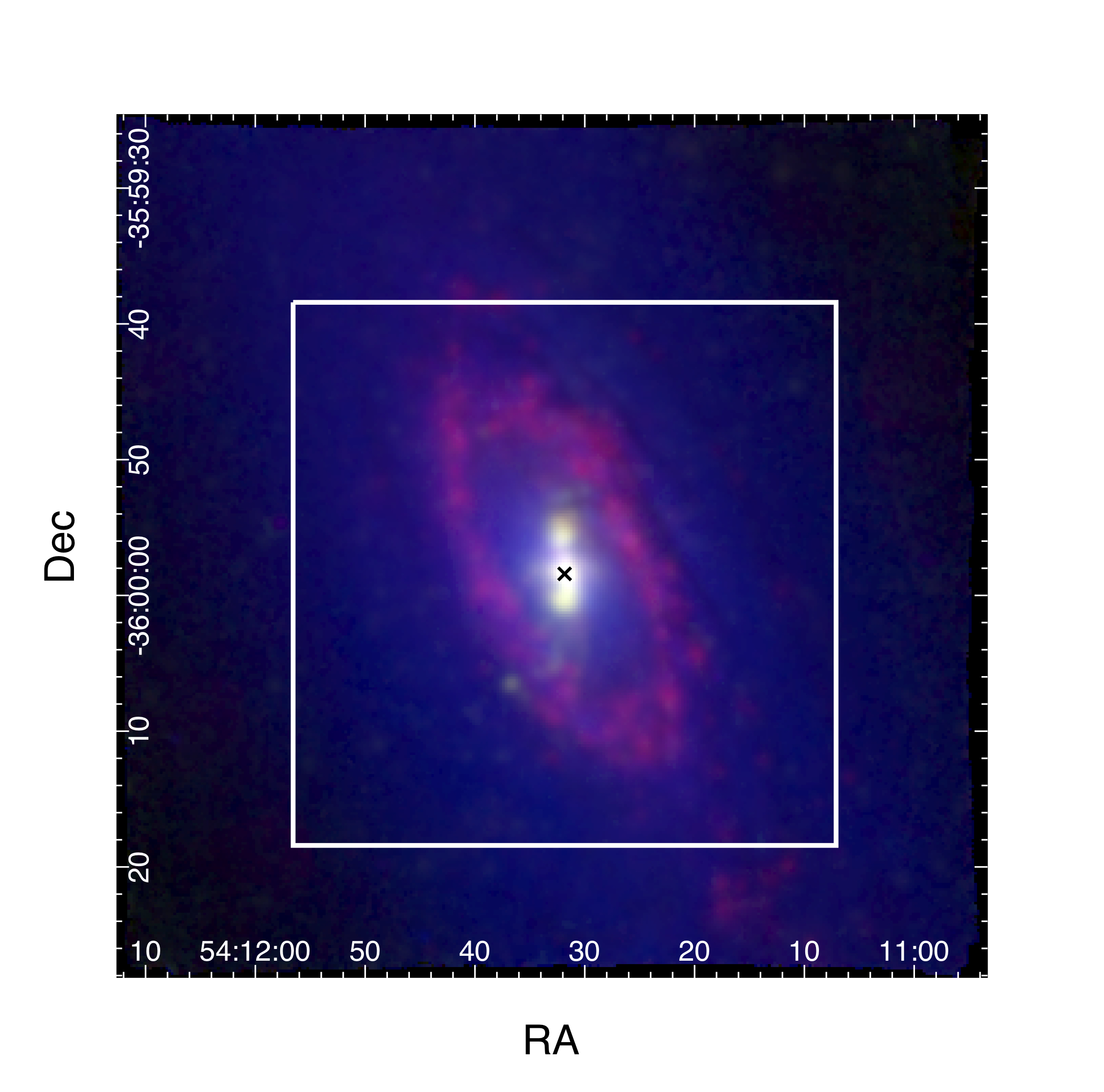}
        \end{subfigure}
        \hfill
        \begin{subfigure}[t]{0.01\textwidth}
        \textbf{c}
        \end{subfigure}
        \begin{subfigure}[t]{0.36\textwidth}\includegraphics[width=\textwidth,trim={2.6cm 0.5cm 2.3cm 0.5cm},clip]{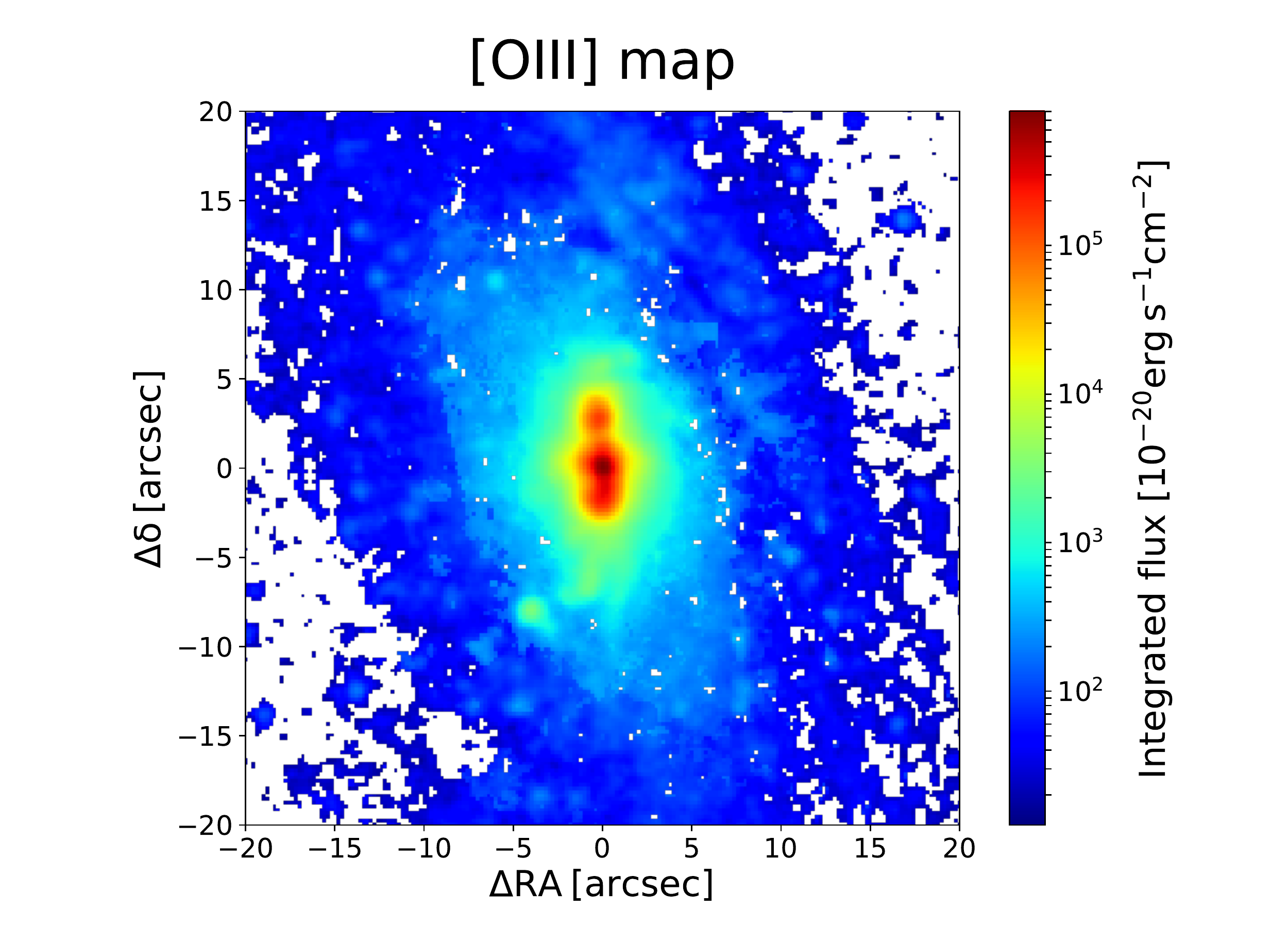}
        \end{subfigure}
        \hfill\null
                \caption{NGC 1386. {\bf (a)} Three-colour image of NGC 1386 ($B$ band in blue, $V$ band in green, and $I$ band in red) obtained with the 2.5 m du Pont Telescope at Las Campanas Observatory for the Carnegie-Irvine Galaxy Survey (CGS, \citealt{Ho:2011aa}). The red box shows the FOV of our MUSE map in panel b, whose side spans $\sim$5.1 kpc. Same as in Fig. \ref{fig:maps_ic5063_1} for {\bf (b)} and {\bf (c)}.}
        \label{fig:maps_n1386_1}

\vspace*{\floatsep}

        \centering
        \hfill
        \begin{subfigure}[t]{0.01\textwidth}
        \textbf{a}
        \end{subfigure}
        \begin{subfigure}[t]{0.36\textwidth}\includegraphics[width=\textwidth,trim={2.6cm 0.5cm 2.3cm 0.5cm},clip]{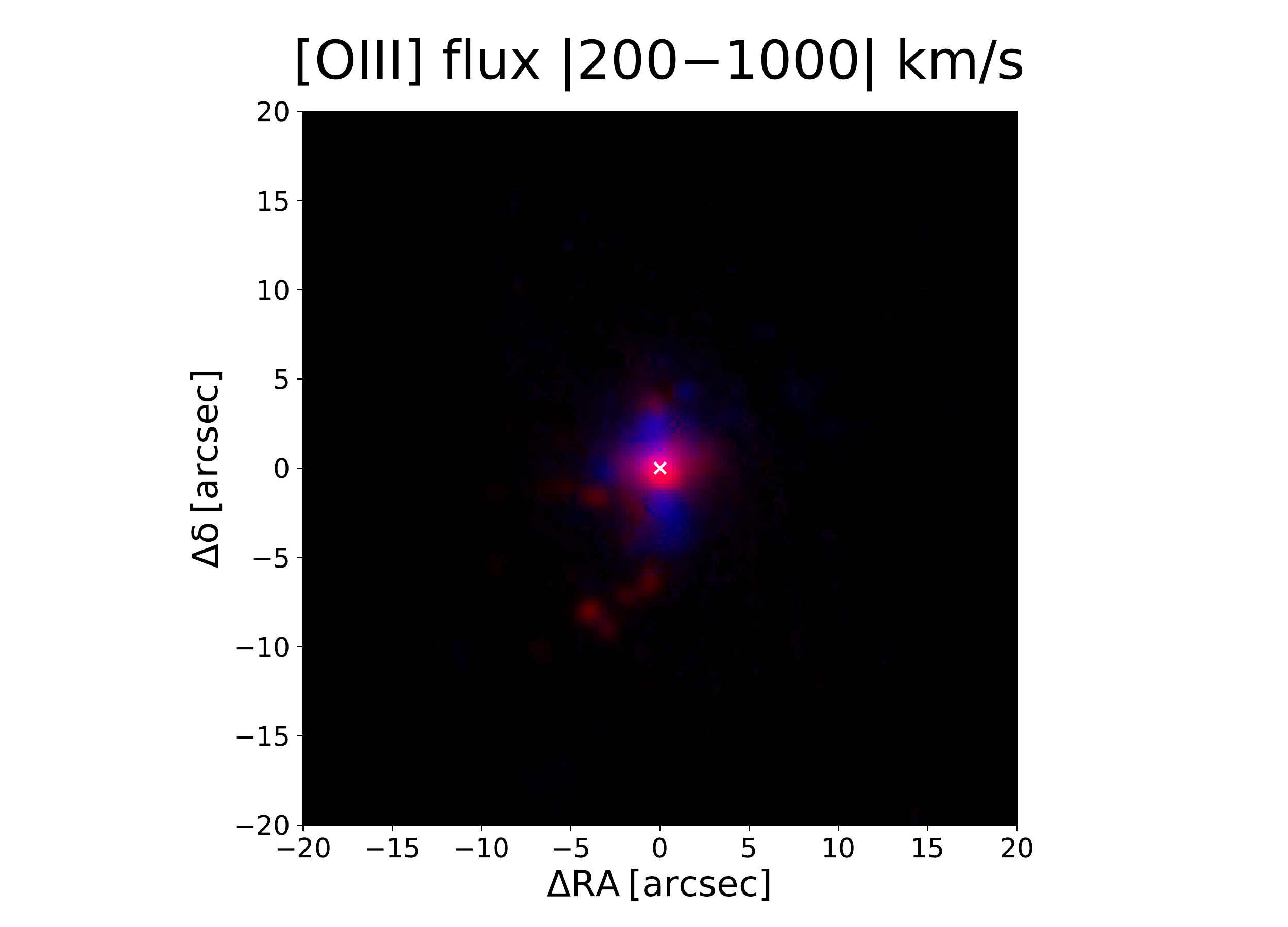}
        \end{subfigure}
        \hfill
        \begin{subfigure}[t]{0.01\textwidth}
        \textbf{b}
        \end{subfigure}
        \begin{subfigure}[t]{0.36\textwidth}\includegraphics[width=\textwidth,trim={2.6cm 0.5cm 2.3cm 0.5cm},clip]{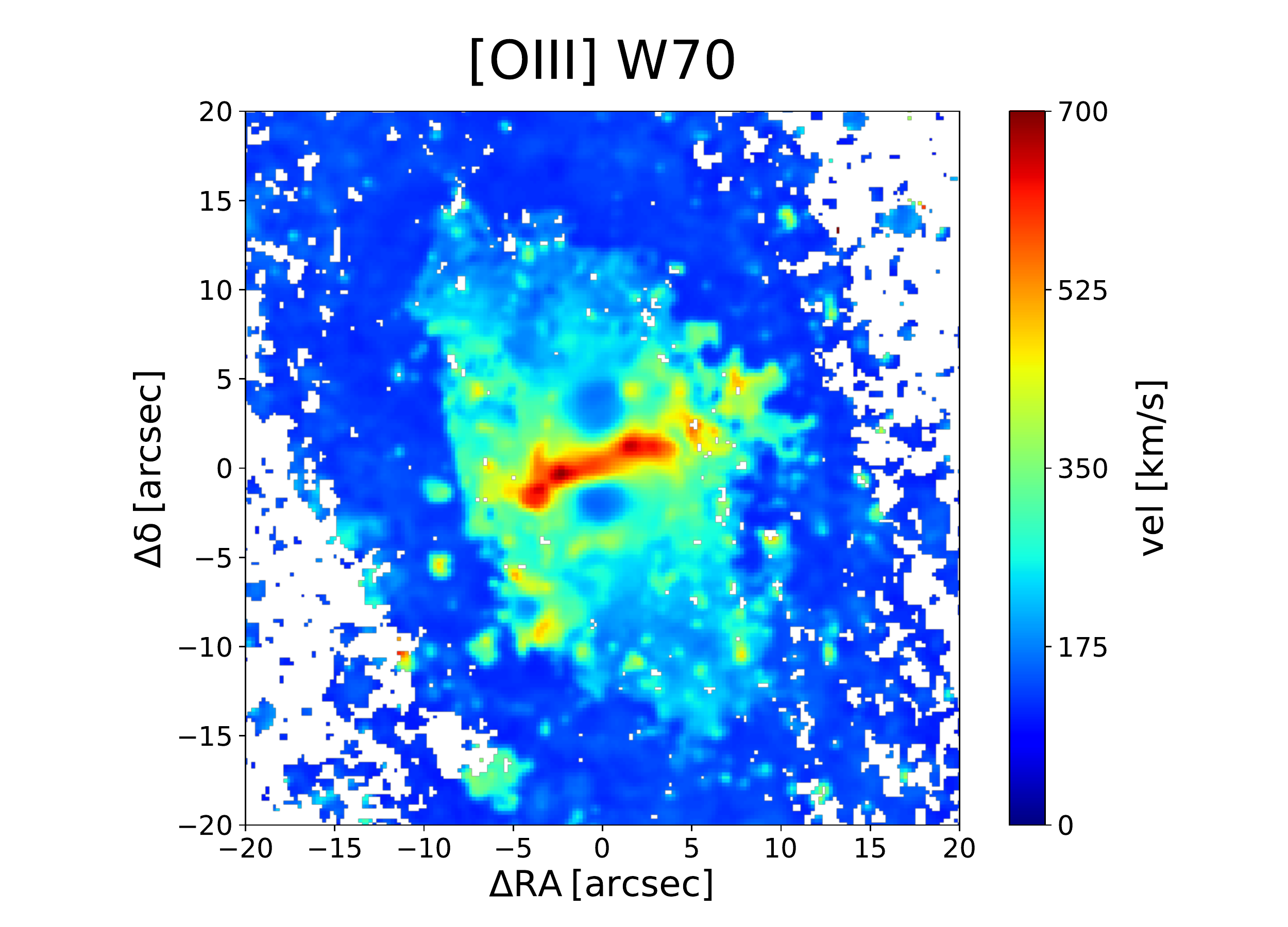}
        \end{subfigure}
        \hfill\null\\
        \hfill
        \begin{subfigure}[t]{0.01\textwidth}
        \textbf{c}
        \end{subfigure}
        \begin{subfigure}[t]{0.36\textwidth}\includegraphics[width=\textwidth,trim={11cm 1.5cm 11cm -1cm},clip]{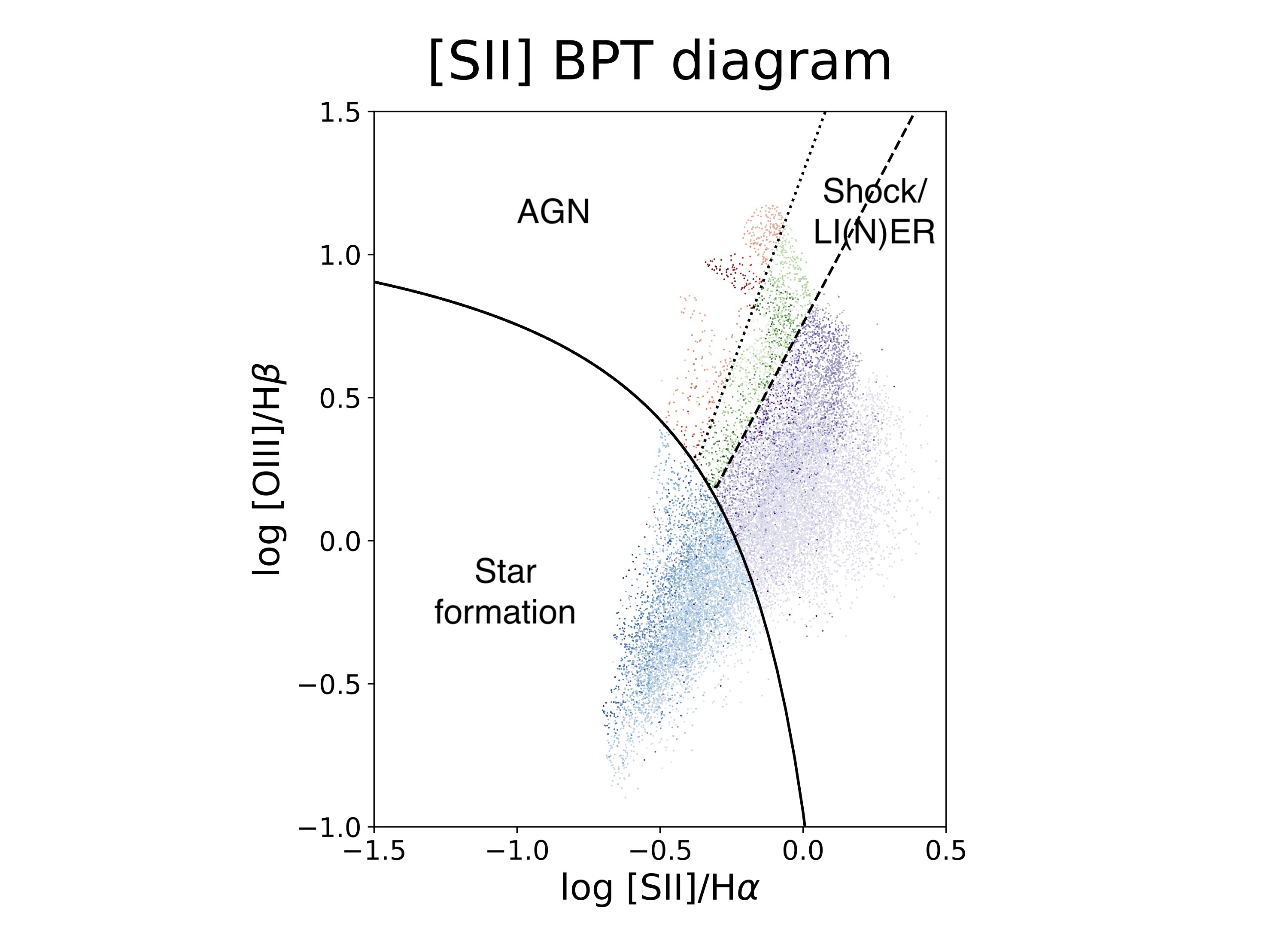}
        \end{subfigure}
        \hfill  
        \begin{subfigure}[t]{0.01\textwidth}
        \textbf{d}
        \end{subfigure}
        \begin{subfigure}[t]{0.36\textwidth}\includegraphics[width=\textwidth,trim={3.7cm 0.5cm 1.3cm 0.5},clip]{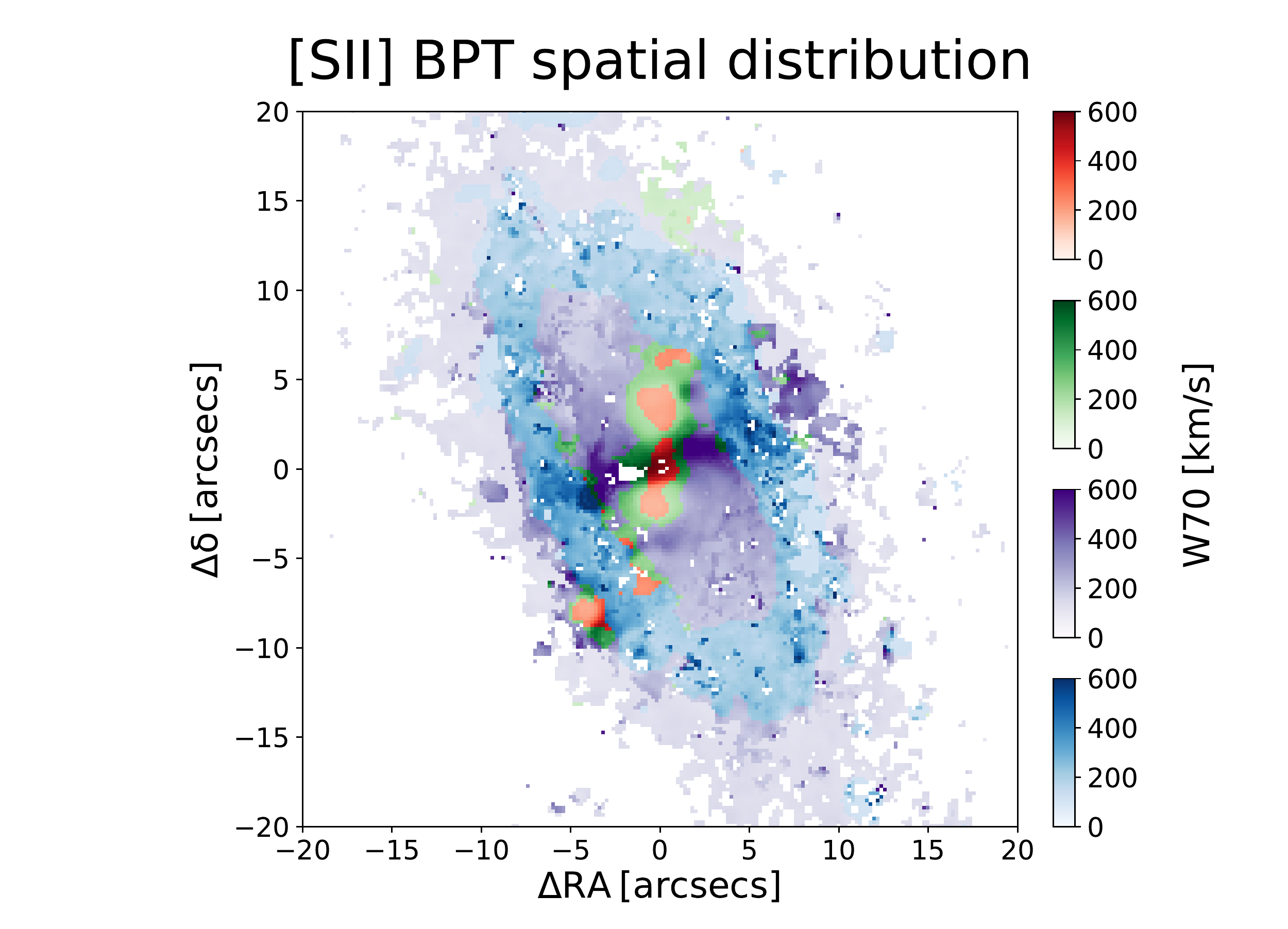}
        \end{subfigure}
        \hfill\null
        \caption{NGC 1386. Same as in Fig. \ref{fig:maps_ic5063_2}.}
        \label{fig:maps_n1386_2}
\end{figure*}

NGC 1386 ($D$\,$\sim$\,16.4 Mpc; 1$''$\,$\sim$\,80 pc) is a radio-quiet ($P_\mathrm{1.4~GHz}$ $\simeq$ $4.6 \times 10^{20}$~W~Hz$^{-1}$, \citealt{Ulvestad:1984aa}) Seyfert 2 spiral galaxy, inclined by $\sim$65$^\circ$ with respect to the line of sight (\citealt{Lena:2015aa}; see also Fig. \ref{fig:maps_n1386_1}a). 
\cite{ferruit2000} reported two elongated  emission-line structures over $\sim$2$''$ to the N and S of the nucleus from [O\,\textsc{iii}] and H$\alpha$+[N\,\textsc{ii}] images, likely due to the AGN ionising radiation field illuminating the gas in the disc (\citealt{Lena:2015aa}). Elongated radio emission, suggestive of a radio jet, is observed to the S of the nucleus across $\sim$1$''$ ($\sim\,$80 pc), and it is also marginally resolved to the N (\citealt{Nagar:1999aa}, \citealt{mundell2009}).
\cite{Lena:2015aa} found broad optical emission-line profiles in the perpendicular (E-W) direction across $\sim$2--3$''$, suggesting the presence of rotation and/or of an outflow in this direction. 


In Figs. \ref{fig:maps_n1386_1} and \ref{fig:maps_n1386_2} we report our MUSE flux and kinematic maps for NGC 1386, respectively.
The red box in Fig. \ref{fig:maps_n1386_1}a shows the $\sim$\,5.1\,$\times$\,5.1 kpc$^2$ FOV of our MUSE maps superimposed on a large-scale image of the entire galaxy.
Fig. \ref{fig:maps_n1386_1}b displays our MUSE three-colour image. 
H$\alpha$ (red) dominates in a prominent circumnuclear star-forming ring.
Inside it, the [O\,\textsc{iii}] emission traces the entire N-S ionisation lobes spanning $\sim$200--300 pc per side, first reported by \cite{ferruit2000}, which further extend to fainter and clumpy emission-line structures towards the SE and NW in an S-shaped pattern.

The bulk of high-velocity [O\,\textsc{iii}] emission, tracing an outflow of ionised gas, is mostly elongated in the N-S direction (Fig. \ref{fig:maps_n1386_2}a), following the [O\,\textsc{iii}] ionisation lobes and the radio jet.
The stellar velocity dispersion across the galaxy is in the range 60$-$130 km/s, therefore the threshold of 200 km/s is appropriate to isolate the contribution of the outflow from that of the rotating disc.

The [O\,\textsc{iii}] line velocity width map (Fig. \ref{fig:maps_n1386_2}b) instead shows a strong and extended (over $\sim$1.5 kpc) enhancement, up to W70 $>$ 600 km/s, almost perpendicular to the ionisation lobes, the ionised outflow, and the putative small-scale radio jet, indicating fast/turbulent motions away from the nucleus in this direction.

The \sii-BPT diagram and associated map for NGC 1386 (Fig. \ref{fig:maps_n1386_2}c,d) show that in this case as well, the gas in the direction of line velocity width enhancement (darker colours in the BPT map), perpendicular to the narrow AGN ionisation lobes (which are clearly visible in the N-S direction in panel d) and radio jet, is dominated by shock or LI(N)ER ionisation.
Unfortunately, the BPT map for NGC 1386 is not visually straightforward and the shock or LI(N)ER-like ionisation does not stand out clearly in this perpendicular direction as it does instead in IC 5063 and NGC 5643. 
The circumnuclear star-forming ring dominates the gas ionisation at about 5$''$ to the east and west of the nucleus, masking the shock or LI(N)ER ionisation, which on the other hand almost entirely fills  the area encompassed within the ring and outside it (except for the AGN lobes).

\section{Discussion: Enhanced line velocity widths perpendicular to radio jets}

\subsection{Incidence of the phenomenon in the MAGNUM survey}
We detect strongly enhanced line widths of [O\,\textsc{iii}] emission line profiles perpendicularly to radio jets and to the [O\,\textsc{iii}] ionisation cones in four galaxies of our sample, IC 5063, NGC 1386, NGC 5643, and NGC 1068.
We note that these objects alone out of the nine we analysed so far from our MAGNUM survey show evidence of a low-power ($\lesssim\,$10$^{44}$~erg~s$^{-1}$) radio jet that interacts with the ISM of the host galaxy given its low inclination with respect to the disc. All the other objects 
in our sample do not show any evidence for such enhanced line velocity widths perpendicular to their ionisation cones and to their high-velocity component tracing the outflowing gas (see e.g. \citealt{Venturi:2018aa} for NGC 1365 and \citealt{Venturi:2017aa} for NGC 4945).

Interestingly, neither in Centaurus A, where a radio jet is present, nor in Circinus, whose radio lobes might be indicative of an (undetected) radio jet, we detect this enhanced W70 feature from MUSE data.
Nevertheless, the nature and structure of the radio emission in these two objects are much different from in those presented in this work. 
The radio lobes of Circinus are roughly perpendicular to the galaxy disc (\citealt{Elmouttie:1998aa}, \citealt{Mingo:2012aa}).
The jet in the radio-loud
Centaurus A is also directed  perpendicular to the galaxy disc, and additionally, it is in a much more evolved state: its outer lobes extend to $\sim$250 kpc (e.g. \citealt{Israel:1998aa}). However, given the high extinction in the central region of these two galaxies (see e.g. \citealt{Mingozzi:2019aa}), we cannot completely exclude perturbed, high-W70 gas in their cores.
Near-IR emission lines  show a velocity dispersion enhancement perpendicular to the jet direction in Centaurus A that is aligned with the galaxy optical major axis (\citealt{Neumayer:2007aa}), but on very small scales ($\sim$1$-$2$''$, i.e. $\lesssim$20$-$40 pc), far different from the kiloparsec-scale extension observed for the galaxies presented in this work. The authors interpreted this small-scale velocity dispersion enhancement as due to an inclined nuclear hot gas disc, and successfully included it in a rotating-disc model. However, in principle it cannot be excluded that the base of the jet might also be responsible for it.
Circinus instead does not show any sign of a central enhanced line velocity width perpendicular to the direction of the ionisation cone and radio lobes even from near-IR data (\citealt{Muller-Sanchez:2011aa}).

In the other galaxies belonging to our MAGNUM survey that do not host a radio jet, we do not observe the strongly enhanced line velocity widths (see e.g. the W70 maps of NGC 4945 in \citealt{Venturi:2017aa} and of NGC 1365 in \citealt{Venturi:2018aa}) that we find in the four jetted objects presented in this work. 
We stress that the W70 enhancement feature does not seem to be associated with the presence of ionisation-cone outflows, which are observed in all MAGNUM galaxies, or with a higher AGN power. 
For instance, the AGN luminosities (as traced by the Swift BAT 14--195 keV hard X-ray luminosity; \citealt{Baumgartner:2013aa}) of NGC 4945 and NGC 1365, which do not host a jet, nor exhibit the perpendicular line-width enhancement phenomenon, are higher than those of NGC 1068 and NGC 5643.
A jet at low inclinations with respect to the galaxy disc seems to be the only feature that the four galaxies studied in this work have in common.



\subsection{Observation of the phenomenon in previous works \label{sec:other_works}} 

Enhanced line velocity widths perpendicular to radio jets (and to the AGN ionisation cones) are observed in a few other local Seyfert galaxies, which we summarise in the following.
We stress that this list is based on our search in the literature and might thus be incomplete.

First, \cite{Couto:2013aa} reported an enhanced velocity dispersion of optical emission lines in a thick band perpendicular to the radio jet and to the \oiii\ emission axis in Arp 102B, but opposite bulk gas velocities were also detected in the dispersion-enhancement direction, which they alternatively interpreted as either due to a small-scale bipolar outflow, to an inner rotating disc, to precession of the jet, or to the lateral expansion of gas in a cocoon around the propagating radio jet. In support to the latter explanation, the results from \cite{Fathi:2011aa} and \cite{Couto:2013aa} indicated that the radio jet impacts on the circumnuclear gas in the galaxy disc.


\cite{Riffel:2014aa, Riffel:2015aa} observed strip-like enhanced velocity dispersion (resembling in shape what we observe in IC 5063 and NGC 1386) in NGC 5929 perpendicular to the radio jet and to the bright emission-line lobes. The feature is found in both ionised gas from \feii\ and Pa$\beta$ and warm molecular gas from H$_2$. Even though the results from their work indicate an interaction of the jet with the gas in the disc, they ascribed the velocity width enhancement to an equatorial outflow from the accretion disc or the dusty torus, supported by the detection of weak extended 0.4 GHz radio emission approximately in the direction of velocity dispersion enhancement by \cite{Su:1996aa}, who attributed it to a flow of relativistic particles launched by the AGN perpendicular to the main radio jet.

The phenomenon was observed also in NGC 2110, where a broadening of velocity dispersion was detected in an elongated area roughly perpendicular to the radio jet and emission-line axis in ionised gas from \oiii\ and \nii\ (\citealt{Gonzalez-Delgado:2002aa}, \citealt{Schnorr-Muller:2014aa}) and warm molecular gas from H$_2$ (\citealt{Diniz:2015aa}). These authors interpreted it as due to gas in the disc disturbed by a nuclear outflow. \cite{Rosario:2010aa} derived an inclination of 20$\degree$ between the jet and the galaxy disc plane in this object.

\cite{Lena:2015aa} observed in the inner $\sim$7$''$ of NGC 1386 with the Gemini Multi-Object Spectrograph (GMOS) Integral Field Unit (IFU) the elongated enhanced velocity dispersion perpendicular to the small radio jet and AGN cones that we fully observe with MUSE on larger scales ($\sim$20$''$). They interpreted it as possibly due to a wind that propagates equatorially away from the dusty AGN torus and rotates about the radiation cone axis.

\cite{Schnorr-Muller:2016aa} observed a band of enhanced velocity dispersion in \nii\ perpendicular to the major axis of the ionised gas emission and to the direction of the bi-polar ionised outflow in NGC 3081. There is tentative evidence for a jet from radio data (\citealt{mundell2009}, \citealt{Nagar:1999aa}), showing marginally resolved emission aligned with the major axis of the ionised gas (thus perpendicular to the velocity dispersion enhancement), although the radio data cannot confirm or rule out a jet.

\cite{Freitas:2018aa} observed a line velocity width enhancement perpendicular to \oiii\ lobes and radio jets in Mrk 79 and Mrk 607, interpreting it as either lateral expansion of the gas due to the passage of a radio jet, and/or expansion of the dusty torus surrounding the nucleus. \cite{Riffel:2013aa} mentioned that the co-spatiality of the ionised bipolar lobes and the radio jet in Mrk~79 indicates an interaction of the jet with the ISM.

\cite{Finlez:2018aa} observed broad optical emission lines (especially \oiii) in NGC 3393 in a thick band perpendicular to radio jet and optical ionisation axis, and also interpreted it as an equatorial outflow from the accretion disc. According to \cite{Finlez:2018aa}, the tight co-spatiality of radio jet and emission-line lobes and the perpendicularity of the edge-on water maser (\citealt{Kondratko:2008aa}) to the jet and the almost face-on galaxy major axis indicate that the jet is launched into the disc of the galaxy.

\cite{Shimizu:2019aa}, \cite{Durre:2019aa}, and \cite{Shin:2019aa} reported elongated fan-shaped (very similar to those we observe in IC 5063 and NGC 1386) enhanced \oiii\ and Balmer lines velocity dispersion (also in \nii, \sii\ and \oi\ in \citealt{Shimizu:2019aa}) perpendicular to the bi-conical ionised outflow and radio jet in NGC 5728, with shock or LI(N)ER-like emission line ratios, as in our case. \cite{Durre:2018aa,Durre:2019aa} also stressed that the jet impacts the ISM in the galaxy disc and its axis is nearly parallel to the plane of the galaxy. Interestingly, \cite{Shimizu:2019aa} reported an elongated velocity dispersion enhancement in the same perpendicular direction and scale in the molecular phase from CO(2-1) as well, whose values are a factor $\sim$3 lower than in the ionised phase. \cite{Durre:2019aa} and \cite{Shin:2019aa} attributed the high dispersion values perpendicular to outflow and jet as due to beam smearing, causing the line of sight to intersect both the approaching and the receding side of the ionisation-cone bi-polar outflow and thus artificially broadening the line. \cite{Shin:2019aa} also mentioned the possibility of an equatorial outflow, making an analogy with the case of NGC 5929 from \cite{Riffel:2014aa} (mentioned above), but rejected it because equatorial radio emission is lacking.

Finally, \cite{Feruglio:2020aa} reported a large \oiii\ velocity dispersion in ESO 428- G 014 perpendicular to the low-power radio jet and to the high bulk velocity outflow detected in CO, H$_2$ and \oiii. Similarly to \cite{Shimizu:2019aa} mentioned above, they also detected an enhancement in the velocity dispersion of CO(2-1) for the molecular gas in the same direction and on the same scale as that in \oiii,\ but smaller by a factor of $\sim$\,6--8. According to \cite{Riffel:2006aa}, the jet is launched at small inclinations into the galaxy disc and impacts its ISM (also \citealt{Falcke:1996aa}).

To summarise, all these galaxies show enhanced line velocity widths perpendicular to their radio jets and ionisation cones/lobes.
We stress that these galaxies also host a normal ionised outflow with high bulk velocity in the direction of the ionisation lobes and jets and that their jets are low power ($\lesssim\,$10$^{44}$~erg~s$^{-1}$) and small scale, extending to a few arcsec (i.e. $\lesssim\,$1 kpc).
Moreover, most of them have a strong brightening of the optical line emission in correspondence to the radio jet hotspots in common, indicating an interaction of the radio jets and the ISM of the host galaxy.

\subsection{Discussion of the origin of the phenomenon}\label{ssec:discussion}
All the systems studied in this work, together with those discussed in Sect. \ref{sec:other_works}, show (compact) low-power radio jets propagating from their nuclei and strongly enhanced gas velocity dispersions on scales of a few kiloparsec in the direction perpendicular to the jet propagation.
Based on the above discussion, we propose that the most likely origin for the observed phenomenon is the radio jet that strongly interacts with the ISM in the galaxy disc during its propagation through it, releasing energy and giving rise to highly turbulent motions in the perpendicular direction. 
As mentioned earlier, the inclination angle of the jets with respect to the galaxy disc plane is indeed low enough to allow strong interaction with the disc ISM, which seems evident even when the jet inclination is not derived.
This scenario is further supported by the shock-like line ratios, which might be associated with such turbulent material, that are detected in the same perpendicular direction (consistent with \citealt{Mingozzi:2019aa}, in which the same MUSE observations were employed). 
The high \sii/\ha\ and \nii/\ha\ observed there (up to values $\sim$0.2$-$0.3 in log, see the BPT diagrams in Figs. \ref{fig:maps_ic5063_2}c,d, \ref{fig:maps_n5643_2}c,d, \ref{fig:maps_n1068_2}c,d, and \ref{fig:maps_n1386_2}c,d and in \citealt{Mingozzi:2019aa}) are indeed reproduced by  shock models (\citealt{Allen:2008aa}) with shock velocities in the range 100$-$1000 km/s (details in \citealt{Mingozzi:2019aa}). In addition, the spectrum of the extended X-ray emission from {\it Chandra} in the region of line-width enhancement in NGC 1068 (the only target out of the four presented with enough X-ray statistics in such region) is also consistent with the presence of shocks.
Moreover, as stressed before, objects in our MAGNUM survey that do host a jet (Centaurus A) or show indication of it (Circinus), but perpendicular to the galaxy disc, exhibit no line-width enhancement perpendicular to the jet or, at most, only on very small scales ($\lesssim\,$20$-$40 pc in Centaurus A). This is consistent with a scenario in which jets launched perpendicular to the galaxy disc have weak or no interaction with the disc ISM, as opposed to those launched close to the disc plane.

Simulations of jets propagating in a clumpy medium (as the ambient gas in galaxies is expected to be) indeed indicate that the effects of jets on the ISM are extremely different depending on their power and on their inclination with respect to the galaxy disc (\citealt{Wagner:2011aa}, \citealt{Mukherjee:2016aa,Mukherjee:2018aa,Mukherjee:2018ab}).
According to such simulations, jets launched perpendicular to the disc will have a very weak impact on the ISM of the galaxy, whereas jets directed at small inclinations (or even up to 45$\degree$) over the disc plane will strongly interact with the clumpy ISM and struggle to proceed through it. Moreover, while high-power jets ($\gtrsim$10$^{45}$~erg~s$^{-1}$) will more easily penetrate  the disc due to their strong ram pressure and impact on it mainly in the proximity of their path, more dramatic jet-ISM interaction will occur in the case of low-power jets ($\lesssim$10$^{44}$~erg~s$^{-1}$), as those hosted in the four sources presented and in those from the literature mentioned in Sect. \ref{sec:other_works}. 
In this circumstance, the jet will propagate extremely slowly through the disc, while at the same time, secondary streams will percolate through the porous interstellar medium, shocking and dispersing clouds in all directions.
This will widely perturb the disc ISM, especially in the direction of minor resistance perpendicular to the disc and primary jet stream, giving rise to strong turbulence in this direction (\citealt{Mukherjee:2018aa,Mukherjee:2018ab} and D. Mukherjee, priv. comm.).
\cite{Mukherjee:2018ab} performed simulations tailored to IC 5063 to explain the properties of the molecular gas outflow along the jet path. These simulations predict that for jet powers $\lesssim\,$10$^{45}$ erg/s, the gas clouds perpendicular to the disc would be visible in \ha, up to kiloparsec scales and with velocity dispersions between a few 10s and a few 100s km/s.
However, with current observations we cannot infer the 3D geometry of the velocity width enhancement in relation to the galaxy disc, if perpendicular to it, as predicted by the above simulations, or in the disc plane.
Nevertheless, the densities we infer for the material exhibiting the velocity width enhancement, of several 100s cm$^{-3}$ for the optically emitting ionised gas (see Table \ref{table:mass} presented later) and of $\ga$0.2$-$0.3 cm$^{-3}$ for the hot X-ray emitting gas in NGC 1068 (Sect. \ref{sec:n1068}), suggest that it either resides in the galaxy disc or (if located perpendicularly to it) originates from the disc rather than being pre-existing galactic halo gas, for which densities of 0.01--0.1 cm$^{-3}$ and 0.001 cm$^{-3}$ would be expected for the two gas phases (e.g. \citealt{Savage:1995aa}, \citealt{Putman:2012aa}).
In summary, while it is unclear whether these simulations can reproduce in detail the observational features discussed in this work, 
they clearly highlight that low-power jets with low inclinations on the galaxy disc can strongly affect the host galaxy ISM.

We consider less likely other explanations alternative to the jet origin for the observed line velocity width enhancement for the following reasons: 
\textit{(i)} We exclude the possibility of beam smearing because the scale on which we and other authors observe the line-width enhancement is much larger than the spatial resolution of the observations.
\textit{(ii)} Some of the above works interpreted the observed feature as due to an equatorial outflow, predicted by some models to originate from the BH accretion disc (e.g. \citealt{Li:2013ab}) or from the dusty torus (e.g. \citealt{Elitzur:2006aa}, \citealt{Mor:2009aa}, \citealt{Elitzur:2012aa}).
We do not exclude that the line-width enhancement observed in our sample and in the other cases we mentioned might also be compatible with an outflow launched radially in the equatorial plane at the base of the jet with a certain opening angle. Projection effects may easily broaden the line profiles, producing the observed line-width enhancement, although we note that a net blue- or red-shifted velocity is only occasionally measured in the high velocity width regions. It is also possible that such an equatorial outflow may interact with the galaxy ISM, losing its speed while promoting turbulence within the disc. However, we suggest that even in this scenario an origin from a jet-ISM interaction producing such equatorial gas flow has to be preferred. The phenomenon of enhanced line velocity widths perpendicular to the AGN ionisation cones is indeed to our knowledge exclusively observed in galaxies hosting a radio jet interacting with the disc, as discussed before.
\textit{(iii)} Finally, the multi-direction outflow scenario due to jet precession seems unlikely because in this case random outflow directions would be expected, while the observed features are systematically perpendicular. Moreover, we stress that the canonical outflow (with high bulk velocity) observed in the direction of the AGN ionisation cones and the material in the perpendicular direction show completely different kinematic properties that point to a different origin of the two, the former being characterised by a coherent velocity field, the latter instead dominated by velocity dispersion and not by a definite net velocity.

Based on all these considerations, 
we consider that the interaction of the jet propagating through the galaxy disc ISM is more likely responsible for the observed phenomenon. 
Unfortunately, the physical details of how the jet could give rise to the observed perpendicular enhanced emission line widths cannot be explained solely through the presented observational data, and thus a complete description of this phenomenon goes beyond the scope of this work.

Finally, as mentioned, in some objects 
the enhanced perpendicular line velocity width is observed not only in the ionised gas, but also in the molecular gas (\citealt{Shimizu:2019aa} and \citealt{Feruglio:2020aa} in 
CO, \citealt{Riffel:2015aa} and \citealt{Diniz:2015aa} in H$_2$), although with lower values than in the ionised phase (a factor 3--8 for CO and 2--3 for H$_2$). We might then speculate that the turbulence and perturbations induced perpendicularly by the radio jet in its propagation through the disc more strongly affect the ionised phase than the denser molecular phase, although additional molecular gas observations are required to assess this issue.



\subsection{Ionised gas mass affected by the phenomenon}

\begin{table*}[tp]
        \caption{Ionised gas mass, kinetic energy, visual extinction, and electron density in the region of enhanced line velocity width (\oiii\ W70\,$>$\,300 km/s), as well as spatial extension of the W70 enhancement (total, not per side), for the four galaxies presented in this work, compared to the length (total, not per side) and power of their jets.}
        \setlength{\extrarowheight}{0.5pt}
        \begin{tabular*}{\textwidth}{l @{\extracolsep{\fill}} c c c c c c c}
                \hline\hline \\[-1em]
                Galaxy name                             &  $M_\mathrm{ion}$  & $M_\mathrm{ion} \sigma_\mathrm{ion}^2/2$  & $A_V$  & $n_\mathrm{e}$  & $R_\mathrm{ion}$ & $R_\mathrm{jet}$  & $P_\mathrm{jet}$  \tablefootmark{(a)}  \\
                    & [10$^6$ M$_\odot$] & [10$^{54}$ erg] & [mag] & [cm$^{-3}$] & [kpc] & [kpc] & [10$^{43}$ erg s$^{-1}$] \\[0.2em]
                \hline \\[-0.8em]
                {\bf IC 5063} & 1.5$_{-0.9}^{+3.0}$ & 0.6$_{-0.2}^{+1.9}$ & 1.7$_{-0.7}^{+0.3}$ & 460$_{-290}^{+670}$ & 7 & 1 & 4.6$_{-1.0}^{+1.3}$ \\[0.4em]
                {\bf NGC 5643} & 0.16$_{-0.05}^{+0.60}$ & 0.07$_{-0.02}^{+0.21}$ & 1.2$_{-0.4}^{+0.7}$ & 470$_{-330}^{+410}$ & 3 & 2 & 1.0$_{-0.4}^{+0.6}$  \\[0.4em]
                {\bf NGC 1068} & 1.5$_{-1.1}^{+6.7}$ & 3$_{-2}^{+8}$ & 0.8$_{-0.3}^{+0.6}$ & 600$_{-500}^{+1800}$ & 1.5 & 0.8 & \hfill 1.8$_{-0.5}^{+0.8}$ \: \tablefootmark{(c)} \\[0.4em]
                {\bf NGC 1386} & 0.07$_{-0.04}^{+0.20}$ & 0.05$_{-0.02}^{+0.09}$ & 0.9$_{-0.4}^{+0.2}$ & 470$_{-340}^{+490}$ & 1.5 & 0.08 & 0.8$_{-0.3}^{+0.5}$ \\[0.4em]
                \hline
        \end{tabular*}
    \tablefoot{
    \tablefoottext{a}{Calculated from the radio power of the jet with the \cite{Birzan:2008aa} relation.}
    \tablefoottext{c}{\cite{Garcia-Burillo:2014aa}.}
        }
\label{table:mass}
\end{table*}

\begin{figure*}
    \centering
    \includegraphics[width=0.8\textwidth,trim={0 9.5cm 0 0},clip]{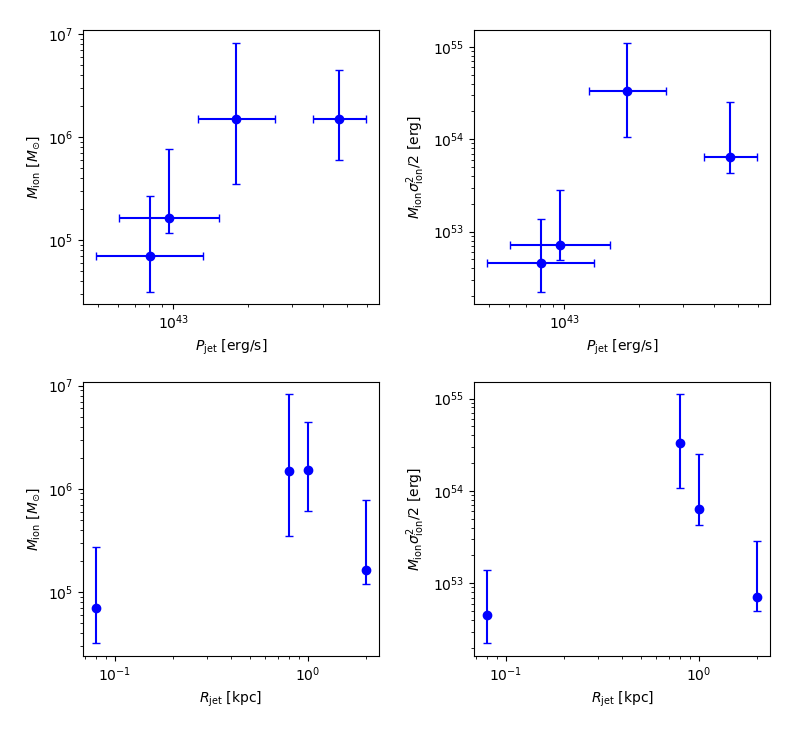}
    \caption{Ionised gas mass $M_\mathrm{ion}$ (left) and gas kinetic energy $M_\mathrm{ion} \sigma_\mathrm{ion}^2/2$ (right) in the region affected by the \oiii\ W70 enhancement ($>\,$300 km/s) vs. jet power $P_\mathrm{jet}$ 
    for the four objects presented, as reported in Table \ref{table:mass}. 
    The jet power resulting from simulations of jet-ISM interactions could be an order of magnitude higher than that reported, obtained from empirical relations (\citealt{Birzan:2008aa}), based on the results from simulations in \cite{Mukherjee:2018aa} for IC 5063. 
    }
    \label{fig:plots_pjet}
\end{figure*}

We estimated the mass of ionised gas that is affected by the phenomenon of velocity width enhancement in the galaxies we presented. We employed \ha\ instead of \oiii\ to obtain the mass because its luminosity does not depend on gas metallicity and on the energy of the ionising photons (e.g. \citealt{Carniani:2015aa}).
We extracted the integrated H$\alpha$ modelled flux from our MUSE maps from the regions with an \oiii\ W70$\,>\,$300~km/s (and a S/N$\,>\,$3) and calculated its luminosity considering the corresponding distances reported in Table \ref{table:fov}.
We corrected the luminosity for extinction by employing a \cite{Calzetti:2000aa} attenuation law for galactic diffuse ISM ($R_V$ = 3.12) and an intrinsic ratio (\ha/\hb)$_0$ = 2.86 (for an electron temperature of $T_\mathrm{e}$ = 10$^4$~K; \citealt{Osterbrock:2006aa}).
We finally calculated the mass of ionised gas through the following relation from \cite{Cresci:2017aa}, which assumes case B recombination in fully ionised gas with electron temperature $T_\mathrm{e}$ = 10$^4$~K,
\begin{equation}
    M_\mathrm{ion}/\mathrm{M}_\odot = 3.2 \times 10^5   \left( \dfrac{L_{\mathrm{H}\alpha}}{10^{40}~\mathrm{erg~s}^{-1}} \right)  \left( \dfrac{n_\mathrm{e}}{100~\mathrm{cm}^{-3}} \right)^{-1} .
\end{equation}
The electron density $n_\mathrm{e}$ was obtained from the \sii\ $\lambda$6716/$\lambda$6731 diagnostic line ratio (\citealt{Osterbrock:2006aa}; still from the spaxels with \oiii\ W70$\,>\,$300~km/s and S/N$\,>\,$3 on the \sii\ lines), assuming a typical value for the temperature of ionised gas of $T_\mathrm{e}$ = 10$^4$~K.
Table \ref{table:mass} reports the mass of ionised gas that we find in the region of line-width enhancement for the four galaxies we analysed in this work. The changes in extinction and electron densities within the considered regions are included in the respective uncertainties on these quantities.
We note that the values we calculated may be considered as upper limits to the ionised gas mass that is affected by the jet perpendicular perturbation because the flux extracted from the integrated line profile may include contributions from unperturbed gas or from gas partaking in the standard high-velocity outflow in the direction of the jet and ionisation cones.
We add that other methods for inferring the ionised gas density, such as that exploiting auroral and transauroral lines or the \cite{Baron:2019aa} method based on the ionisation parameter, give higher electron densities by up to an order of magnitude than those obtained through the \sii\ doublet ratio method we employed (\citealt{Davies:2020aa}). When these alternative methods are used, the resulting masses would be up to an order of magnitude lower than those we obtain. However, as indicated by the high \sii/\ha\ line ratios observed, the gas in the region perpendicular to the jet is characterised by low ionisation and the \sii-ratio method likely traces the ionised gas density properly.

We also estimated the kinetic energy of the gas in the same region, $M_\mathrm{ion} \sigma_\mathrm{ion}^2/2$, by employing the \oiii\ velocity dispersion $\sigma_\mathrm{ion}$ in the enhancement region.
We note that the kinetic energy determined using the \ha\ line is marginally different (because its W70 is not identical to that of \oiii, see Fig. A1 in the appendix), but still consistent with the uncertainties.

We compared the inferred masses and kinetic energies with the extension and power of the jet to determine if any correlation exists between them.
The jet power, when not found in literature, was calculated by employing Eq. 16 from \cite{Birzan:2008aa}, which relates the cavity (jet) power and the 1400 MHz radio luminosity of the source in objects showing cavities in their X-ray haloes filled by radio emission.
For NGC 5643 we considered the radio power $P_\mathrm{8.4~GHz}$ $\sim$ 5.5$\,\times\,$10$^{20}$~W~Hz$^{-1}$ given in \cite{leipski2006} and for NGC 1386 the flux density $S_\mathrm{8.4~GHz}$ $\sim$ 10.3 mJy from \cite{mundell2009} (corresponding to $P_\mathrm{8.4~GHz}$ $\sim$ 3.3$\,\times\,$10$^{20}$~W~Hz$^{-1}$ at the distance of the galaxy). We assumed a power-law spectrum ($\propto \! \nu^{-\alpha}$) with $\alpha$~=~1 to rescale to the 1.4 GHz power involved in the \cite{Birzan:2008aa} relation. 
For NGC 1068, we employed the value given by \cite{Garcia-Burillo:2014aa}, who also employed the \cite{Birzan:2008aa} relation. 
For consistency, we also calculated the jet power for IC 5063 using the \cite{Birzan:2008aa} relation, finding $P_\mathrm{jet}$ $\sim$ 5$\,\times\,$10$^{43}$ erg/s, although we note that higher values are reported by \cite{Morganti:2015aa} (5$-$9$\,\times\,$10$^{43}$ erg/s) and \cite{Mukherjee:2018aa} (between 10$^{44}$ and 10$^{45}$ erg/s), the latter from jet-ISM interaction simulations. 
As commented in \cite{Mukherjee:2018aa}, the values obtained from simulations for IC 5063 are about an order of magnitude higher than those obtained from empirical relations (i.e. \citealt{Birzan:2008aa}, \citealt{Cavagnolo:2010aa}) between radio power and cavity power derived for classical evolved radio jets in haloes of galaxies or clusters, which may not apply to jets propagating into the ISM of a galactic disc.
Based on this, we stress that in addition to IC 5063, the jet powers calculated for the other three galaxies, reported in Table \ref{table:mass} and Fig. \ref{fig:plots_pjet}, might also be an order of magnitude higher when resulting from jet-ISM simulations.

Although the values reported in Table \ref{table:mass} can be considered only as indicative as pointed out above, we note that higher masses and kinetic energies of the ionised gas in the line-width enhancement region are roughly associated with more powerful jets (Fig. \ref{fig:plots_pjet}), suggesting that more powerful jets are able to affect larger quantities of ISM and reinforcing the possibility that jets are responsible for the observed phenomenon. The same would hold also for the jet length if we excluded NGC 5643, which exhibits a longer jet given its power compared to the other targets. 

In order to infer whether the jet is energetic enough to power the observed features, we must compare the total kinetic energy, $E_\mathrm{jet}$, produced during its travelling time, $t_\mathrm{jet}$, with the kinetic energy of the material affected by the line-width enhancement, $M_\mathrm{ion} \sigma_\mathrm{ion}^2/2$. By assuming that the currently measured jet power $P_\mathrm{jet}$ is representative for its mean value over its travelling time, we have $E_\mathrm{jet}$ = $P_\mathrm{jet} t_\mathrm{jet}$.
\cite{Mukherjee:2018ab} estimated a jet travelling time of $\sim\,$0.4 Myr for the case of IC 5063. 
Using Eq. A1 from \cite{Mukherjee:2018ab} and the same parameters they employed for IC 5063\footnote{$\zeta$ = $10^{-6}$, ratio between jet density and density of the ambient medium into which the jet is propagating; $\chi$ = 4.8, jet proper density parameter (ratio of jet rest mass energy to enthalpy); $\Gamma$ = 4, jet bulk Lorentz factor.}, we can estimate the jet travelling time for the other remaining three objects in our sample. By considering the jet lengths given in Table \ref{table:mass} (divided by 2 to obtain the distance travelled by the jet per side), we obtain $t_\mathrm{jet}$ $\sim$ 0.8, 0.3 and 0.03 Myr for NGC 5643, NGC 1068, and NGC 1386, respectively. We stress that different values from those adopted for IC 5063 for the quantities involved in the equation may apply to these other three objects. 

By dividing the kinetic energy of the line-width-enhanced perpendicular material, $M_\mathrm{ion} \sigma_\mathrm{ion}^2/2$, by $P_\mathrm{jet} t_\mathrm{jet}$, we found values far lower than 1, in the range 10$^{-4}$--10$^{-2}$. This indicates that the jets 
are easily capable, even with a low efficiency of energy transfer, to inject the required energy into the ISM.
An even lower efficiency would be needed if the jet powers are one order of magnitude larger than those considered, which are derived from empirical relations, 
and/or if the \sii-ratio method employed underestimates the gas density (and thus overestimates its kinetic energy), as discussed before.

\section{Conclusions}
We presented flux, kinematics, and excitation (BPT) maps of the ionised gas of the nearby Seyfert galaxies IC 5063, NGC 5643, NGC 1068, and NGC 1386 obtained with the optical and near-IR integral field spectrograph MUSE at the VLT as part of our MAGNUM survey.
All these galaxies host a low kinetic power ($\lesssim$\,10$^{44}$ erg~s$^{-1}$) radio jet on scales $\lesssim\,$1 kpc aligned with the AGN ionisation cones axis, which has low inclinations with respect to the galaxy disc ($\sim$45$\degree$ at most) and shows evidence of interaction with the disc ISM.
The results of the work are summarised below.

We find that the bulk of the high-velocity gas (in the range $\pm$|200--1000| km/s) is directed as the jet and AGN ionisation cones, as expected for outflows in Seyfert galaxies.
However, we detect a strong (up to W70$\,\gtrsim\,$800$-$1000 km/s) and extended ($\gtrsim\,$1 kpc) emission-line velocity width enhancement perpendicular to the direction of the AGN ionisation cones and jets, with fairly symmetric line profiles and without a coherent velocity shift on each side of the nucleus.
Moreover, we find that the excitation of the gas in this perpendicular region is consistent with the presence of shocks, that together with the broadness of the line profiles might be associated with turbulent gas.
Other recent works observed the same peculiar phenomenon of enhanced line widths perpendicular to ionisation cones and jets in nearby Seyferts that host low-power jets showing evidence of interaction with the galaxy disc ISM. 

We consider the interaction between the jet and the galaxy disc, perturbing the disc material during the jet propagation through it, as the most likely origin for the observed phenomenon. 
We favour this over alternative proposed interpretations, such as beam smearing, equatorial outflows from the accretion disc or the dusty torus, and multi-direction outflows due to jet precession for the following reasons: 

First, the perpendicular extended line velocity width enhancement is observed exclusively in galaxies hosting a (low-power) jet whose inclination happens to be low enough over the galaxy disc to have significant interaction with its ISM, as also indicated by recent simulations. This suggests that the jets are likely responsible for the observed phenomenon.

Second, the scales on which the phenomenon occurs ($\gtrsim\,$1 kpc, i.e. several arcsec) are well resolved by MUSE, which excludes beam smearing.

Third, the very broad line profiles might be compatible with an outflow launched in the equatorial plane with a wide angle, considering projection effects, although a high net blue- or red-shifted velocity is only occasionally measured in the regions with high velocity width. However, even in this case, we favour an origin from jet-ISM interaction to produce this equatorial gas flow rather than an accretion disc or torus wind. The phenomenon of enhanced line velocity widths perpendicular to the AGN ionisation cones is indeed to our knowledge exclusively observed in galaxies hosting a radio jet interacting with the disc.
    
Fourth, the observed enhanced line velocity widths are systematically (roughly) perpendicular to the high-velocity ionisation-cone outflow, which excludes the multi-direction outflow scenario caused by jet precession because random outflow directions would be expected in this case.
    Furthermore, the different kinematic properties of the gas in the two directions, that is, broad and symmetric in one case and narrower, asymmetric and with a spatially coherent velocity in the other, also disfavours a common origin.

We find that the jets are powerful enough to provide the kinetic energy of the ionised gas observed in the line-width enhancement region and that higher masses and kinetic energies of the line-width-enhanced gas tend to be associated with more powerful jets. 
Our results demonstrate that low-power jets are capable of affecting the host galaxy, in line with current cosmological simulations and recent observational works.
However, a larger sample with high-quality (MUSE-like) integral field spectroscopic data
would be needed to identify more sources showing the phenomenon presented in this work and to better investigate the above trends.
A similar study focused on molecular gas would also be needed to assess to which extent the phenomenon affects the molecular phase, given that a few works have reported enhanced line velocity widths perpendicular to AGN ionisation cones and jets even in H$_2$ and CO.

\begin{acknowledgements}
We acknowledge support from ANID programs FONDECYT Postdoctorado 3200802 (G.V.), Basal-CATA AFB-170002 (G.V. and E.T.), FONDECYT Regular 1190818 (E.T.) and Anillo ACT172033 (E.T.).
G.C. and Al.M. acknowledge support from PRIN MIUR project ``Black Hole winds and the Baryon Life Cycle of Galaxies: the stone-guest at the galaxy evolution supper'', contract $\#$2017PH3WAT.
M.P. is supported by the Programa Atracción de Talento de la Comunidad de Madrid via grant 2018-T2/TIC-11715. 
R.M. acknowledges ERC Advanced Grant 695671 ``QUENCH'' and support by the Science and Technology Facilities Council (STFC).
This research has made use of the services of the ESO Science Archive Facility.
The scientific results reported in this article are based in part on data obtained from the Chandra Data Archive (ObsID: 344).
This research has made use of software provided by the Chandra X-ray Center (CXC) in the application package CIAO.
This research has made use of NASA’s Astrophysics Data System Bibliographic Services.
This research has made use of the NASA/IPAC Extragalactic Database (NED), which is funded by the National Aeronautics and Space Administration and operated by the California Institute of Technology.
We acknowledge the usage of the HyperLeda database (\url{http://leda.univ-lyon1.fr}; \citealt{Makarov:2014aa}).
This research has made use of ``Aladin sky atlas'' developed at CDS, Strasbourg Observatory, France (\citealt{Bonnarel:2000aa}).
This research made use of the Python packages SciPy \citep{scipy2020}, NumPy \citep{numpy2020}, IPython \citep{ipython2007} and Matplotlib \citep{matplotlib2007}.
This research made use of Astropy (\url{http://www.astropy.org}), a community-developed core Python package for Astronomy \citep{astropy:2013, astropy:2018}.
The National Radio Astronomy Observatory is a facility of the National Science Foundation operated under cooperative agreement by Associated Universities, Inc.
The Australia Telescope Compact Array is part of the Australia Telescope National Facility which is funded by the Australian Government for operation as a National Facility managed by CSIRO.
We thank Raffaella Morganti for the fruitful discussion on this work and for having provided the ATCA radio data of IC 5063. We thank Dipanjan Mukherjee for the useful insights regarding jet-ISM simulations. We thank Tanya Urrutia for her contribution to the earliest work on the MAGNUM project.
\end{acknowledgements}

\bibliographystyle{aa}
\bibliography{linewidth_perpend_jets_langedit_v2_PROD_arxiv.bib}

\begin{appendix}
\section{\ha\ W70 maps}
\begin{figure*}[b]
    \centering
    \hfill
    \includegraphics[width=0.365\textwidth,trim={2.6cm 0.5cm 2cm 0.5cm},clip]{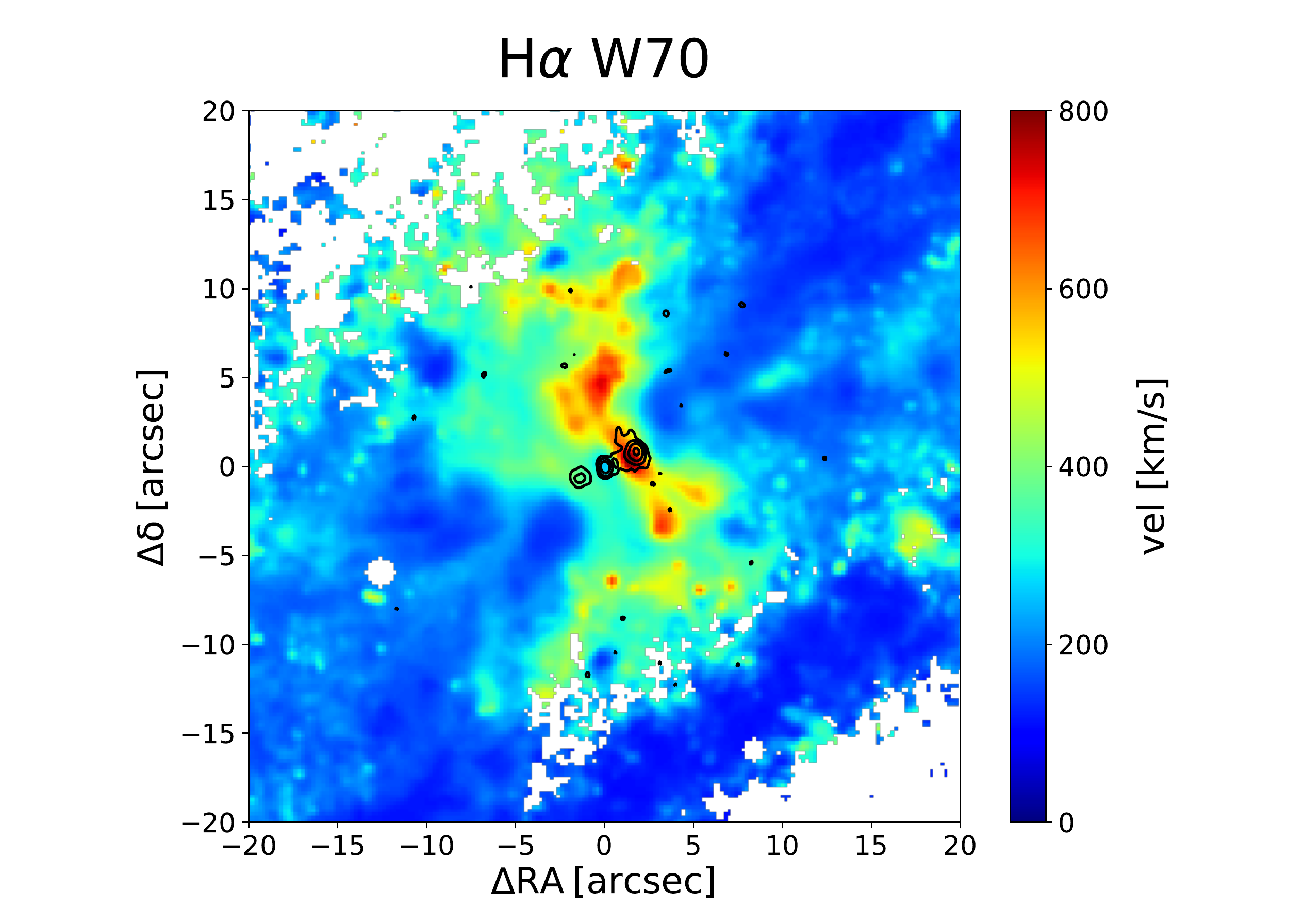}
    \hfill
    \includegraphics[width=0.365\textwidth,trim={2.6cm 0.5cm 2cm 0.5cm},clip]{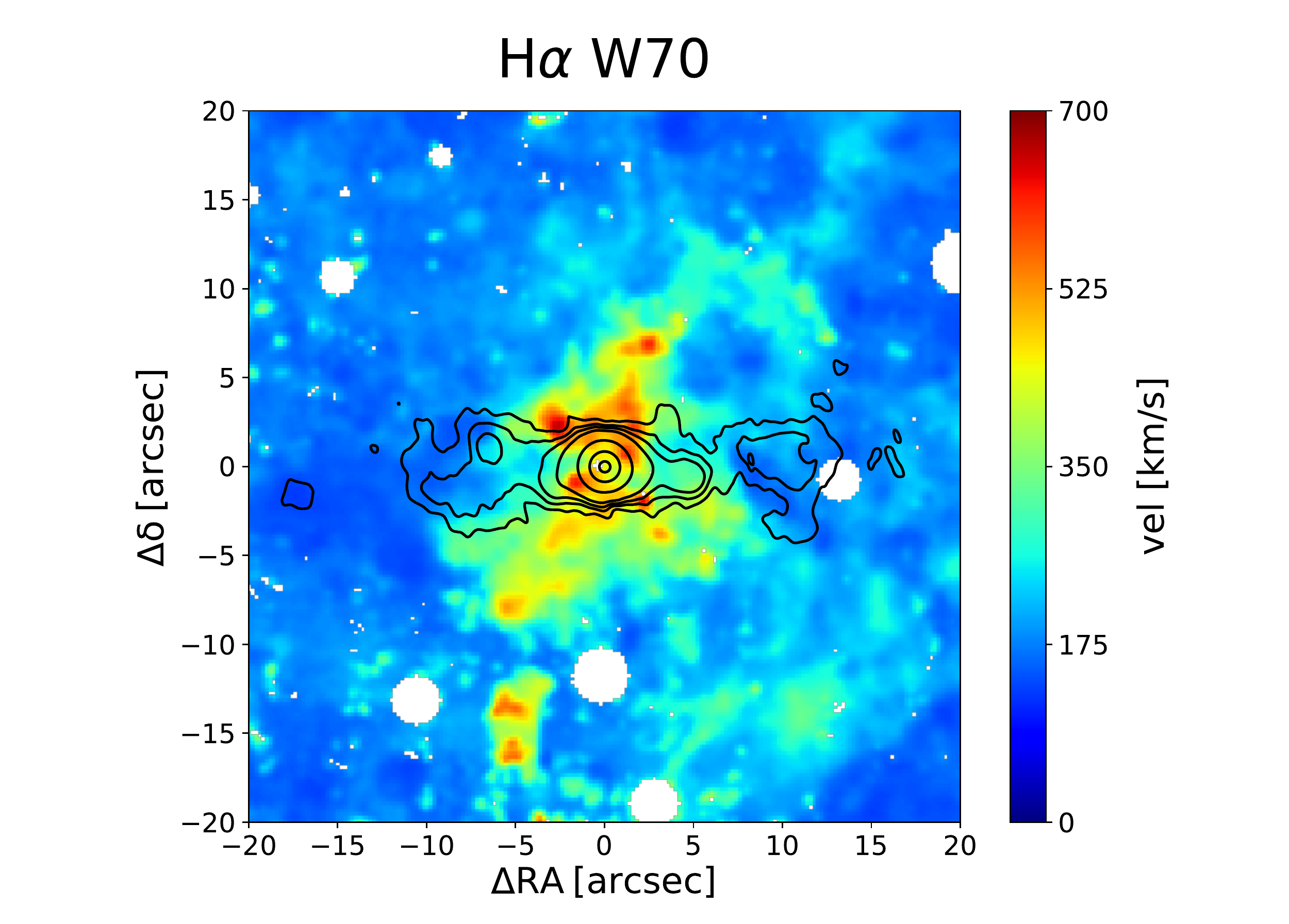}
    \hfill\null\\
    \vspace*{\floatsep}
    \hfill
    \includegraphics[width=0.365\textwidth,trim={2.6cm 0.5cm 2cm 0.5cm},clip]{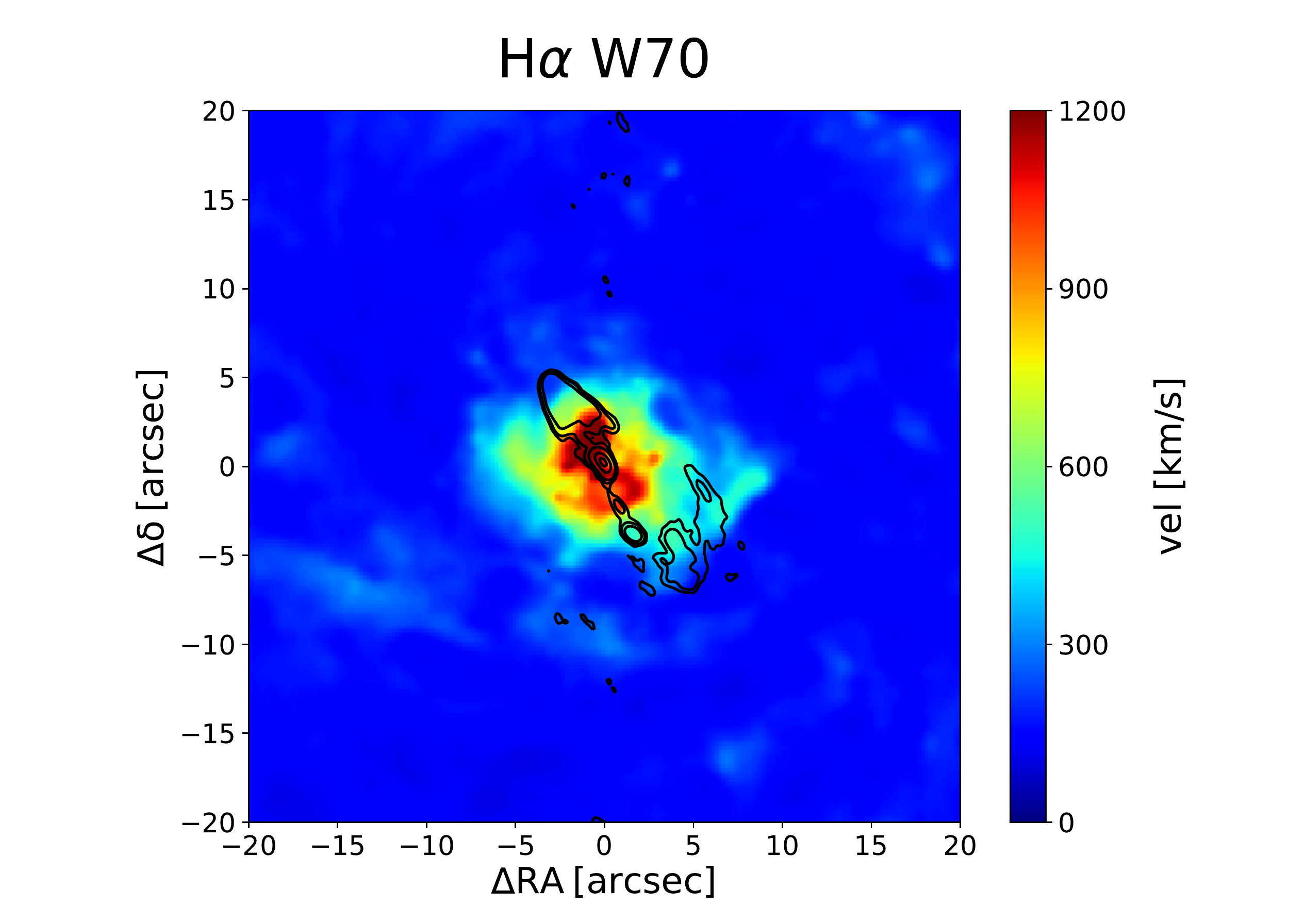}
    \hfill
    \includegraphics[width=0.365\textwidth,trim={2.6cm 0.5cm 2cm 0.5cm},clip]{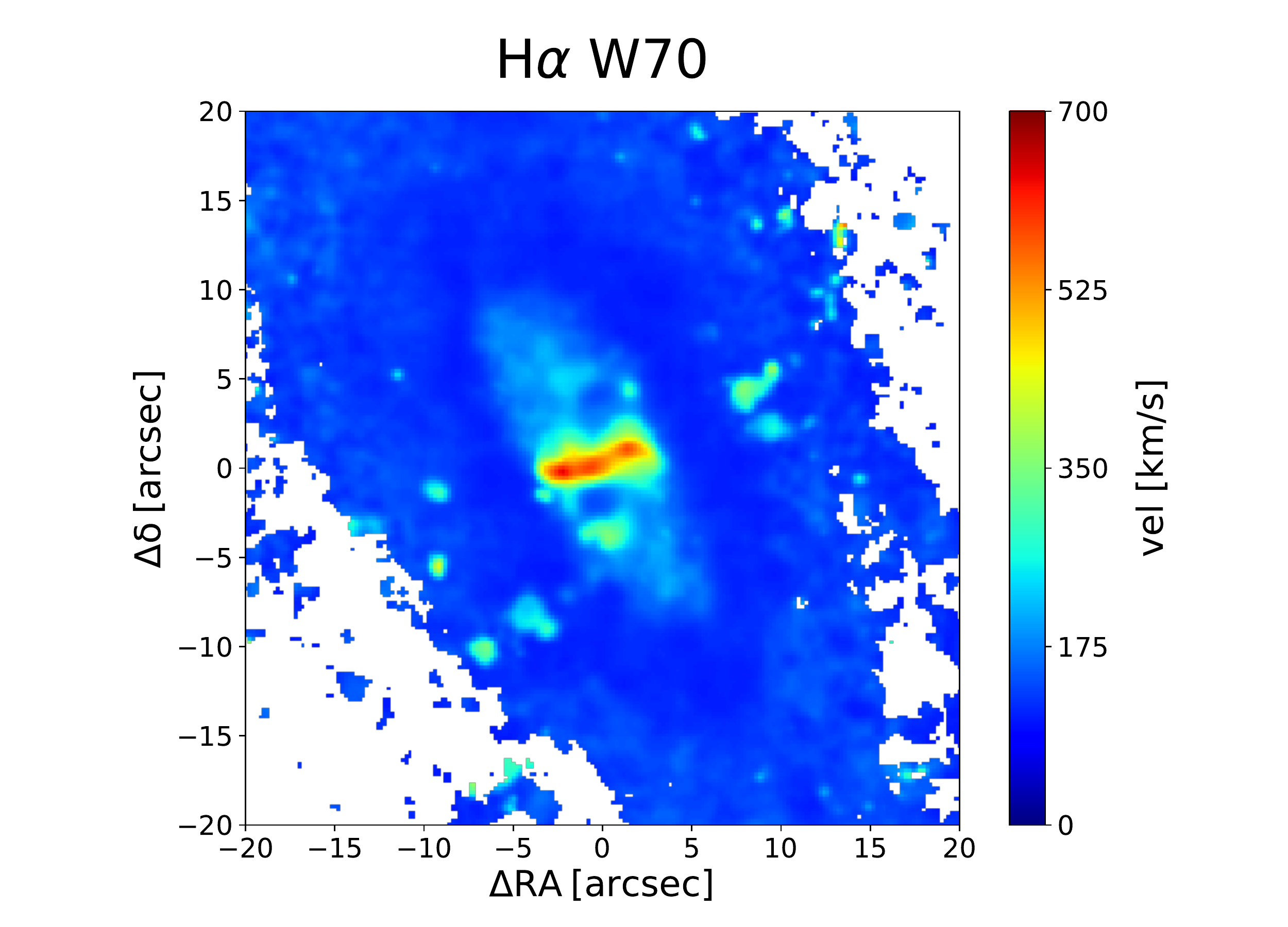}
    \hfill\null
    
    \caption{H$\alpha$ W70 of IC 5063, NGC 5643, NGC 1068, and NGC 1386 (from top left to bottom right). The radio contours are superimposed (see main text for more details on reported radio data). The maps are zoomed in the central 40$''$\,$\times$\,40$''$ as in the \oiii\ W70 maps of each respective galaxy (Figs. \ref{fig:maps_ic5063_2}b, \ref{fig:maps_n5643_2}b, \ref{fig:maps_n1068_2}b, and \ref{fig:maps_n1386_2}b).}
    \label{fig:w70_ha}
\end{figure*}

\begin{figure*}
    \centering
    \includegraphics[width=0.365\textwidth,trim={2.4cm 0.5cm 2cm 0.5},clip]{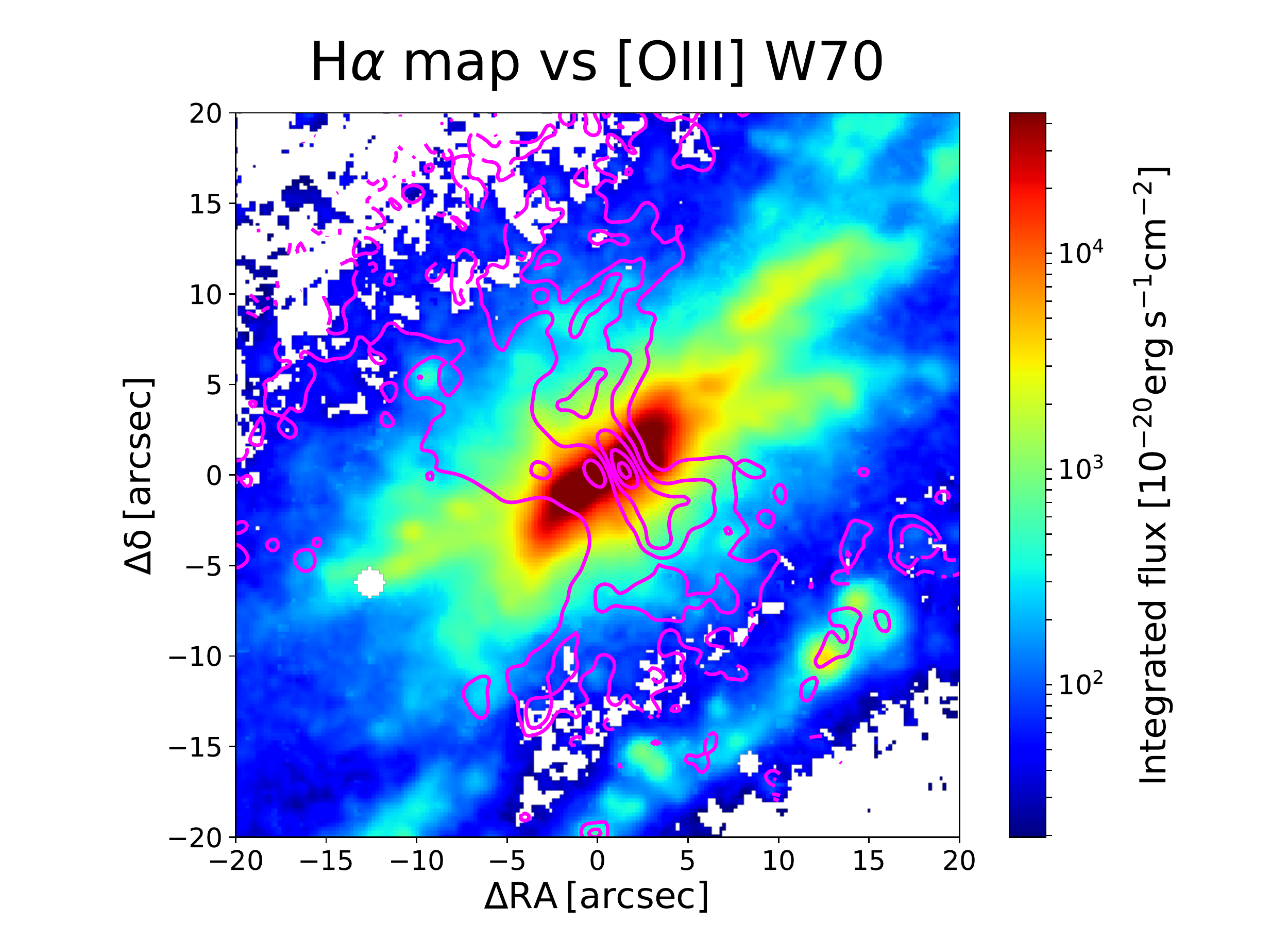}
    \caption{Map of \ha\ emission of IC 5063. The \oiii\ W70 contours are superimposed. The map shows that the stripe of \ha\ star-forming (as indicated by the BPT diagram in Fig. \ref{fig:maps_ic5063_2}c,d) clumps to the SW of the nucleus are in the direction of the high-W70 material, approximately at its edge.}
    \label{fig:ic5063_posfeedb}
\end{figure*}

In Fig. \ref{fig:w70_ha} we report the MUSE maps of the \ha\ W70 for the four galaxies presented.
We also display in Fig. \ref{fig:ic5063_posfeedb} the \ha\ emission of IC 5063 (also reported in the three-colour image Fig. \ref{fig:maps_ic5063_1}b). The \oiii\ W70 contours (from Fig. \ref{fig:maps_ic5063_2}b) are superimposed. This shows that the stripe of star-forming \ha\ clumps to the SW of the nucleus is located at the edge of the W70 enhancement, along its axis. As mentioned in Sect. \ref{ssec:ic5063}, this may represent a candidate of star formation induced by the high-W70 material (positive feedback) and requires further investigation.

\end{appendix}

\end{document}